\documentclass{svproc}
\usepackage{url}
\usepackage{subcaption}

\usepackage{algorithm}
\usepackage{algpseudocode}
\usepackage{mathtools}
\usepackage[symbol]{footmisc}
\newcommand{\R}{\ensuremath{\mathbb{R}}}
\usepackage{placeins}
\usepackage{pxfonts}
\usepackage{cite}
\usepackage{hyperref}

\begin{document}
\mainmatter              
\title{InfluenceNet: AI Models for Banzhaf and Shapley Value Prediction}
\titlerunning{InfluenceNet}  
%
\author{Benjamin Kempinski\inst{1,2} \and Tal Kachman\inst{1,2,}}
\authorrunning{Kempinski B. et al.} 
%
\tocauthor{Benjamin Kempinski, Tal Kachman}
\institute{Radboud university, Nijmegen,  Netherlands
\and
Donders Institute for Machine Learning and Neural Computing, Nijmegen, Netherlands}

\maketitle              

\begin{abstract}
Power indices are essential in assessing the contribution and influence of individual agents in multi-agent systems, providing crucial insights into collaborative dynamics and decision-making processes. While invaluable, traditional computational methods for exact or estimated power indices values require significant time and computational constraints, especially for large $(n\ge10)$ coalitions. These constraints have historically limited researchers' ability to analyse complex multi-agent interactions comprehensively. To address this limitation, we introduce a novel Neural Networks-based approach that efficiently estimates power indices for voting games, demonstrating comparable and often superiour performance to existing tools in terms of both speed and accuracy. This method not only addresses existing computational bottlenecks, but also enables rapid analysis of large coalitions, opening new avenues for multi-agent system research by overcoming previous computational limitations and providing researchers with a more accessible, scalable analytical tool.This increased efficiency will allow for the analysis of more complex and realistic multi-agent scenarios.

\keywords{Game Theory, Voting games, Power indices, Neural Networks}
\end{abstract}
\section{Introduction}
Power indices are fundamental tools in cooperative game theory and political science, providing crucial insights into the distribution of influence within complex decision-making systems~\cite{979e58a8-277d-347d-b16a-71b264c25a0d,WINTER20022025}. From political coalitions and corporate boardrooms to collaborative networks and distributed systems, these mathematical measures, particularly the Shapley-Shubik and Banzhaf power indices, help quantify the relative voting power of the individual participants. By analysing these indices, researchers can identify key decision-makers, understand intricate power dynamics, and optimize voting structures for enhanced fairness and efficiency.

However, the practical application of power indices has long been hindered by significant computational challenges. Traditional calculation methods exhibit exponential complexity~\cite{0c1bb2a2-f3c9-32ac-96d1-731cd3f439a9,10.1007/s10458-009-9078-9}, making exact computations impracticable for large-scale systems.\footnote[2]{For example: given a simple binary voting system the number of possible coalitions is equal to $2^n$. So, while such a system of 10 agents would have $2^{10}=1,024$ potential coalitions, the same system with 20 agents would have $2^{20}=1,048,576$ different potential coalitions.} While researchers have developed approximation methods like Monte-Carlo approximations~\cite{10.1007/s10458-009-9078-9} and Maximum Sample Reuse ~\cite{pmlr-v206-wang23e}, these approaches still struggle to balance computational efficiency with accuracy when analysing complex, real-world systems.

Recent advances in machine learning, particularly in Neural Networks (NN), present promising opportunities to overcome these computational barriers. NN have demonstrated remarkable success in approximating complex functions and identifying patterns in high-dimensional spaces across various domains~\cite{2500c2e5-e926-3cfe-83d3-bbb49b56b64e,8286426,SCHMIDHUBER201585}. However, their application to power index calculation presents unique challenges, as NN traditionally show limitations when processing tabular data~\cite{Shwartz-Ziv,grinsztajn2022}. This tension between potential utility and known limitations raises essential questions about the viability of NN in this domain.

In this work, we demonstrate that NN can effectively approximate both Banzhaf and Shapley-Shubik power index values for large-scale coalitions $(n\ge10)$, while significantly reducing the computational requirements compared to existing methods. Our approach, inspired by Leong and Shoarm's Marginal Contribution Networks~\cite{Marginal_Contribution_Nets}, not only overcomes NN's traditional limitations in handling tabular data but also provides robust and reliable estimates of power indices in previously intractable scenarios.

{\bf Our contributions}: We introduce how a simple NN architecture can perform power index approximation, demonstrating its effectiveness in handling large-scale coalition voting systems. We further provide comprehensive empirical evidence showing that our approach achieves superior computational efficiency while maintaining acceptable accuracy levels, with explicit characterization of bias tendencies and their limitations. Finally, we present a practical framework that enables researchers to analyse and evaluate large coalition systems that were previously computationally prohibitive, opening new avenues for understanding complex collaborative dynamics. This work significantly expands the practical applicability of power indices in analysing real-world voting systems and cooperative arrangements. By providing efficient and reliable computational tools, we enable researchers and practitioners to better understand and optimize decision-making processes in large-scale collaborative environments.

\section{Related Work}

We investigae using neural networks to predict power indices, including Shapley values and Banzhaf indices, in the marginal contribution network class of coalitional games~\cite{ieong2005marginal,elkind2009tractable}. Game theory studies interactions and teams between participants~\cite{arin2004cooperation,aumann1988value,albers1979example} with cooperative game theory examining how teams of players may cooperate to achieve certain outcomes~\cite{owen2013game,fudenberg1991game,maschler2020game,ambrus2009coalition,chalkiadakis2011computational}, with many applications~\cite{driessen2013cooperative}, including security~\cite{agah2006security}, network analysis~\cite{aditya2021coalitional,suri2008determining,narayanam2010shapley,bachrach2013sharing,michalak2013efficient,diazortiz2023usingcooperativegametheory,daphnee_2024}, voting~\cite{fatima2007randomized,bachrach2016analyzing,zick2011shapley}, logics and reasoning~\cite{karmakar2024expected,bachrach2013proof,livshits2021shapley} robotics~\cite{remman2021robotic,banarse2019body,aslam2024cooperative,martin2023multi}, data selection and valuation~\cite{ghorbani2019data,patel2021game}, market analysis~\cite{shapley1969market,bachrach2010honor,acuna2018cooperation,lewenberg2015bitcoin,wolak1988measuring,blocq2014shared,aminadav2011rebuilding}, interpretability~\cite{rozemberczki2022shapley} team formation~\cite{sie2009influence,gaunt2016training,meulbroek2019forming,mash2017form,bachrach2020negotiating}. The concept of \textit{power index} quantifies the importance of each player~\cite{shapley1953value,banzhaf1965weighted,deegan1978new,straffin1988shapley}.  Calculating these indices exactly is computationally expensive in large games~\cite{conitzer2004computing,bachrach2009power,michalak2013efficient}, motivating approximation algorithms\cite{fatima2008linear,castiglione2018efficient, deng2019approximation}. Some early work explored using machine learning for related tasks like coalition formation \cite{eichberger2021machine}, with some earlier research on using neural networks to predict power indices in very specific games~\cite{cornelisse2022neural}.

We focus on the prominent family of marginal contribution networks~\cite{bach2019marginal}. There are many classes of coalitional games, that model different relations between players. Graph games represent interactions between players positioned on a graph \cite{jackson2008economic}, with whose structure defines which players can directly influence each other. More restricted are Network Flow Games~\cite{granot1992some,resnick2009cost,reijnierse1996simple,bachrach2007computing,aziz2010monotone,potters2006nucleolus} which involve players controlling edges or nodes in a network, where the value of a coalition is related to the maximum flow that can be achieved through the network given the players' resources \cite{shapley1955method}. Similarly, path interdiction games model scenarios where players try to disrupt paths in a network by removing edges or nodes, so the value of a coalition  represents the damage they can inflict or the connectivity they can maintain \cite{wood1993deterministic}. In contrast, in Marginal Contribution Networks (MCNs) \cite{bach2019marginal} the value of a coalition is determined by the marginal contributions of its members, considering their connections in the network. This framework is particularly relevant as it captures how a player's influence depends on their local network structure. Some interesting cases are influence games \cite{aral2010viral}, where players try to maximize their spread in a social network, and games on hypergraphs \cite{breuer2019power} that  allow for higher-order interactions. 

\section{Notations and preliminary}
We describe basic concepts from game theory in this work, which are required for the rest of the paper\cite{10.1007/s10458-009-9078-9, Marginal_Contribution_Nets}.

\subsection{Marginal contribution games}
Let $G = (R, L)$ be a coalition game where $L = {1, \ldots, m}$ is the set of agents and $R = {r_1, \ldots, r_n}$ is a set of rules. Each rule $r_i \in R$ takes the form:
\begin{center}
$r_i: Pattern_i \rightarrow v_i$
\end{center}
Where $v_i \in \mathbb{R}$ represents the value assigned to a coalition matching $Pattern_i$. A coalition $C \subseteq L$ of agents that follow the rule pattern receives its value $v_i$. The total value of a group of agents is defined to be the sum over the values of all the rules that the coalition fills. The coalition rules are composed of two parts: positive requirements ("must-have") $req$, representing agents who are required to take part in the coalition, and negative requirements ("must not have") $ban$, representing agents banned from taking part in the coalition~\cite{GRECO201719,Marginal_Contribution_Nets}.
For example, given the following rule set: 
\begin{center}
$\{ a \wedge b \} \rightarrow 3$ \\
$\{ a \wedge \neg c \} \rightarrow 1$ \\
$\{ b \wedge \neg c \} \rightarrow 2$ \\
\end{center}
Then $v(\{a\})=1, v(\{b\})=2, v(\{a,b\})=3+1+2=6$.

\subsection{Power index calculation in cooperative games}
A simple cooperative game with $m$ agents is defined by a characteristic function that maps any group (coalition) of agents to a real value $v : \{0,1\}^m \rightarrow \R $.
A coalition $C \subseteq L$ is said to be winning if $v(C)=1$ and losing if $v(C)=0$.
An agent $j$ is considered "critical" if the agent's removal from that coalition would turn it into a losing coalition. Critical agents significantly affect the coalition and are, as such, extensively researched. Two common approaches to measuring this power are the Banzhaf and the Shapley-Shubik indices. 

The Banzhaf power index assesses the ratio of critical roles in coalition to all coalitions where an agent $j$ appears. The Banzhaf index is given by $\beta = \beta(v) = \beta_1(v),\beta_2(v),\dots, \beta_n(v)$ where:

\begin{equation}
    \beta_j = \frac{1}{2^{m-1}} \smashoperator[r]{\sum_{C \subseteq N \setminus \{j \in C\}}}[ v(C \cup \{j \}) - v(C)] 
\end{equation}

The Shapley-Shubik index assesses the critical permutation $(\pi)$ of the agent ${i}$ in a coalition $C$, out of all sets of all permutations $(\Pi)$. 
Given a coalition $C$, defined as $C_\pi(j) = \{j | \pi(j) < \pi(j)\}$, the Shapley-Shubik index is given by $Sh(v) = (Sh_1(v),Sh_2(v),\dots, Sh_n(v))$ where:

\begin{equation}
     Sh_i(v) = \frac{1}{n!} \smashoperator[r]{\sum_{\pi\in\Pi}}[v(C_\pi(j) \cup \{j \}) - v(C_\pi(j))
\end{equation}

The key distinction lies in their underlying assumptions: while the Banzhaf index assumes that all coalitions are equally likely to occur, Shapley-Shubik assumes that all permutations of coalitions are equally likely to occur. This leads to different power assessments that can provide complementary insights into complex decision-making systems.

\section{Methodology}
Our research employs a comprehensive approach to approximating power indices using NN, focusing on generating diverse and representative datasets that capture the complexity of real-world voting systems. 

\subsection{Dataset generation strategies}
\label{data_generation}
We generate numerous datasets representing various voting coalitions to train our NN models. We employed three distinct strategies for generating random coalition rules to ensure diversity and robustness in our training data: 
\begin{itemize}
    \item Uniform Random Sampling: We randomly assign agents to the rules (both required and banned) using a uniform random distribution.
    \item Coin Flip Assignment: Each agent is assigned to the required or the banned group with equal probability. This method introduces a more structured randomness to the system while decreasing the number of non-participating agents.
    \item Probabilistic mixture of Gaussian: we sample coalition rules from a mixture of Gaussian distributions. This method allows more nuanced and potentially more realistic rule generations to capture more complex voting scenarios.
\end{itemize}

The strategies result in varying distributions and entropy, providing rich training data for the NN to learn. For detailed implementation, see Algorithms \ref{alg:uniform_random}, \ref{alg:coin_flip_random},  and \ref{alg:MOG_random}.

To further enrich the dataset's complexity, we developed three distinct methods for scoring rules: 
\begin{itemize}
    \item Uniform Weights: All rules are equally important, creating a baseline scenario.
    \item Low-variance Gaussian distribution: Rule weights are sampled from a Gaussian distribution with a small variance, introducing subtle differences between rule importance.
    \item High-variance Gaussian distribution: Rule weights are sampled from a Gaussian distribution with significant variance, creating more pronounced differences between rule importance.
\end{itemize}
 
\begin{algorithm}[H]
\caption{Random uniform rules}
\begin{algorithmic}[1]
\Require Dataset size $k$, Number of coalitions $n$, Number of Agents $m$, Probability $p$
\Ensure Coalition matrix $R$
\For{$i = 1$ to $k$}
    \State $X \gets \{x_{ab}\}_{a=1,b=1}^{n,m}$ where $x_{ab} \sim \mathcal{U}(0,1)$ \Comment{Sample uniform random array}
    \State $Y \gets \{y_{ab}\}_{a=1,b=1}^{n,m}$ where $y_{ab} \sim \mathcal{U}(0,1)$ \Comment{Sample uniform random array}
    \State $req_i \gets \text{round}(X, p) := \begin{cases} 1 & \text{if } x_{ab} \geq p \\ 0 & \text{if } x_{ab} < p \end{cases}, \forall x_{ab} \in X$
    \State $ban_i \gets \text{round}(Y, p) := \begin{cases} 
    0 & \text{if } x_{ab} = 1 \\
    1 & \text{if } x_{ab} = 0 \text{ and } y_{ab} \geq p \\
    0 & \text{if } y_{ab} < p
    \end{cases}, \forall y_{ab} \in Y$
    \State $R_i =( req_i,ban_i)$ 
\EndFor
\State \Return $R$
\end{algorithmic}
\end{algorithm}

\begin{algorithm}[H]
\caption{Coin-flip rules}
     \begin{algorithmic}[1]
     \Require Dataset size $k$, Number of coalitions $n$, Number of Agents $m$, number of coins $c$ \\
     \Ensure Coalition matrix $R$  
             \State $req, ban \gets \text{zeros}(m)$
        \For{$i=1$ to $k$}
            \For{$j=1$ to $n$}
                \State $X \gets \{x_{a}\}_{a=1}^{c}$ where $x_{a} \sim \mathcal{U}\{1,2,\dots,m\}$ \Comment{Sample uniform random indices}
                \State $Y \gets \{y_{a}\}_{a=1}^{c}$ where $y_{a} \sim \mathcal{U}(0,1)$ \Comment{Sample uniform random number}
                \For{$l = 1$ to $c$}
                    \State $req_{X_l}  := \begin{cases} 1 & \text{if } y_{l} = 1 \\ 0 & \text{if } y_{l} = 0 \end{cases}$
                    \State $ban_{X_l}  := \begin{cases} 1 & \text{if } y_{l} = 0 \\ 0 & \text{if } y_{l} = 1  \end{cases}$
                \EndFor
            \EndFor
        \State $R_i =( req_i,ban_i)$ 
        \EndFor
    \State \Return $R$ 
    \end{algorithmic}
    \label{alg:coin_flip_random}
 \end{algorithm}

\begin{algorithm}[H]
\caption{Probabilistic mixture of Gaussian}
     \begin{algorithmic}[1]
    \Require Dataset size $k$, Number of coalitions $n$, Number of Agents $m$, Distribution shape $\alpha$, Distribution rate $\beta$ , Probability $p$\\
    \Ensure Coalition matrix $R$

        \For{$i=1$ to $k$}
            \State $\mu_i, \sigma_i \sim \Gamma(\alpha,\beta)$ \Comment{Sample mean and std from a gamma distribution}
            \State$ X \sim \mathcal{N}(\mu_i, \sigma_i^2)^{n x m}$ where $A \in \R^{nxm}$ \Comment{Sample array from a Gaussian distribution}
            \State$ Y \sim \mathcal{N}(\mu_i, \sigma_i^2)^{n x m}$ where $A \in \R^{nxm}$ \Comment{Sample array from a Gaussian distribution}
            \State $req_i \gets \text{round}(X, p) := \begin{cases} 1 & \text{if } x_{ab} \geq p \\ 0 & \text{if } x_{ab} < p \end{cases}, \forall x_{ab} \in X$
        \State $ban_i \gets \text{round}(Y, p) := \begin{cases} 
        0 & \text{if } x_{ab} = 1 \\
        1 & \text{if } x_{ab} = 0 \text{ and } y_{ab} \geq p \\
        0 & \text{if } y_{ab} < p
        \end{cases}, \forall y_{ab} \in Y$
   
    \State $R_i =( req_i,ban_i)$ 
        \EndFor
    \State \Return $R$
    \end{algorithmic}
\label{alg:MOG_random}
 \end{algorithm}

\subsection{Power index calculations}
In order to train the NN to predict the power indices, we needed to label our generated data. As the complexity of calculating the exact Banzhaf and Shapley-Shubik power indices are $O(2^m)$ and $O(m!)$, respectively, making such calculations for large number of agents $m$ unfeasible, we used a Monte-Carlo approximation instead. Our method included sampling 10,000 random coalitions and compare:
\begin{itemize}
    \item Banzhaf power index: Whether a rule is fulfilled/broken by the addition of an agent $a$. The sum of critical scenarios is divided by the total number of samples, to get $\beta_a$
    \item  Shapley-Shubik index: What the marginal contribution of agent $a$ to each coalition  is. The agent is added to all possible coalitions where he is not already a member. The average of these marginal contributions gives us $Sh_a$
\end{itemize}

See algorithms ~\ref{alg:banzhaf_calc} and ~\ref{alg:shapley_calc} for exact implementation details.

Each of our final datasets would be a three-dimensional tensor of shape $k, n, (2*m+1)$, where:
\begin{itemize}
    \item  $k$ are the number of datapoints
    \item $n$ the number of coalitions/rules
    \item $m$ the number of agents (having the first $m$ positions represent $req$ indices,  while the second $m$ positions represent the $ban$ indices in each coalition
    \item Each rule includes a rule value/score (as described in subsection \ref{data_generation})
\end{itemize}

Each datapoint in the dataset received a label in the form of an $1xm$ array, containing either the Banzhaf or Shapley-Shubik values for the $m$ agents in the coalition. Since NN require large number of examples in order to learn patterns well~\cite{NIPS2006_5da713a6}, we tried different number of datapoints for model training. We found that best results were achieved when using a dataset size of $k=200,000$ for all datasets.

\begin{algorithm}[H]
\caption{Monte-Carlo Banzhaf Power Index Approximation}
\begin{algorithmic}[1]
\Require Set of agents $A$, Set of rules $R$ with positive and negative requirements, Rule weights $W$ (where $W_r \in \R^{|R|}$ for each rule $r \in R$), and a number of simulations $N$
\Ensure Approximate weighted Banzhaf power index for each agent $\beta$
\ForAll{agent $a \in A$}
    \State $P_a \leftarrow 0$ \Comment{The weighted pivotal count of each agent $a$ starts as 0}
\EndFor
\For{$i = 1$ to $N$}
    \State $C \sim U(2^{|A|})$ \Comment{Generate a random coalition C}
    \ForAll{agent $a \in A$}
        \State $C' \leftarrow C \setminus \{a\}$ \Comment{Create a new coalition $C'$ by removing $a$ from $C$}
        \If{$v(C') \neq v(C)$} 
            \State \( \Delta_r \gets \{ r \in R \mid \text{status of } r \text{ changes} \} \) \Comment{Take all rules whos status changed}
            \State $\Delta_{wa} \leftarrow \sum_{r \in \Delta_r} W_r$ \Comment{Calculate the total weights of all rules whos status changed}
            \State $P_a \leftarrow P_a + \Delta_{wa}$ \Comment{Update the weighted pivotal count of agent $a$}
        \EndIf
    \EndFor
\EndFor
\ForAll{agent $a \in A$}
\State $\beta(a) = \frac{P_a}{N \cdot \sum_{r \in R} W_r}$ \Comment{Calculate weighted Banzhaf index of $a$}
\EndFor
\State \Return $\beta(A)$ \Comment{Weighted Banzhaf index for each agent}

\end{algorithmic}
\label{alg:banzhaf_calc}
\end{algorithm}

\begin{algorithm}[H]

\caption{Monte-Carlo Shapley-Shubik Power Index Approximation}
\begin{algorithmic}[1]
\Require Set of agents $A$, Set of rules $R$ with positive and negative requirements, Rule weights $W$ (where $W_r \in \R^{|R|}$ for each rule $r \in R$), and a number of simulations $N$
\Ensure Approximate weighted Shapley-Shubik indices for each agent
\ForAll{agent $a \in A$}
    \State $P_a \leftarrow 0$
\EndFor
\For{$i = 1$ to $N$}
    \State $\pi \sim \Pi(A)$ \Comment{Generate a random permutation $\pi$ of agents}
    \State $C \leftarrow \emptyset$ \Comment{Initialize empty coalition $C$}
    \State $coalition\_won \leftarrow \text{false}$ \Comment{Flag to track if coalition has been made winning}
    \ForAll{$a \in \pi$}
        \State $C' \leftarrow C \cup a$ \Comment{Add $a$ to coalition $C$}
        \If{$v(C') \neq v(C)$ and $\neg coalition\_won$} 
            \State $coalition\_won \leftarrow \text{true}$
            \State $\Delta_r \gets { r \in R \mid \text{status of } r \text{ changes} } $ \Comment{Take all rules whose status changed}
            \State $\Delta_{wa} \leftarrow \sum_{r \in \Delta_r} W_r$ \Comment{Calculate the total weights of all rules whose status changed}
            \State $P_a \leftarrow P_a + \Delta_{wa}$
        \EndIf
        \State $C \leftarrow C'$
    \EndFor
\EndFor
\ForAll{agent $a \in A$}
    \State $Sh(a) = \frac{P_a}{N \cdot \sum_{r \in R} W_r}$ \Comment{Calculate the weighted Shapley-Shubik index of agent $a$}
\EndFor
\State \Return $Sh(A)$ \Comment{Weighted Shapley-Shubik indices for each agent}
\end{algorithmic}
\label{alg:shapley_calc}
\end{algorithm}

\subsection{Model training}
We trained multiple NN models, one on each dataset configuration with different number of coalitions $n$, number of agents $m$, method of data generation, rule value and value threshold $p$. The threshold $p$ is kept as a measure for additional variations in sparsity/density of the rules in the dataset, as seen in algorithms \ref{alg:uniform_random},\ref{alg:coin_flip_random},\ref{alg:MOG_random}.

To predict the power index of each dataset (i.e. matrix regression) we used a feedforward NN MSE regression. The architecture is composed of three hidden layers, each with ReLU activation function and 20\% dropout. Input and output layers are according to the shape of the largest input matrix and maximal number of agents, respectively. The smaller matrices received right-side zero padding in order to allow model comparison between the different datasets shapes. The models hidden layers where of dimensions $[512, 256, 128]$. 

All datasets were split into 'train':'test' with a ratio of 80:20. All models were trained on a single 'train' dataset using MSE as success metric. Following  training, all models were evaluated on all other 
'test' datasets. 

\subsection{Graph analysis}
To gain additional insights into the structural properties of voting systems,   we represented each voting system as an undirected graph~\cite{easley2010networks,HELLMAN201822}. Thus, we could use established graph-theoretical metrics to analyse coalitions dynamics and power structures. In our representation, agents served as nodes, while edges represented meaningful relationships derived from coalition patterns. 

We analyzed several graph-theoretical properties to characterize voting system dynamics:
\begin{itemize}
    \item Maximum clique size: Identifies the largest group of mutually connected agents
    \item Average degree: The average number of edges per node.
    \item Variance of degrees: the variance in the number of edges per node
    \item Average clustering of a graph
    \item Max betweenes: the maximal length of the sum of all pair-shortest paths.
\end{itemize}

\section{Results}
\subsection{Uniform random distribution analysis}
\begin{figure}
    \centering
    \begin{subfigure}[b]{0.32\textwidth}
        \includegraphics[width=\textwidth]{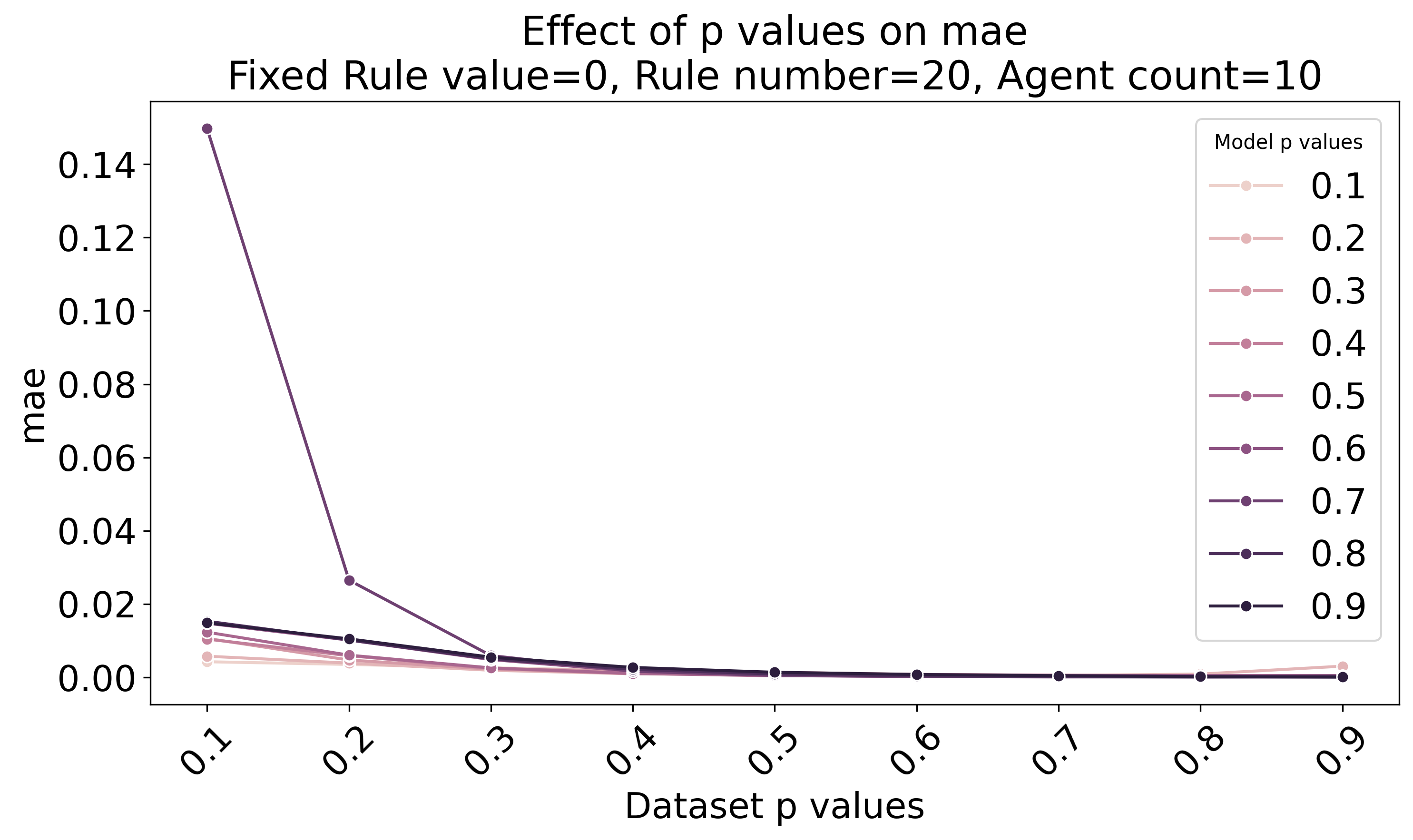}
        \caption{n=10 agents, uniform rules}
        \label{fig:uniform_subfig1}
    \end{subfigure}
     \hfill
    \begin{subfigure}[b]{0.32\textwidth}
        \includegraphics[width=\textwidth]{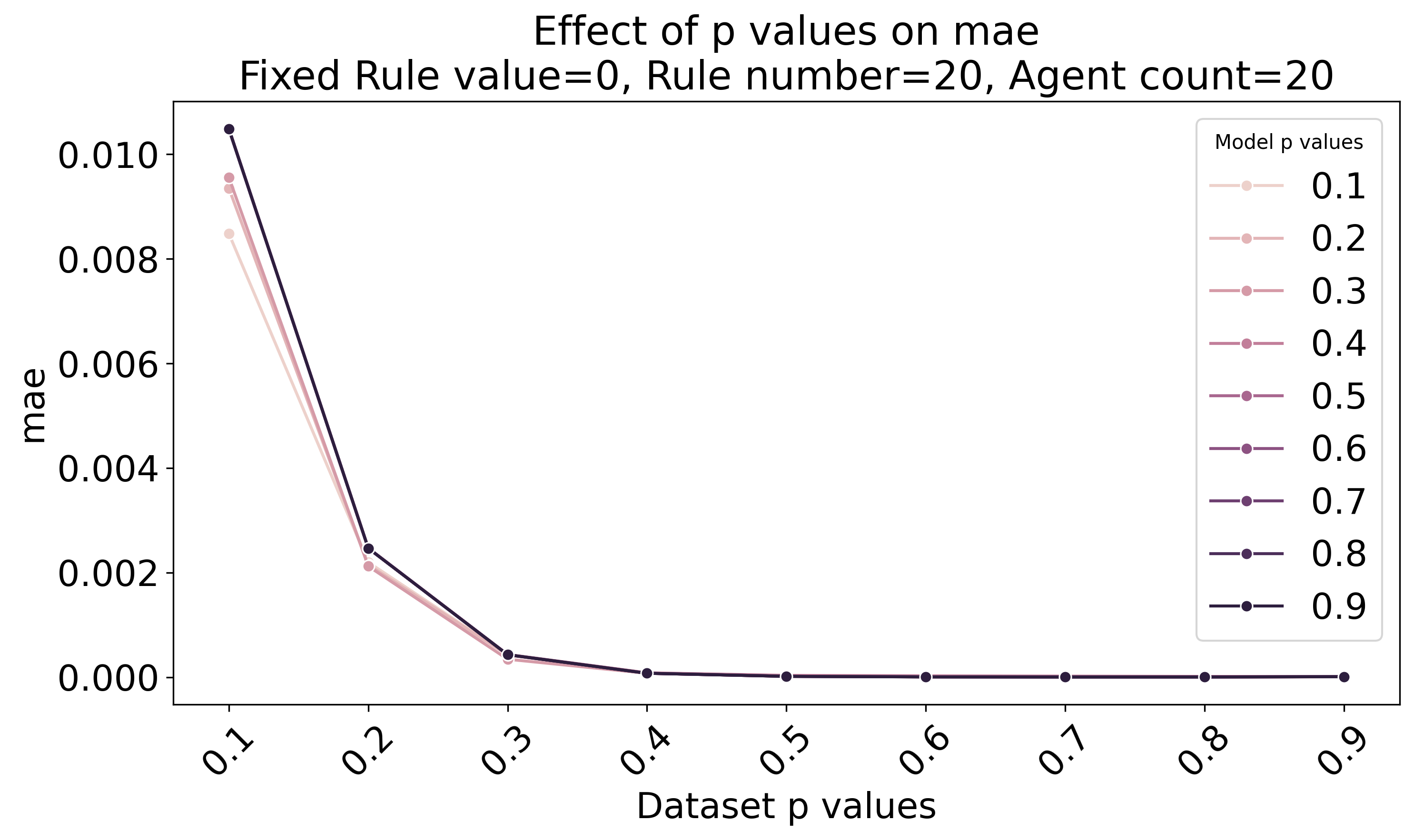}
        \caption{n=20 agents, uniform rules}
        \label{fig:uniform_subfig2}
    \end{subfigure}
     \hfill
    \begin{subfigure}[b]{0.32\textwidth}
        \includegraphics[width=\textwidth]{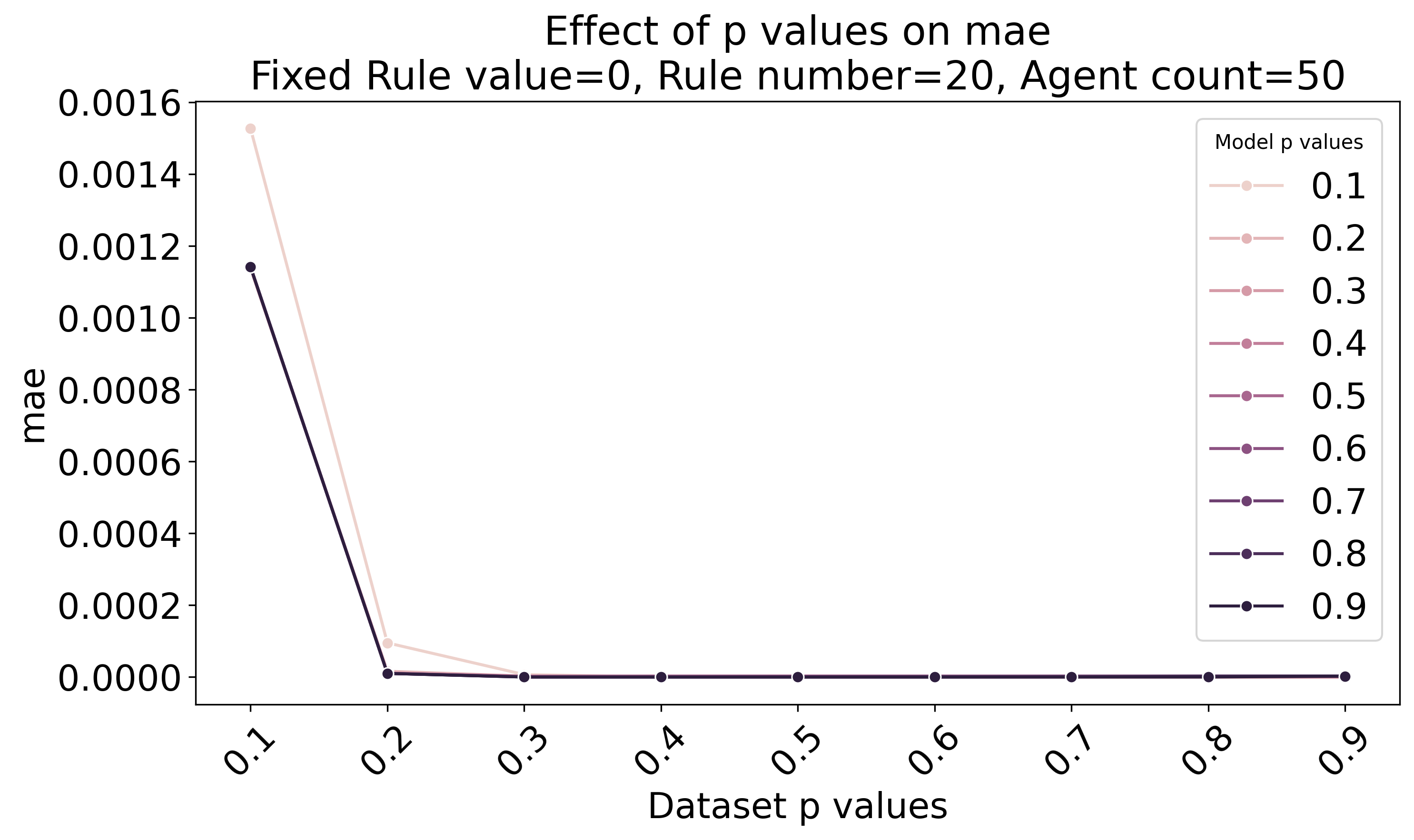}
        \caption{n=50 agents, uniform rules}
        \label{fig:uniform_subfig3}
    \end{subfigure}
    \\[\baselineskip]    
    \begin{subfigure}[b]{0.32\textwidth}
        \includegraphics[width=\textwidth]{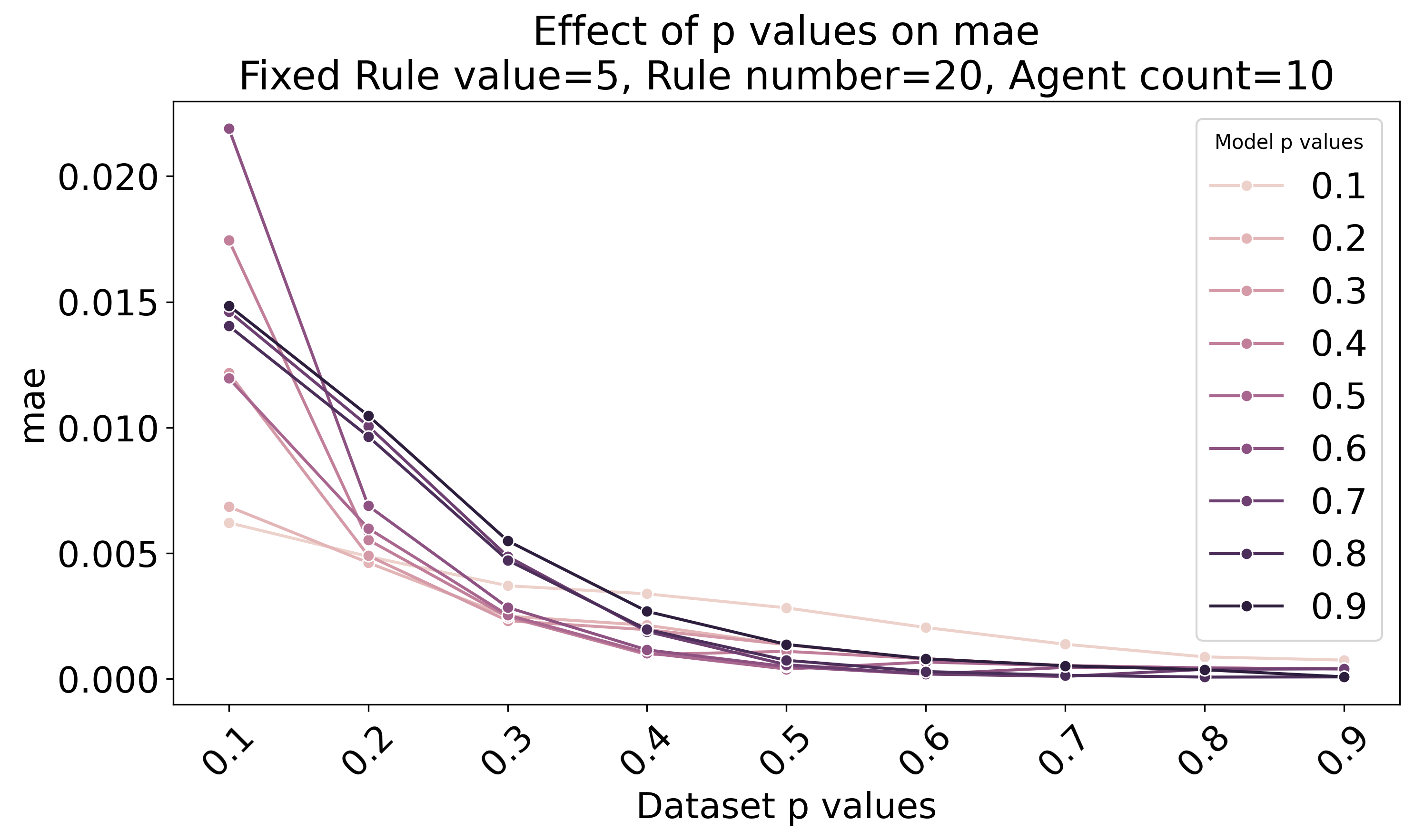}
        \caption{n=10 agents, Low-variance Gaussian rules ($\mu = 5$)}
        \label{fig:uniform_subfig4}
    \end{subfigure}
     \hfill
    \begin{subfigure}[b]{0.32\textwidth}
        \includegraphics[width=\textwidth]{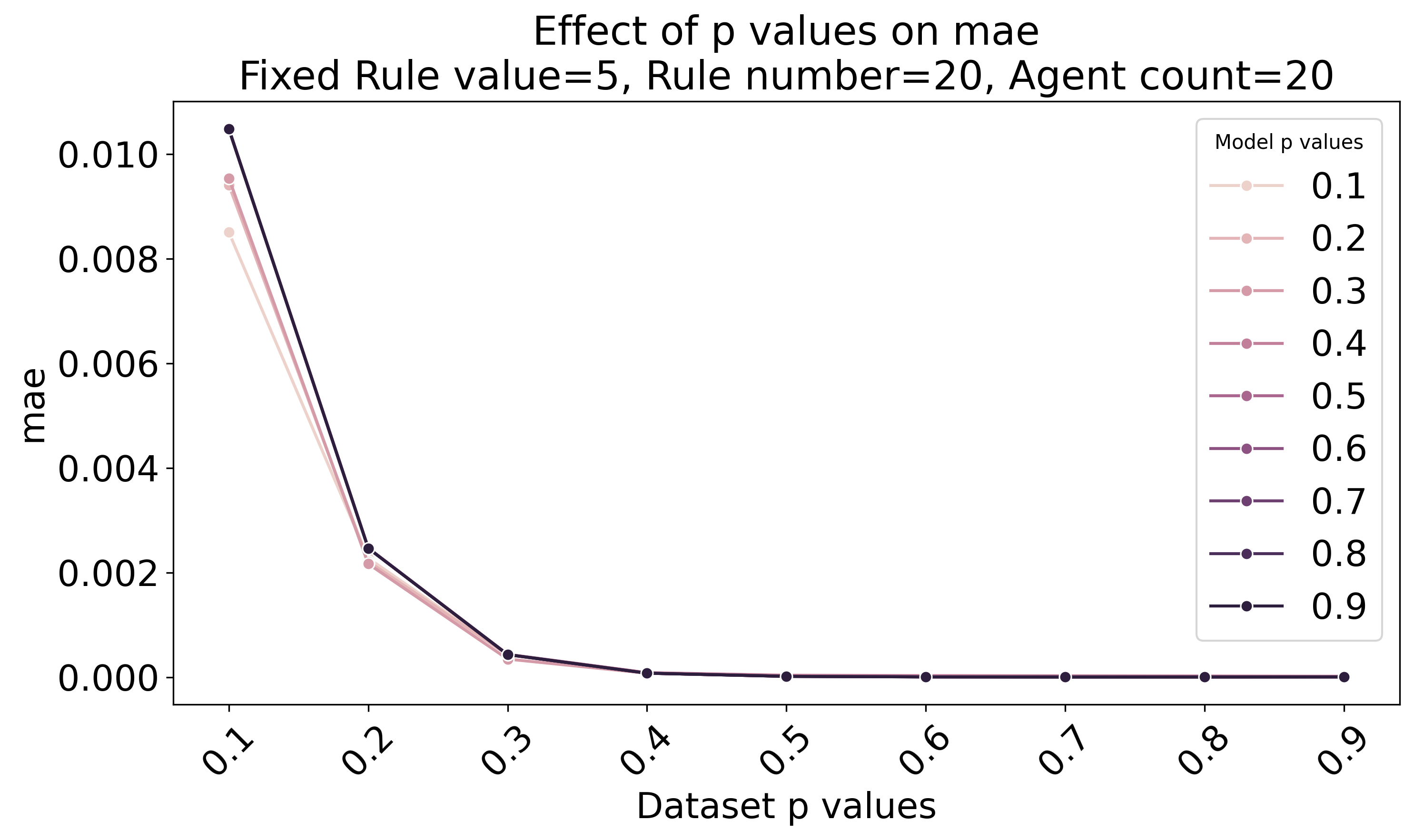}
        \caption{n=20 agents, Low-variance Gaussian rules ($\mu = 5$)}
        \label{fig:uniform_subfig5}
    \end{subfigure}
     \hfill
    \begin{subfigure}[b]{0.32\textwidth}
        \includegraphics[width=\textwidth]{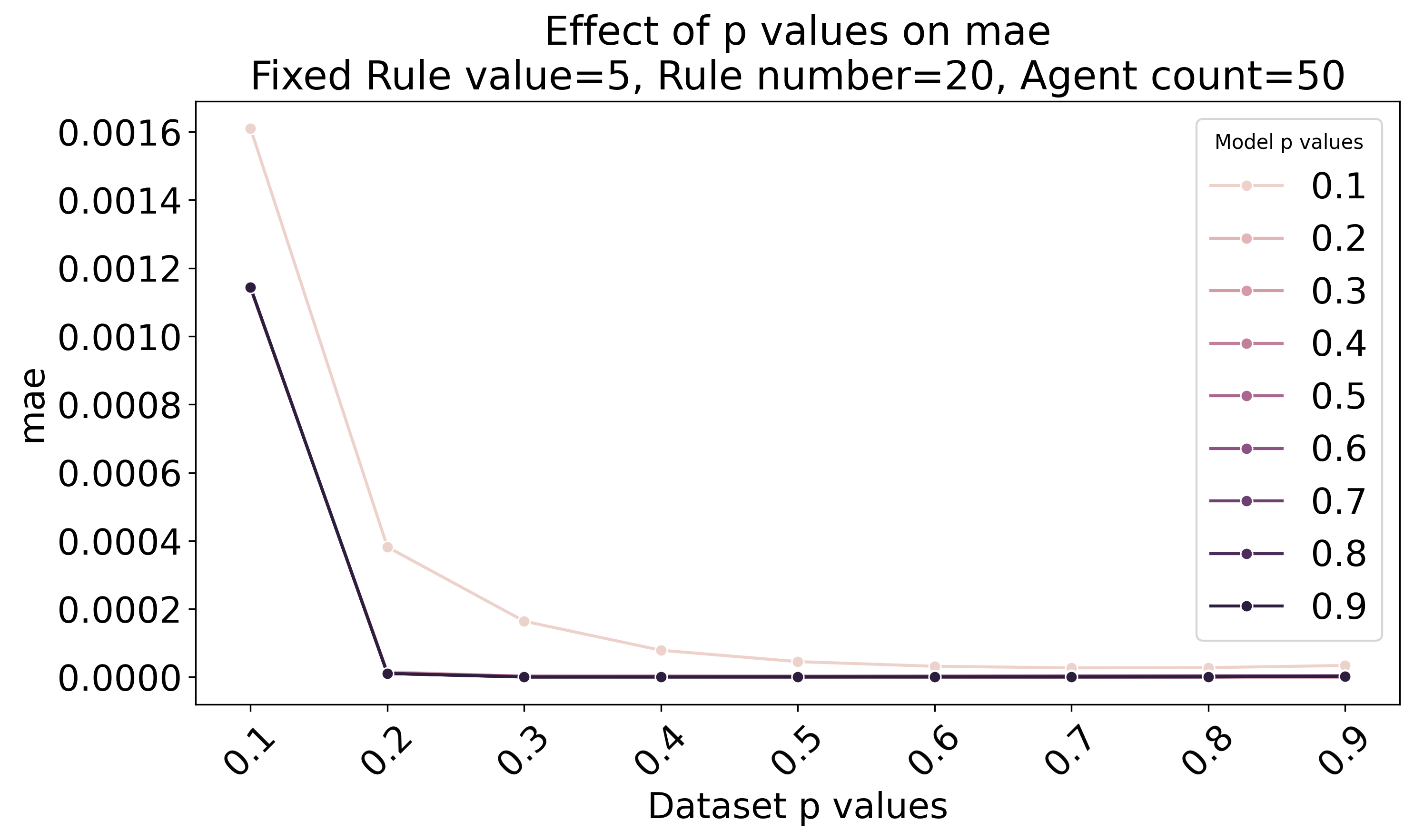}
        \caption{n=50 agents, Low-variance Gaussian rules ($\mu = 5$)}
        \label{fig:uniform_subfig6}
    \end{subfigure}
    \\[\baselineskip]
        \begin{subfigure}[b]{0.32\textwidth}
        \includegraphics[width=\textwidth]{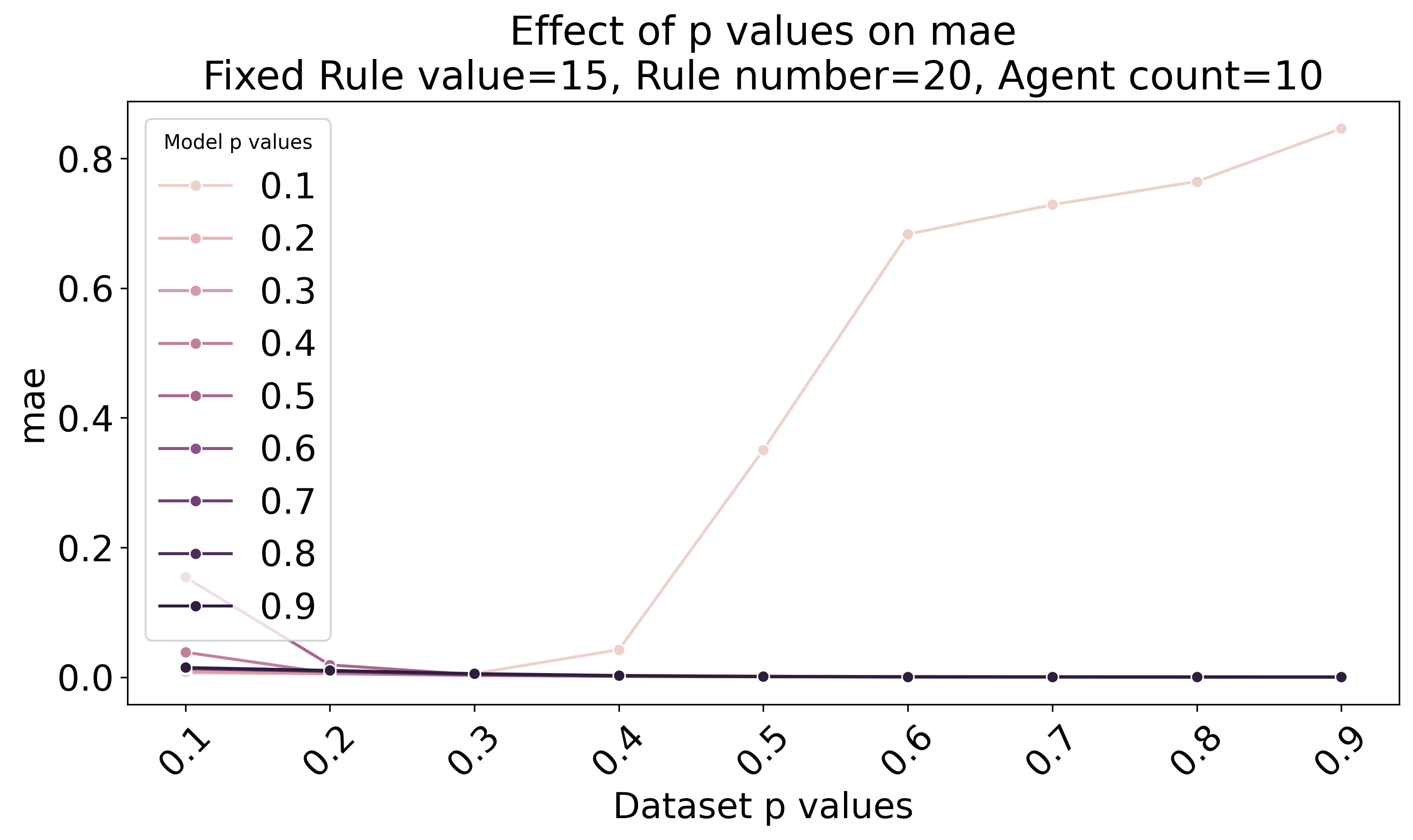}
        \caption{n=10 agents, high-variance Gaussian rules ($\mu = 15$)}
        \label{fig:uniform_subfig7}
    \end{subfigure}
     \hfill
    \begin{subfigure}[b]{0.32\textwidth}
        \includegraphics[width=\textwidth]{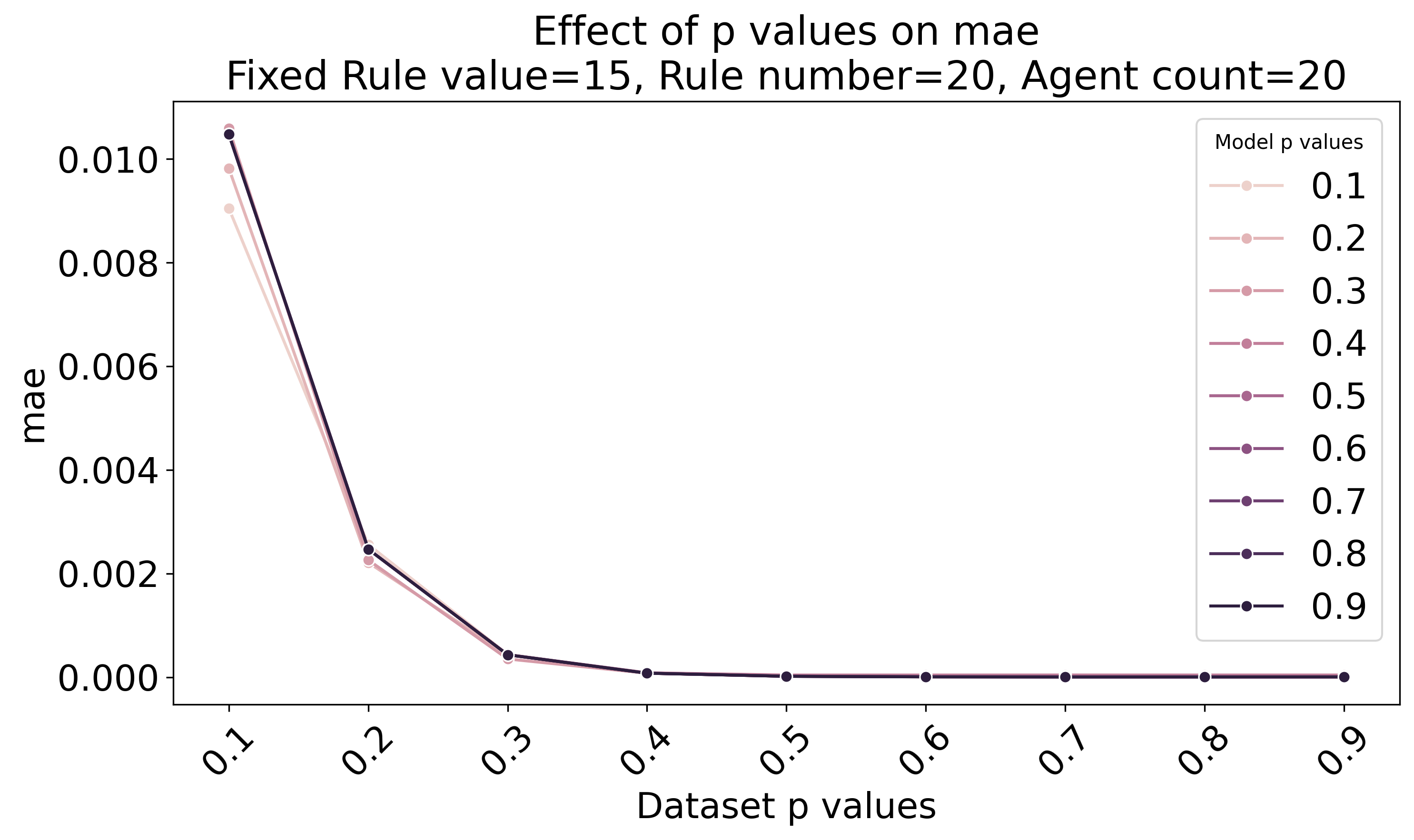}
        \caption{n=20 agents, high-variance Gaussian rules ($\mu = 15$)}
        \label{fig:uniform_subfig8}
    \end{subfigure}
     \hfill
    \begin{subfigure}[b]{0.32\textwidth}
        \includegraphics[width=\textwidth]{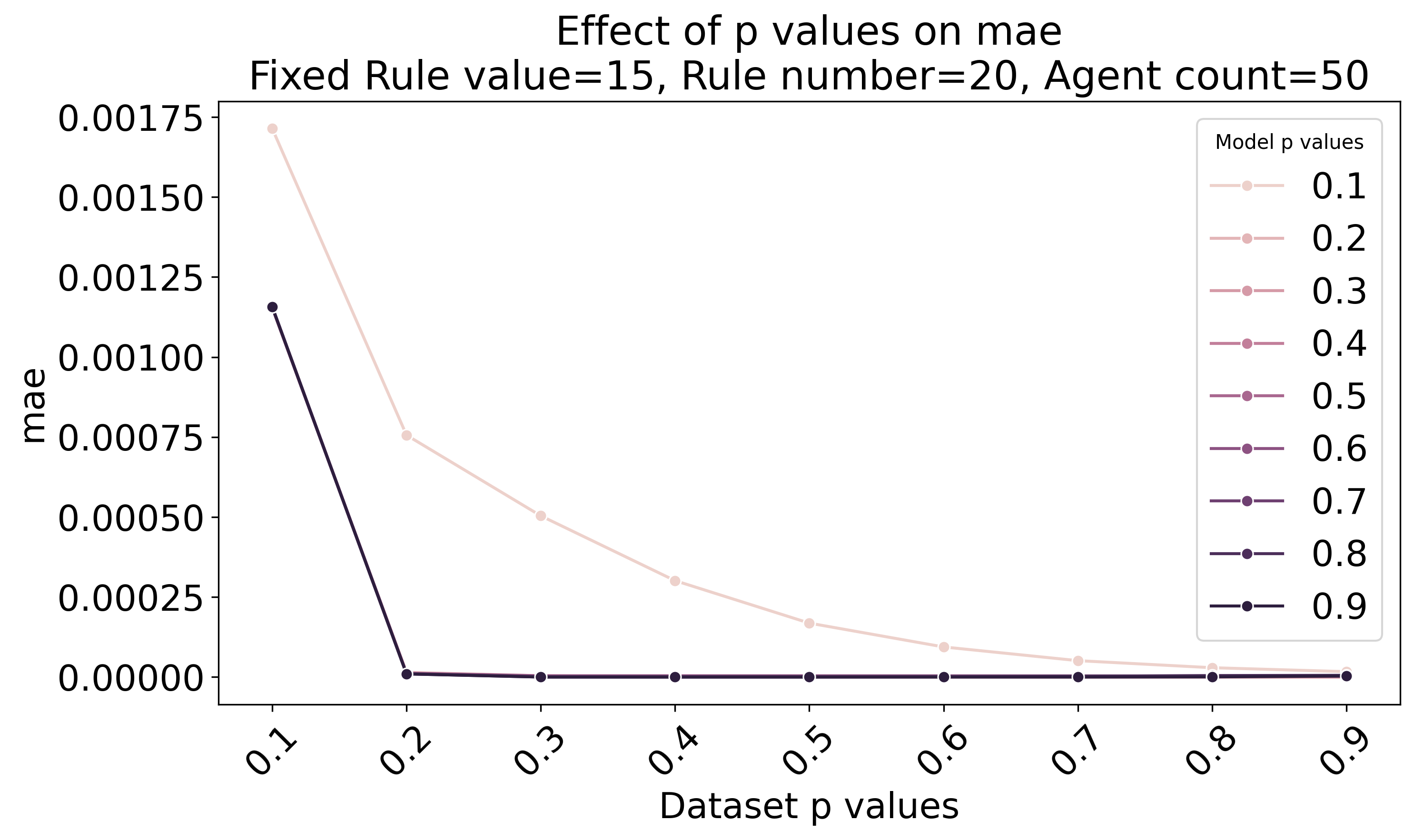}
        \caption{n=50 agents, high-variance Gaussian rules ($\mu = 15$)}
        \label{fig:uniform_subfig9}
    \end{subfigure}
    \caption{Models performance on uniform random datasets Mean Absolute Error (MAE) across different dataset configurations The x-axis represents the sparsity threshold (p), and the y-axis shows the MAE values. Subplots show variations across: (A-C) uniform rule values with 20 rules and 10, 20, and 50 agents respectively; (D-F) low-variance Gaussian distribution (mean=5) with 20 rules across 10, 20, and 50 agents; (G-I) high-variance Gaussian distribution (mean=15) with 20 rules for 10 and 20 agents. Each line within subplots represents a model trained with a different p value while maintaining consistent coalition parameters.}
    \label{fig:uniform_cross_p_data_model_same_config_MAE}
\end{figure}

Our analysis focused on training and evaluating the NN models' predictive performance across various dataset configurations and examining how parameter modifications affected inference accuracy. The Mean Absolute Error (MAE) results, presented in Figure~\ref{fig:uniform_cross_p_data_model_same_config_MAE}, demonstrate consistently low error values across all test sets. We observed that models trained on sparse coalition rules generally outperformed those trained on denser coalitions when evaluated on sparse datasets. This difference is very apparent in datasets with low number of agents, and minimal when comparing agent-rich datasets ( in which case the MAE differences $\leq 0.001$). Regardless of other configurations, all dense rules have very low power indices, due to the high number of constraints on them, which makes their number of cases in which agents can be of greater importance over the other possible agents less likely ~\cite{GRECO201719}.

\begin{figure}
    \centering
    \begin{subfigure}[b]{0.32\textwidth}
        \includegraphics[width=\textwidth]{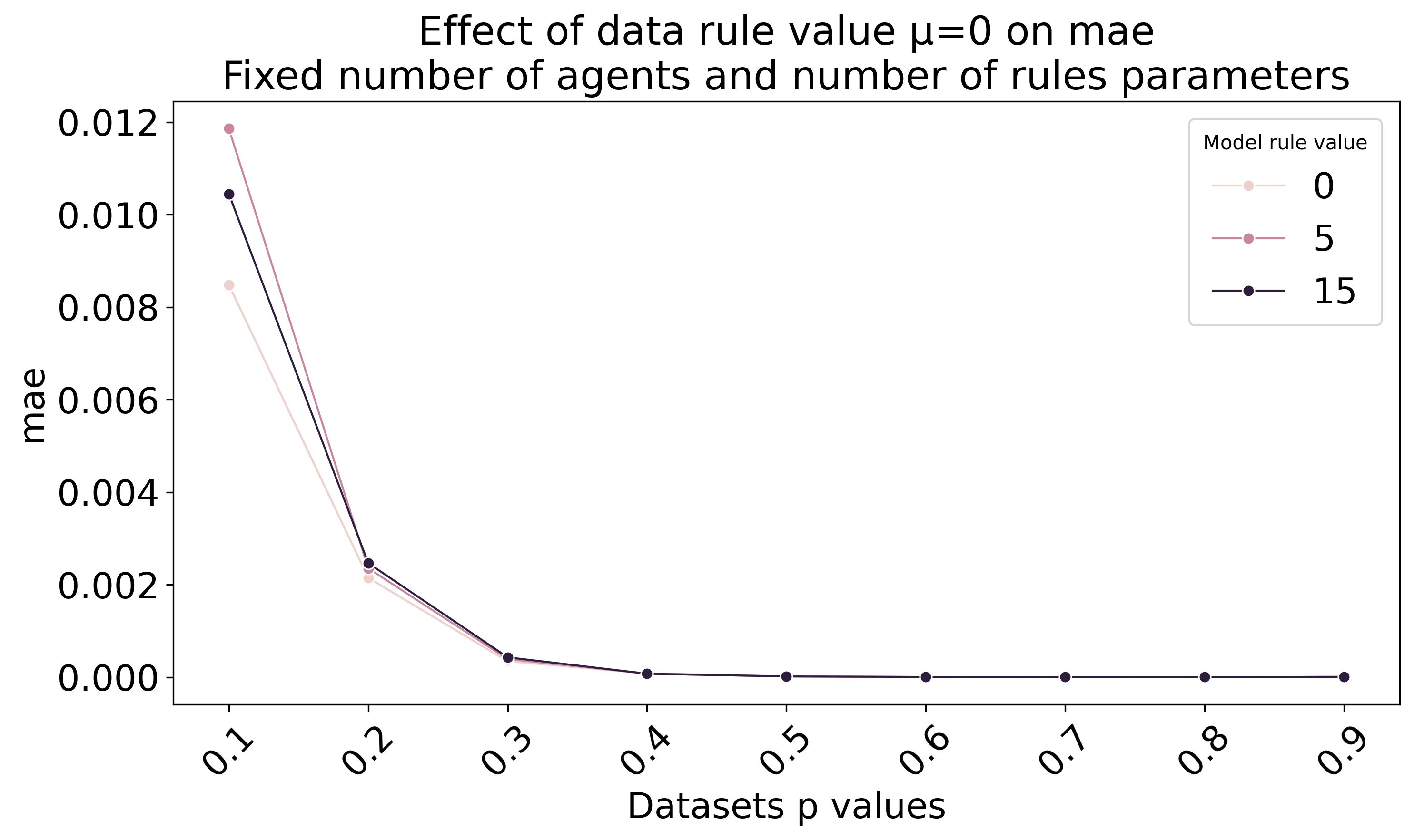}
        \caption{Uniform rule value distribution}
        \label{fig:delta_rule_value_subfig1}
    \end{subfigure}
         \hfill
    \begin{subfigure}[b]{0.32\textwidth}
        \includegraphics[width=\textwidth]{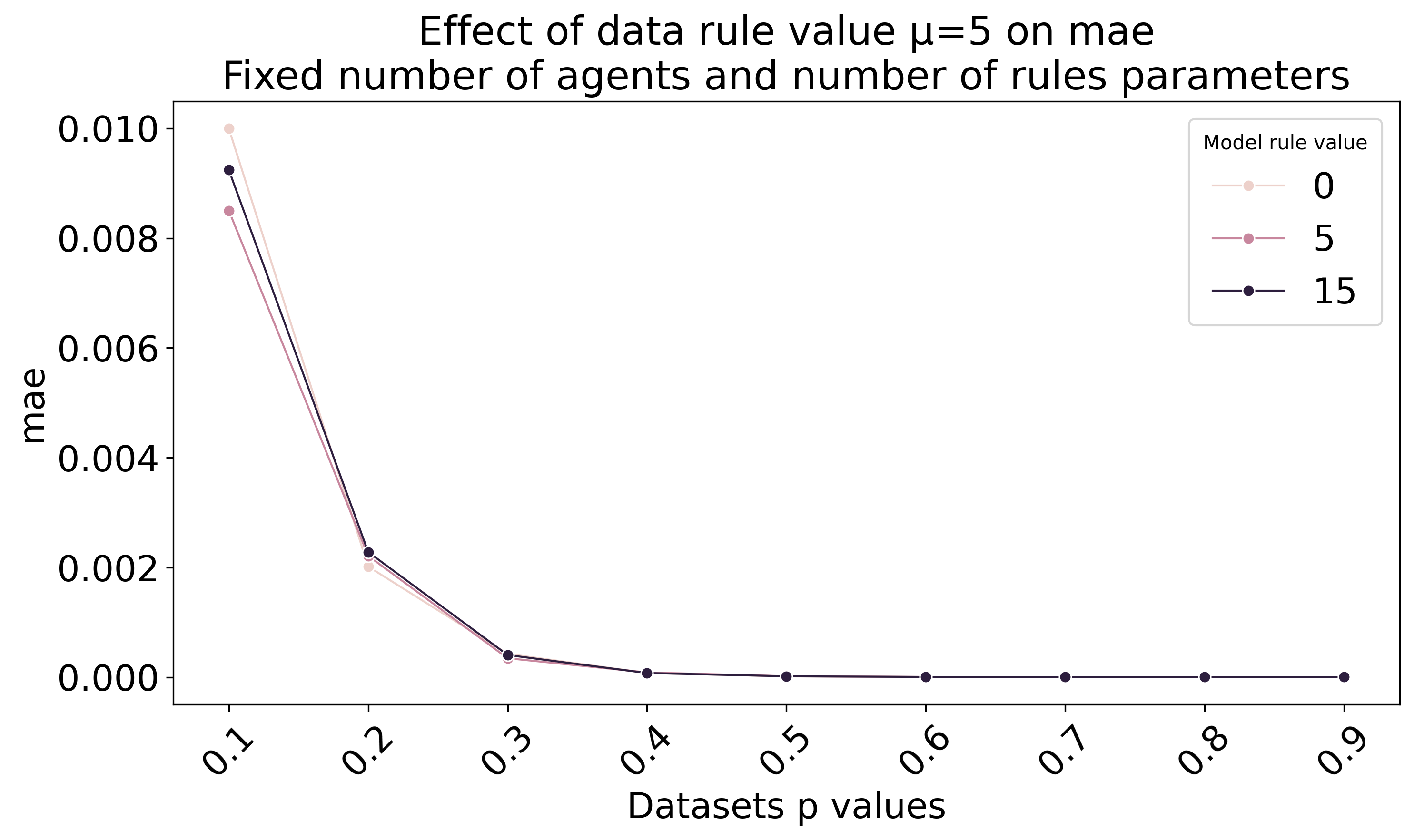}
        \caption{Low-variance $(\mu=5)$ rule value distribution}
        \label{fig:delta_rule_value_subfig2}
    \end{subfigure}
         \hfill
    \begin{subfigure}[b]{0.32\textwidth}
        \includegraphics[width=\textwidth]{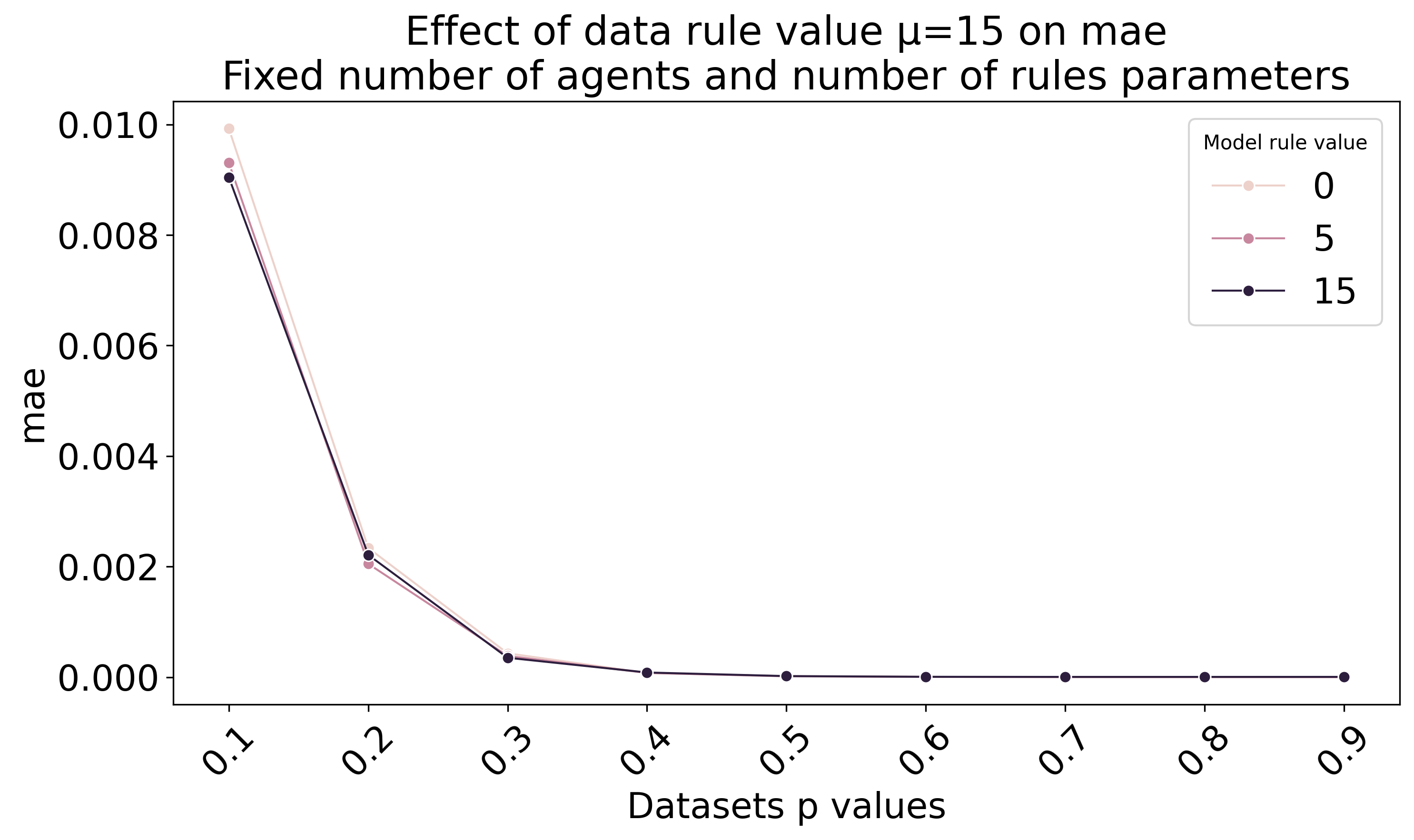}
        \caption{High-variance $(\mu=15)$ rule value distribution}
        \label{fig:delta_rule_value_subfig3}
    \end{subfigure}
    \caption{Impact of rule value distribution on model performance using Mean Absolute Error (MAE). Comparison across (A) uniform rule value, (B) low-variance Gaussian rule value, and (C) high-variance Gaussian rule value distributions. All configurations maintain consistent number of rules and agents. The x-axis represents the sparsity threshold (p), and the y-axis shows MAE values.}
    \label{fig:uniform_data_model_different_rule_value_MAE}
\end{figure}

\begin{figure}
    \centering    
    \begin{subfigure}[b]{0.32\textwidth}
        \includegraphics[width=\textwidth]{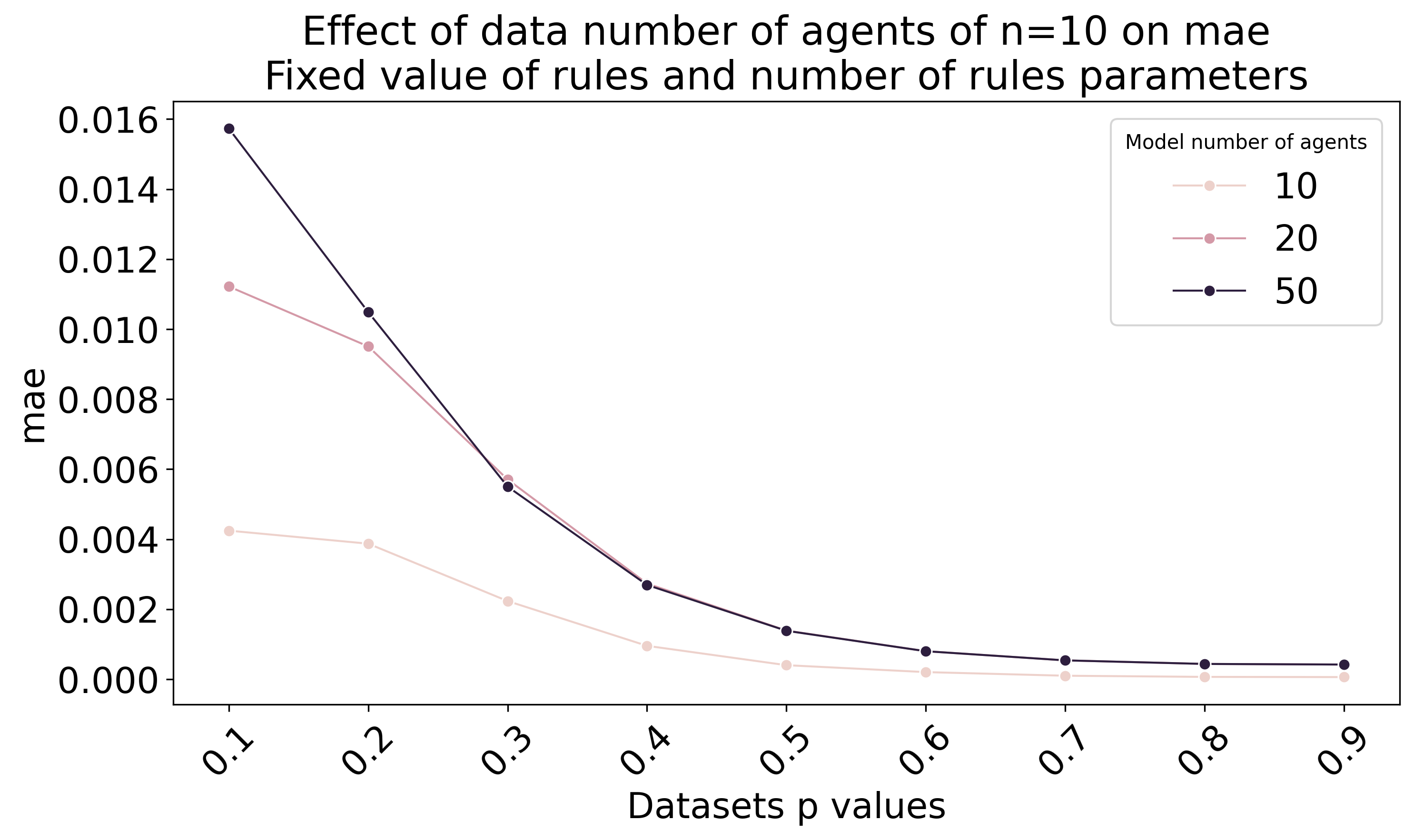}
        \caption{Performance on 10-agent datasets}
        \label{fig:delta_agent_subfig1}
    \end{subfigure}
         \hfill
    \begin{subfigure}[b]{0.32\textwidth}
        \includegraphics[width=\textwidth]{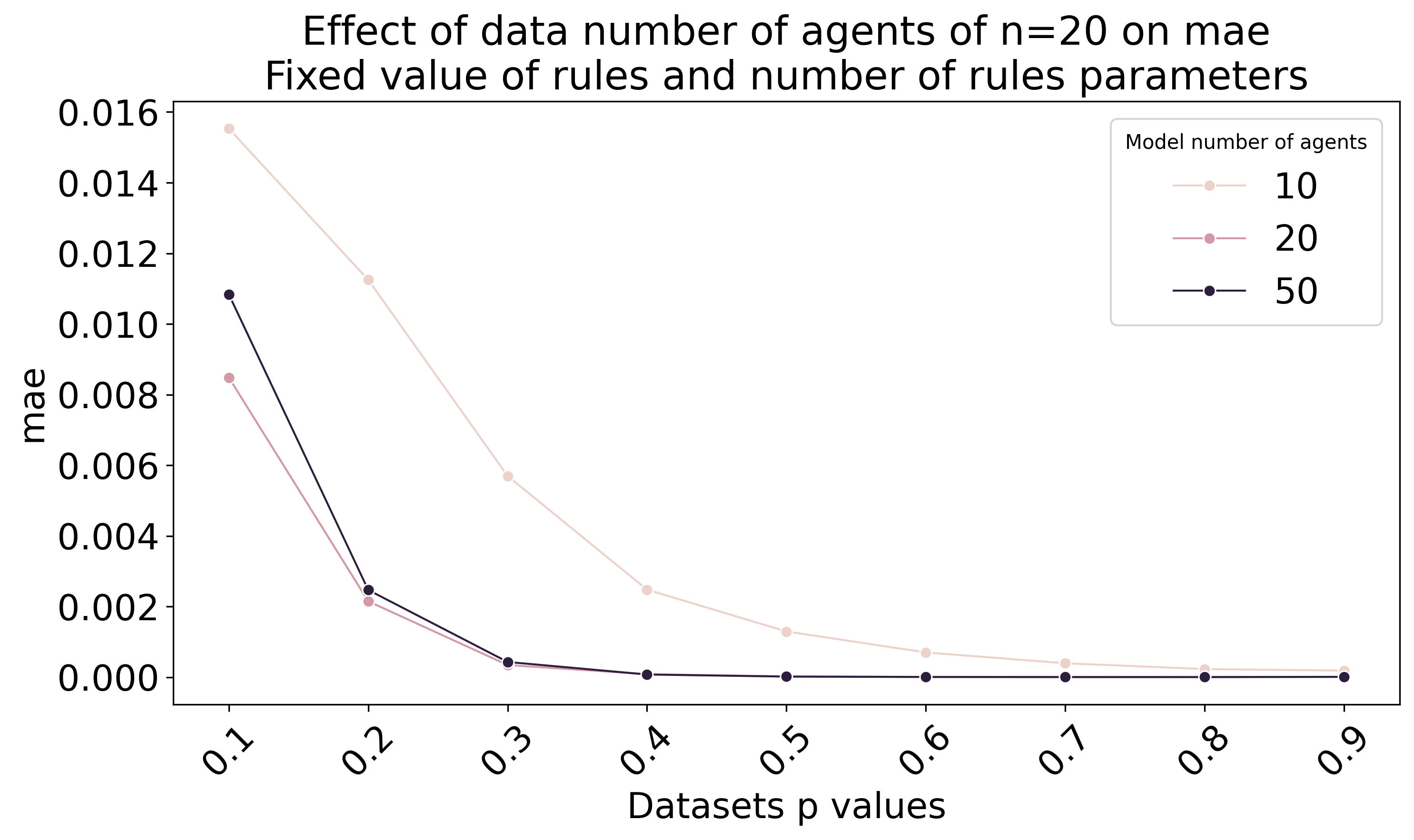}
        \caption{Performance on 20-agent datasets}
        \label{fig:delta_agent_subfig2}
    \end{subfigure}
     \hfill
    \begin{subfigure}[b]{0.32\textwidth}
        \includegraphics[width=\textwidth]{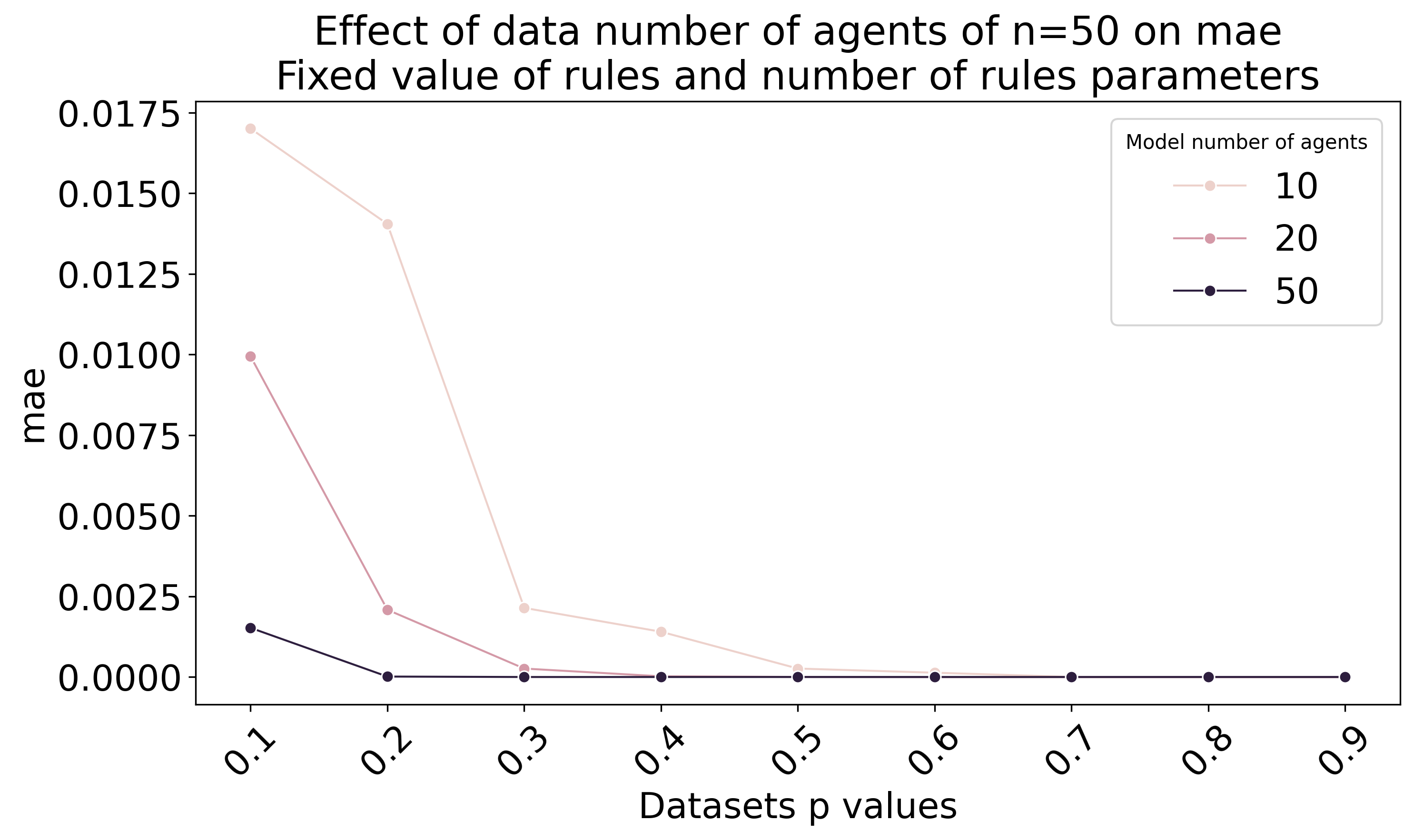}
        \caption{Performance on 50-agent datasets}
        \label{fig:delta_agent_subfig3}
    \end{subfigure}
    \caption{Cross-model comparison of Mean Absolute Error (MAE) for different numbers of agents. Results show model performance on datasets with (A) 10 agents, (B) 20 agents, and (C) 50 agents, while maintaining consistent rule values and number of rules across all configurations. The x-axis represents the sparsity threshold (p), and the y-axis shows MAE values.}
    \label{fig:uniform_data_model_differnet_number_of_agnets_MAE}
\end{figure}

The models exhibited robust performance when tested against configurations that differed from their training conditions. As shown in Figures~\ref{fig:uniform_data_model_differnet_number_of_agnets_MAE} and~\ref{fig:uniform_data_model_different_rule_value_MAE}, variations in the rule values resulted in only negligible changes in model performance. When comparing between different number of agents. In the latter, the differences became more noticeable, and as seen in figures~\ref{fig:delta_rule_value_subfig1} and ~\ref{fig:delta_rule_value_subfig3}, models trained datasets with large number of agents performed worse on datasets with a low number of agents, and vice versa for models trained on datasets with low number of agents when evaluated on datasets with large number of agents. The former having less erros compared to the latter.

Regarding rule value configurations, models trained on datasets with non-uniform rule values evaluated on datasetbwith varying rule values underperformed more compared to the models who trained on varying rule value and evaluated on uniform rule value. unlike the differences in number of agents, the rule value variance was significantly lower (highest MAE difference was $\leq 0.004$), as illustrated in Figure~\ref{fig:uniform_data_model_different_rule_value_MAE}. This pattern suggests that while rule value differences affect the models ability to predict correctly outside its training configurations, the NN models is robuster to rupe value xhange, and less so to the size of coalition as depicted in the number of agents.

\FloatBarrier 

\begin{figure}
    \centering
    \begin{subfigure}[b]{0.32\textwidth}
        \includegraphics[width=\textwidth]{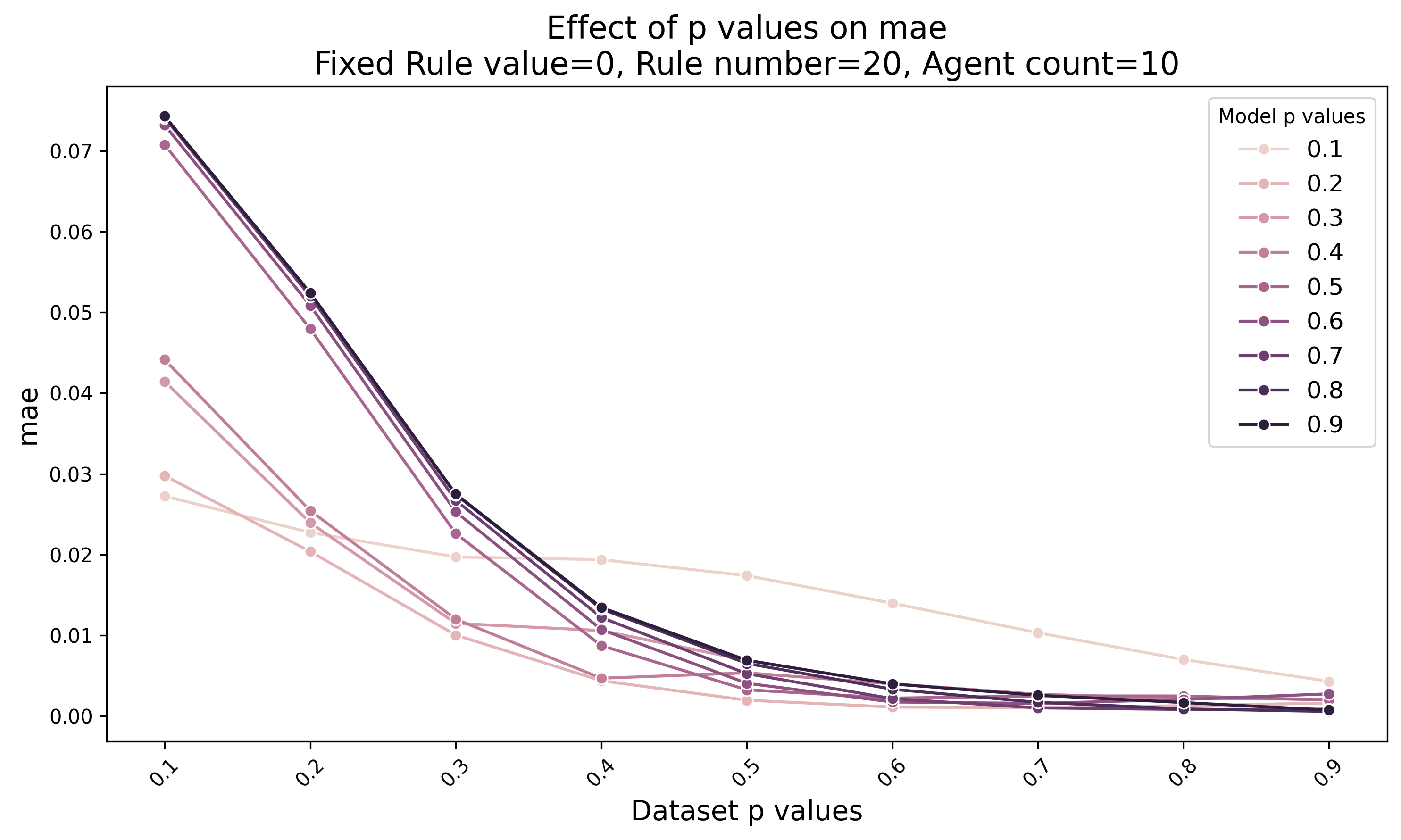}
        \caption{n=10 Agents}
        \label{fig:unpadded_delta_num_agent_subfig1}
    \end{subfigure}
         \hfill
    \begin{subfigure}[b]{0.32\textwidth}
        \includegraphics[width=\textwidth]{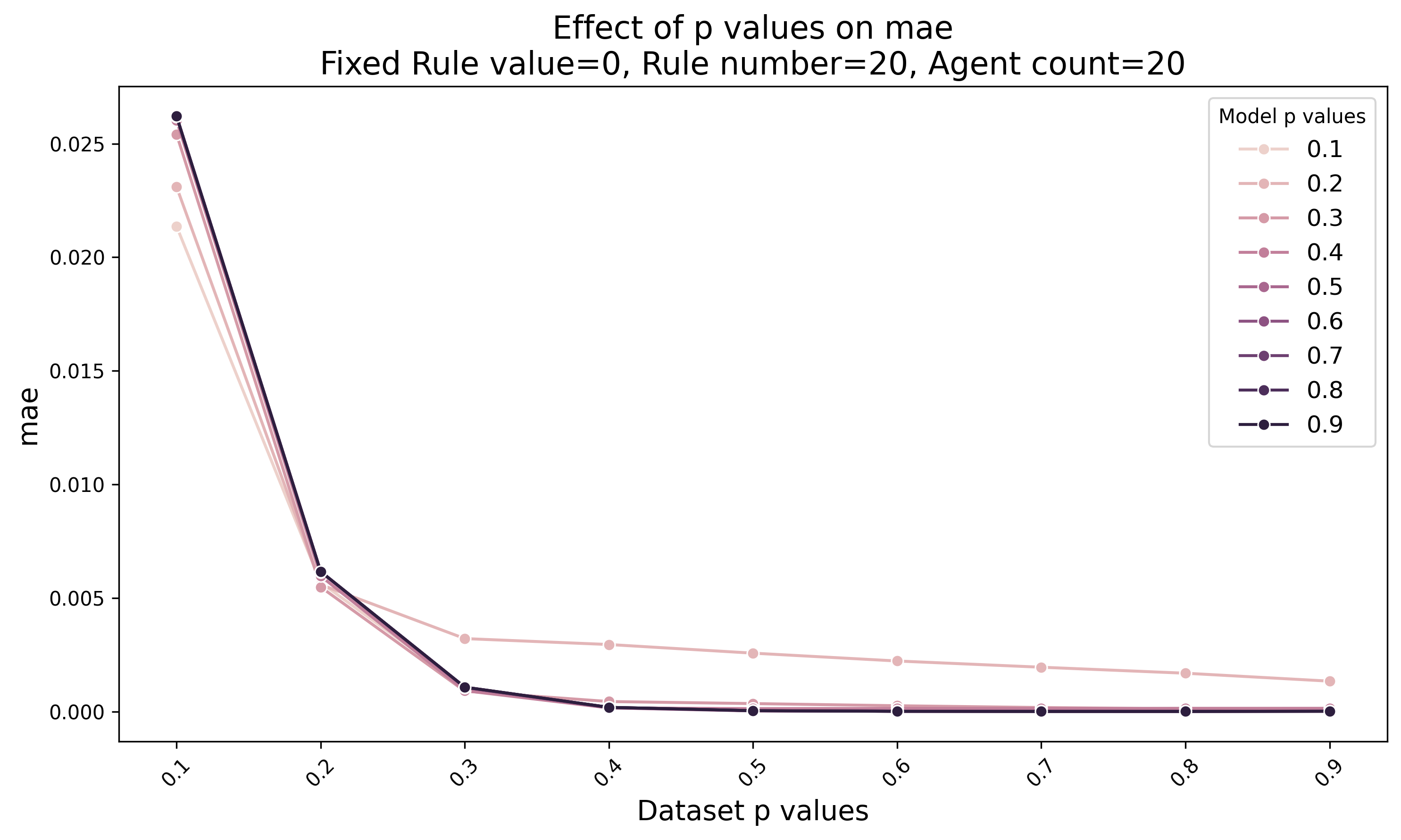}
        \caption{n=20 Agents}
        \label{fig:unpadded_delta_num_agent_subfig2}
    \end{subfigure}
         \hfill
    \begin{subfigure}[b]{0.32\textwidth}
        \includegraphics[width=\textwidth]{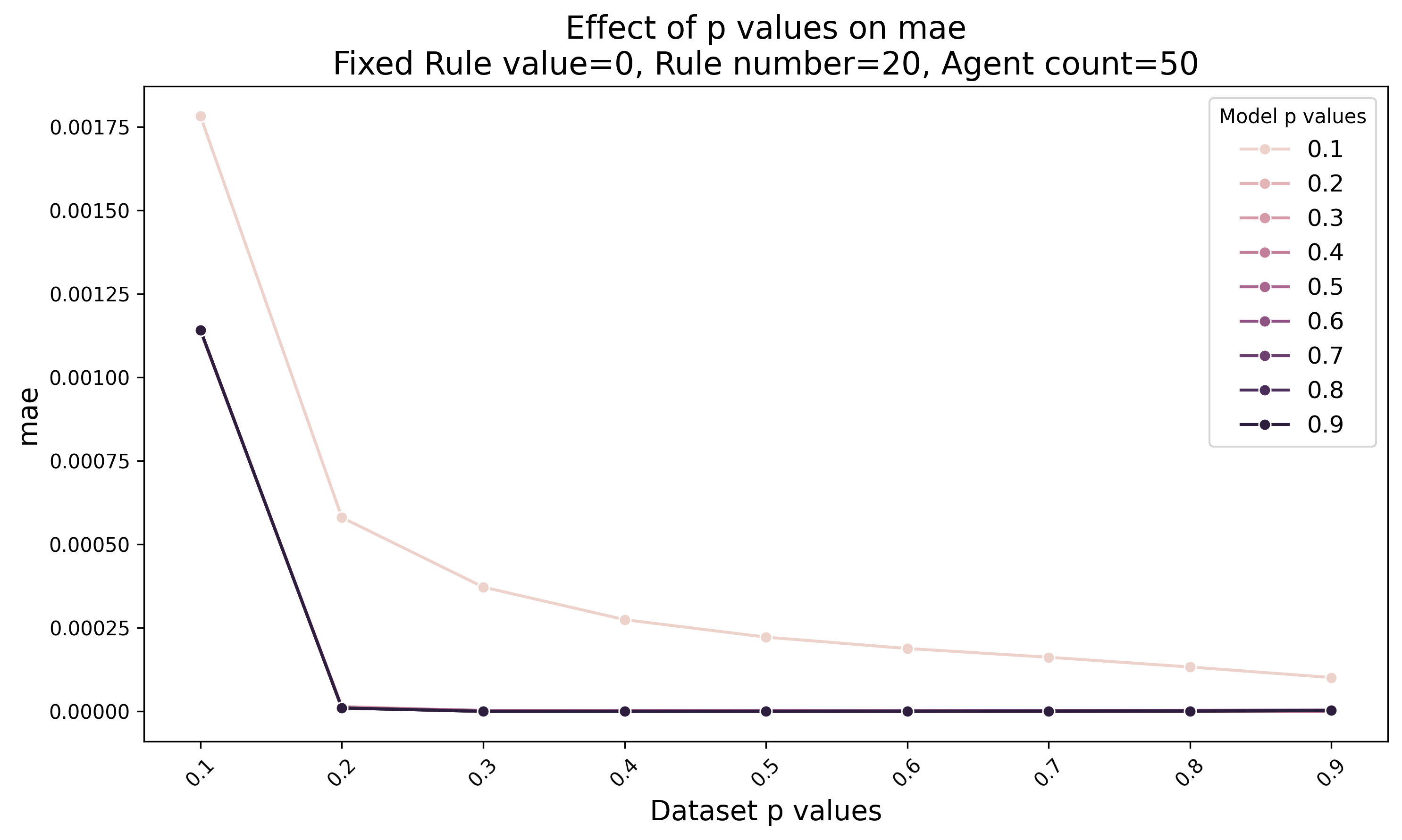}
        \caption{n=50 Agents}
        \label{fig:unpadded_delta_num_agent_subfig3}
    \end{subfigure}
    \caption{Impact of number of agents distribution on model performance using Mean Absolute Error (MAE). Comparison across (A) n=10, (B) n=20 (C) n=50 agents. All configurations maintain consistent number of rules and rule value, but are unpadded and so have different dimensions. The x-axis represents the datasets sparsity threshold (p), and the y-axis shows MAE values.}
    \label{fig:uniform_data_model_different_agent_num_unpadded_MAE}
\end{figure}

To validate that Padding doesn't fundamentally alter our findings, we have compared the model results using unpadded datasets. While this prevents direct comparison across a different number of agents), it allowed us to examine how the p-thresholds affect the  model, and also perform general comparisons between the different dataset parameters. Such Figure~\ref{fig:uniform_data_model_different_agent_num_unpadded_MAE} shows model performance across unpadded datasets of different number of agents.  Though padding improves overall model consistency, particularly for n=10 agents which showed the poorest results, two key patterns persist: sparser datasets consistently underperform compared to denser ones, and larger agent populations yield better results despite greater variance across p thresholds.

\subsection{Cross method distribution analysis}
For detail analysis as seen in figures ~\ref{fig:uniform_cross_p_data_model_same_config_MAE},~\ref{fig:uniform_data_model_different_rule_value_MAE} and ~\ref{fig:uniform_data_model_differnet_number_of_agnets_MAE} for the coinflip random and MoG methods (As described in algorithms \ref{alg:coin_flip_random} and \ref{alg:MOG_random}, respectively) see Appendices ~\ref{appendix:coinflip} and ~\ref{appendix:mog}

\subsection{Graph analysis}
\begin{figure}
    \centering
        \begin{subfigure}[b]{0.45\textwidth}
        \includegraphics[width=\textwidth]{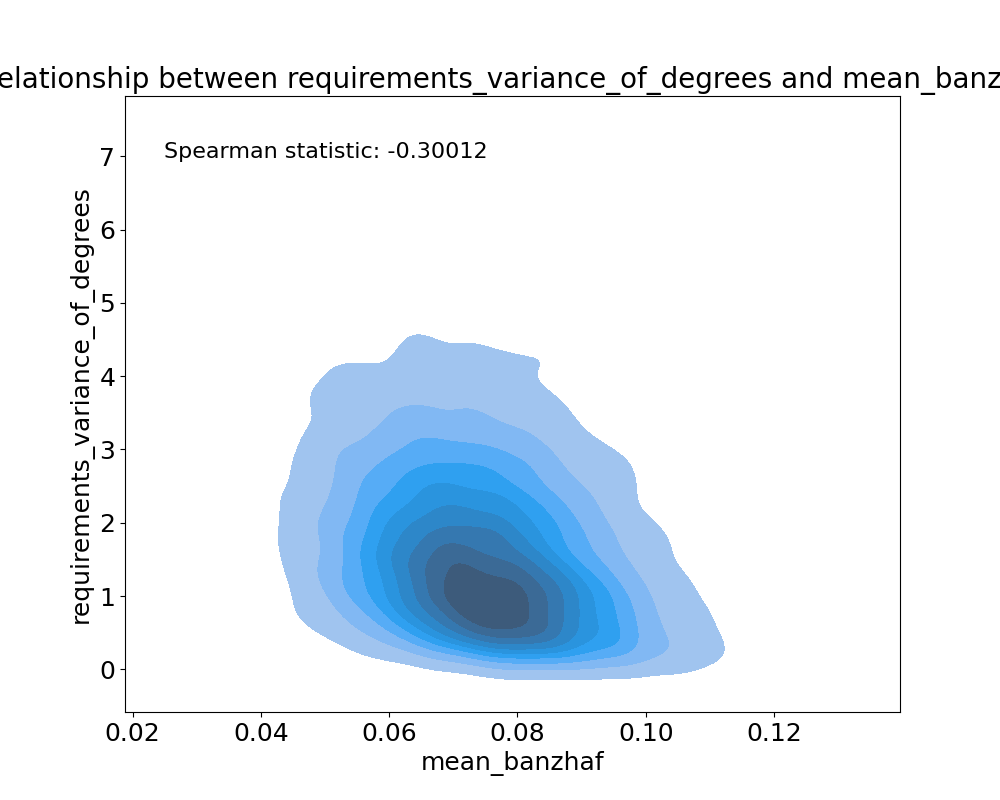}
        \caption{Mean Banzhaf and average degree of requirements}
        \label{fig:mean_vs_avg_degree_req}
    \end{subfigure}
         \hfill
    \begin{subfigure}[b]{0.45\textwidth}
        \includegraphics[width=\textwidth]{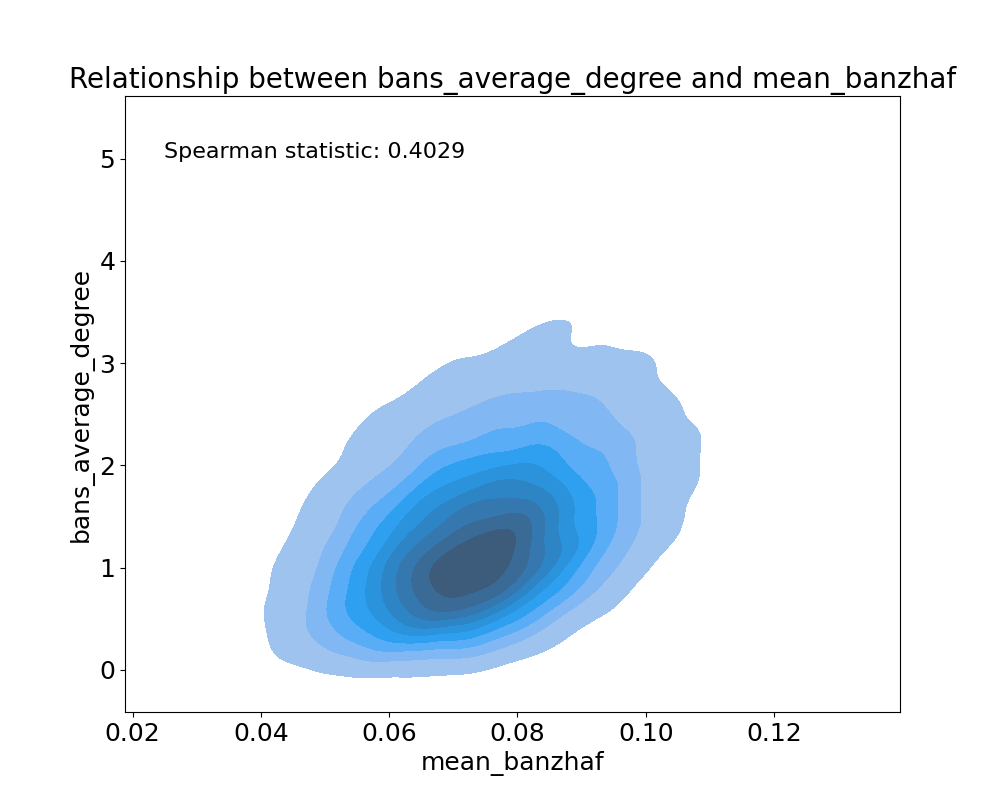}
        \caption{Mean Banzhaf and average degree of bans}
        \label{fig:mean_vs_avg_degree_ban}
    \end{subfigure}
        \\[\baselineskip]
    \begin{subfigure}[b]{0.45\textwidth}
        \includegraphics[width=\textwidth]{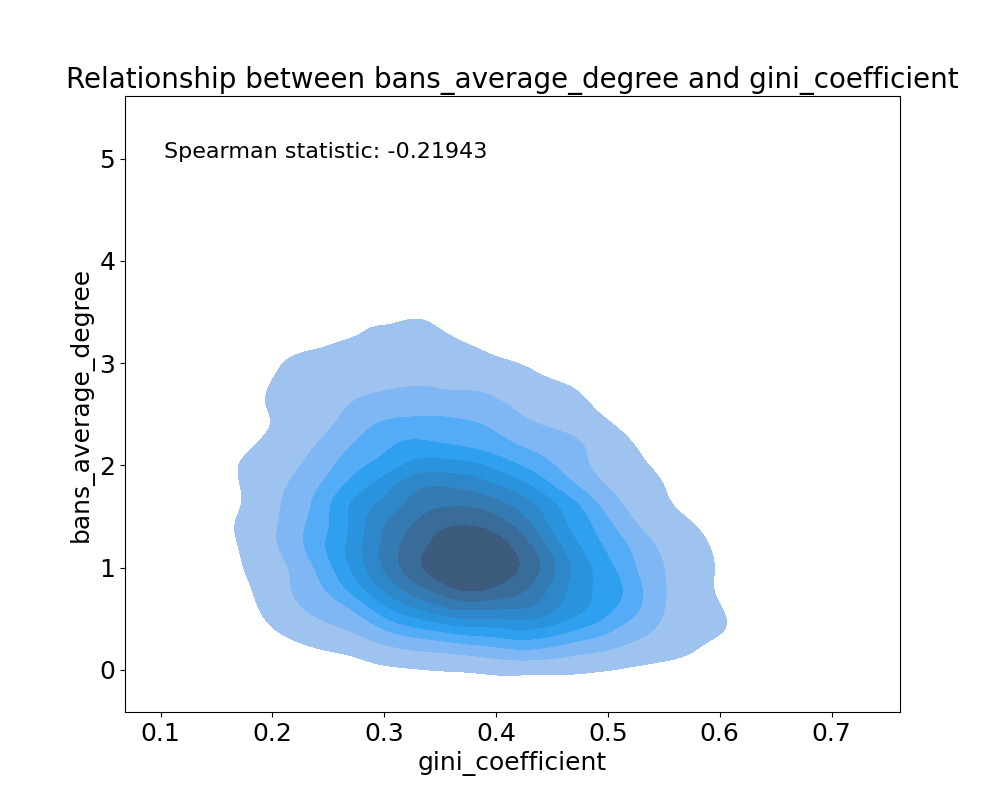}
        \caption{Banzhaf Gini coefficient and average degree of Bans}
        \label{fig:gini_vs_avg_degree_ban}
    \end{subfigure}
    \hfill
        \begin{subfigure}[b]{0.45\textwidth}
        \includegraphics[width=\textwidth]{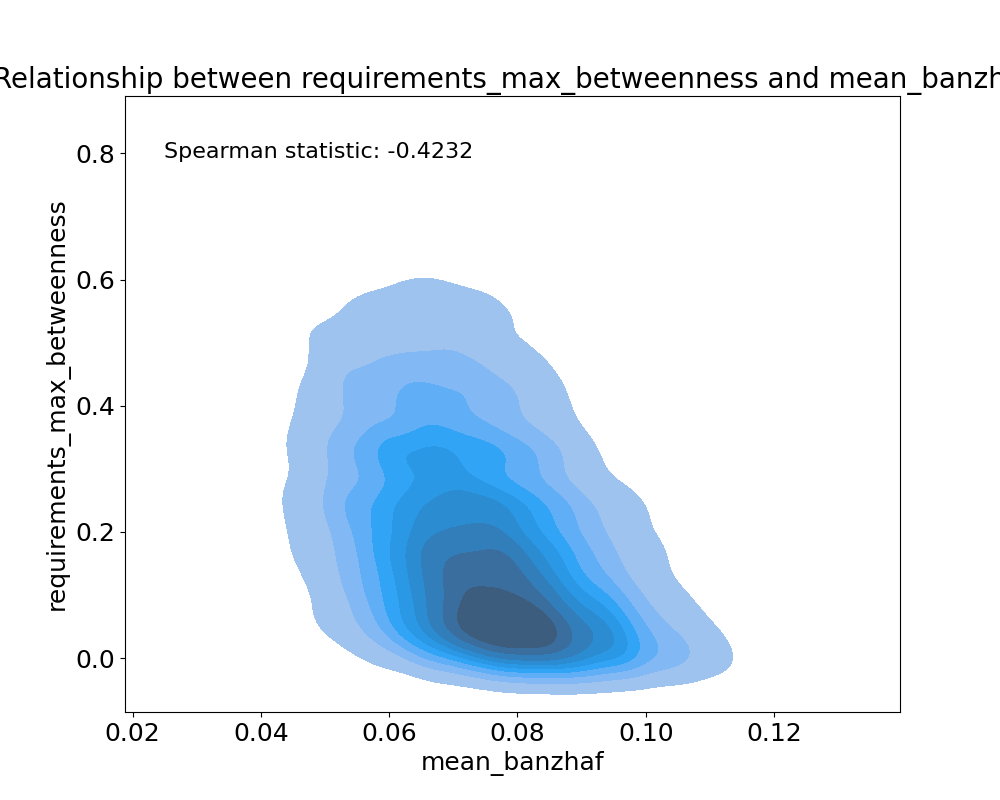}
        \caption{Mean Banzhaf and max betweenness of requirements}
        \label{fig:mean_vs_max_betweeness}
    \end{subfigure}
    
    \caption{Correlation comparison of coalition Banzhaf distribution and the coalitions as a graph statistics. All figures were based on coalitions with p threshold of 0.1, n=10 agents and uniform rule value. Notice the similar strength and reverse sign comparing the mean Banzhaf values and the 'req' and 'ban' average degree in figure (A) and (B), respectively.}
    \label{fig:graph_banzhaf_plot}
\end{figure}

Having the ability to measure the power index of large coalitions, we decided to explore how voting coalitions can be described as graphs, and see how they correlate with different power index stats. We converted the requirements and the bans into separate undirectional graphs, and measured the Spearman correlation between the power index statistics and the graph statistics, as seen in figure~\ref{fig:graph_banzhaf_plot}. We find that the strongest correlations (highest absolute $\rho$ values) were achieved in sparser coalitions (i.e. p-threshold $\leq 0.3$). This is in accordance with these datasets having the greatest variance in their coalition distribution and points of freedom, whereas the denser datasets are with higher constraints, reducing the power of the individual agents.

We filtered out the statistically insignificant results, and kept all cases with Spearman $\rho>0.2 $ and p value$\leq0.05$. Then we measured which attributes are the most prevalent across all datasets turned to graphs. The results in Table~\ref{table:banzhaf_graph_summary_statistics} show the percentage of degree of significant correlations that occurred in all datasets.

\begin{table}[!ht]
    \centering
    \caption{Table showing the repeated significant correlation statistics between graphs and Banzhaf values}
    \label{table:banzhaf_graph_summary_statistics}
    \begin{tabular}{|l|l|l|}
    \hline
    \textbf{Banzhaf Statistics} & \textbf{Graph Statistics} & \textbf{Significant in \% of datasets} \\ \hline
        Mean Banzhaf & Average clustering of requirements & 44.44 \\ \hline
        Mean Banzhaf & Average degree requirements & 44.44 \\ \hline
        Mean Banzhaf & Max betweenness of requirements & 38.89 \\ \hline
        Mean Banzhaf & Variance degrees of requirements & 33.33 \\ \hline
        Mean Banzhaf & Size of largest clique of requirements & 33.33 \\ \hline
        Variance Banzhaf & Average degree of requirements & 29.63 \\ \hline
        Banzhaf Gini coefficient & Average degree of bans & 22.22 \\ \hline
        Mean Banzhaf & Average degree of bans & 12.96 \\ \hline
        Variance Banzhaf & Max betweenness of requirements & 12.96 \\ \hline
    \end{tabular}
\end{table}
\FloatBarrier

\section{Discussion}
Our research demonstrates that NN can effectively approximate power indices for large-scale voting systems, offering a significant advancement in computational efficiency while maintaining high accuracy. This approach addresses a fundamental challenge in cooperative game theory and political science, where traditional computational methods often become intractable for larger systems. By reducing computation time from 15 and 50 minutes runtime for the Monte-Carlo approximations (Generating the n=10 and n=50 agent matrices, respectively) to just 8 minutes model training time for each of the models for comparable results, our method makes the analysis of complex voting systems more accessible for researchers and practitioners.

The NN models exhibited several notable characteristics that challenge conventional assumptions about power index computation. First, contrary to initial expectations, the models showed improved accuracy when dealing with larger agent sets ($n=50$) despite the increased computational complexity these scenarios typically present. This counter-intuitive finding suggests that larger systems may contain more regular patterns that neural networks can leverage for prediction. The increased dimensionality appears to provide richer information for power index estimation, potentially capturing underlying mathematical structures that become more pronounced at scale.

Our analysis of sparse-versus-dense coalition rules revealed interesting patterns in model behaviour. Models trained on dense coalition rules demonstrated inferior performance when evaluated on sparse datasets, suggesting that these configurations contain more distinct and learnable patterns that models trained on denser dataset aren't presented with. However, the reverse more prominent differences indicate that the models aren't robust in either direction of different coalition densities. This limited robustness is particularly important for real-world applications, where voting systems may vary significantly in their constraint density. further research is required to improve this.

The integration of graph-theoretical analysis with power index prediction provides novel insights into coalition dynamics. Strong correlations between graph metrics (such as average degree and clustering coefficients) and power indices, particularly in sparse coalitions (p $\leq$\ 0.3), suggest that network structure plays a crucial role in determining voting power. These findings could inform the design of voting systems by helping predict how structural changes might affect power distribution.

The computational efficiency gains are most pronounced for larger coalition sizes, which posed the greatest challenges for traditional methods. While Monte-Carlo approximations showed exponential increases in computation time with coalition size, our NN approach maintained relatively consistent training times regardless of the number of agents. This scalability makes our method particularly valuable for analysing previously intractable large-scale voting systems, such as those found in corporate shareholding structures or international political bodies.

However, several limitations and areas for future research should be acknowledged:
\begin{itemize}
    \item NN rely on existing previous labels. For such large rule set ($m\geq20$) and number of agents ($n\geq50$), the initial approximation needed in order to generate the labels is a severe bottleneck.
    \item while we show the good performance given this NN architecture, better results may be achieved using more sophisticated NN architecture.
    \item Our current approach focuses on static voting systems; extending the methodology to handle dynamic coalitions where relationships between agents evolve over time represents an important direction for future work.
    \item The impact of different rule value distributions on model performance needs further study.
\end{itemize}

Future work should focus on overcoming current limitations by developing specialized architectures for large-scale voting systems, integrating temporal components to account for dynamic coalition formation and exploring advanced power index approximation techniques for training data generation. The impact of this research extends beyond theoretical considerations, offering practical tools for analysing power dynamics in a wide range of applications, including corporate governance systems with complex ownership structures, international political organizations with diverse voting procedures, and social networks where influence plays a crucial role in decision-making.

This research represents a significant step toward making power index analysis more accessible and practical for large-scale applications, while opening new avenues for understanding complex voting systems through the lens of machine learning.

\section{Acknowledgment}
We thank Omer Shechter from Reichman University for his productive discussions and assistance with running some of the computations.


\clearpage
\appendix
\section{Coin-flip results}
\label{appendix:coinflip}
\begin{figure}[h!] 
    \centering
    \begin{subfigure}[b]{0.3\textwidth}
        \includegraphics[width=\textwidth]{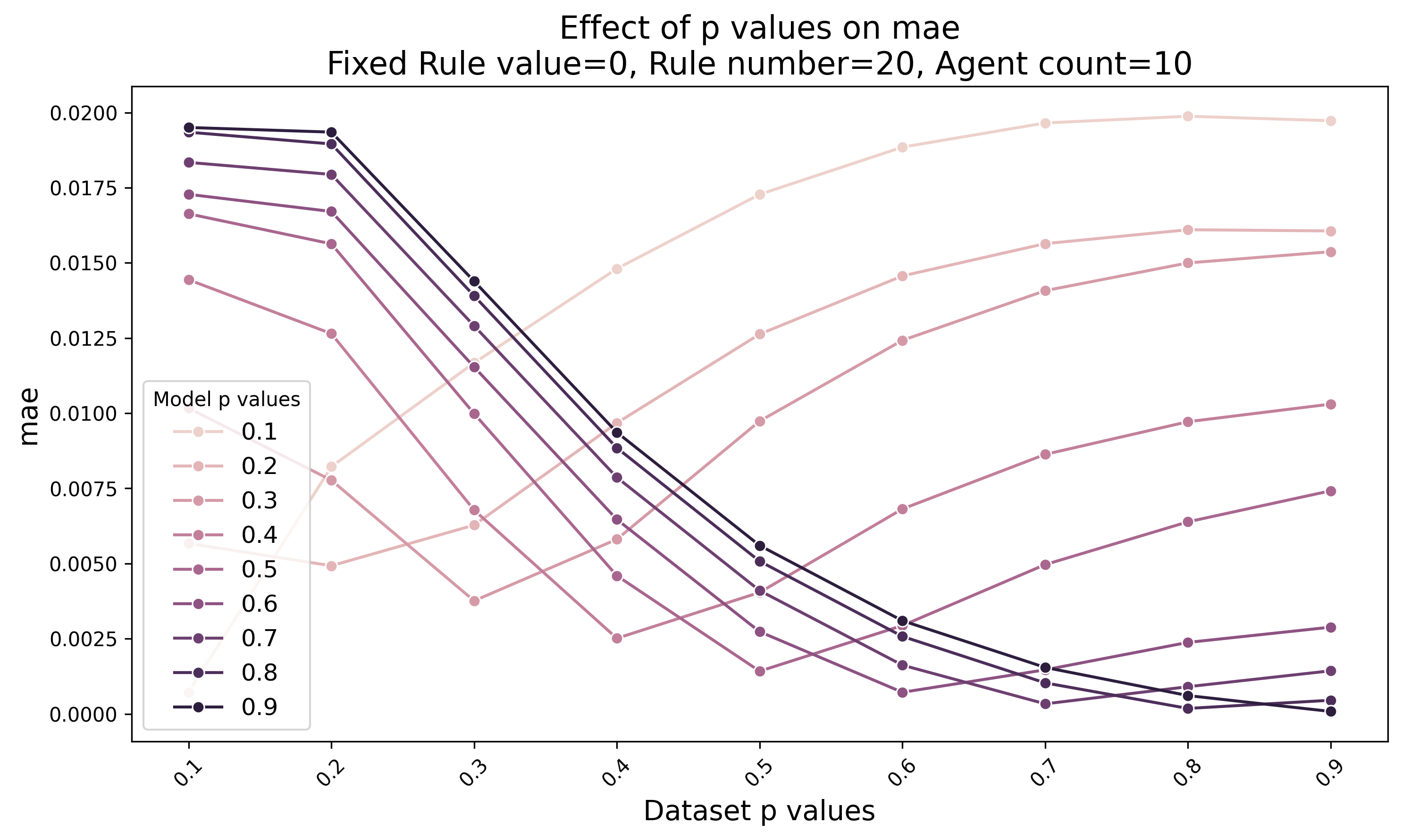}
        \caption{n=10 agents, uniform rules}
        \label{fig:coinflip_subfig1}
    \end{subfigure}
     \hfill
    \begin{subfigure}[b]{0.3\textwidth}
        \includegraphics[width=\textwidth]{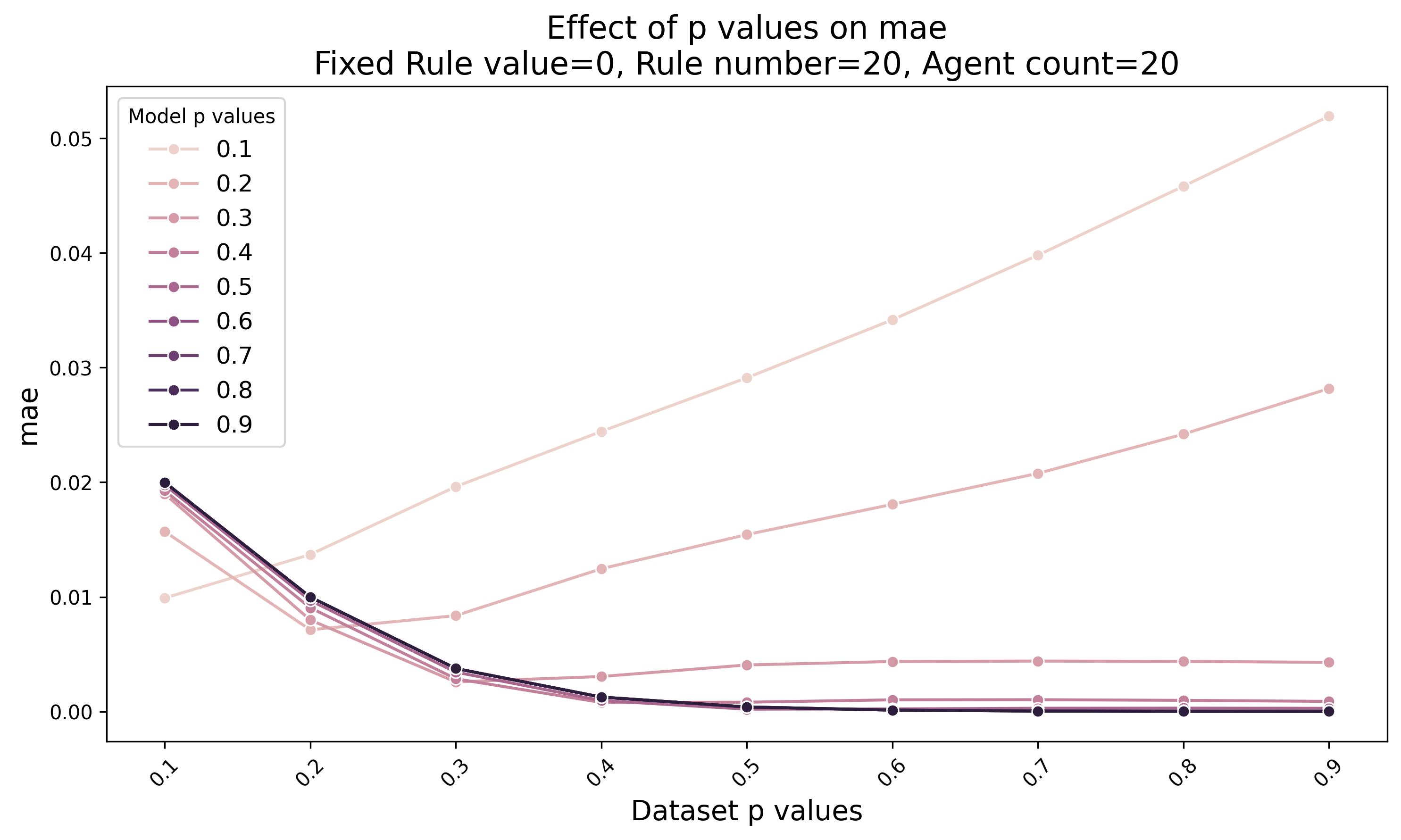}
        \caption{n=20 agents, uniform rules}
        \label{fig:coinflip_subfig2}
    \end{subfigure}
     \hfill
    \begin{subfigure}[b]{0.3\textwidth}
        \includegraphics[width=\textwidth]{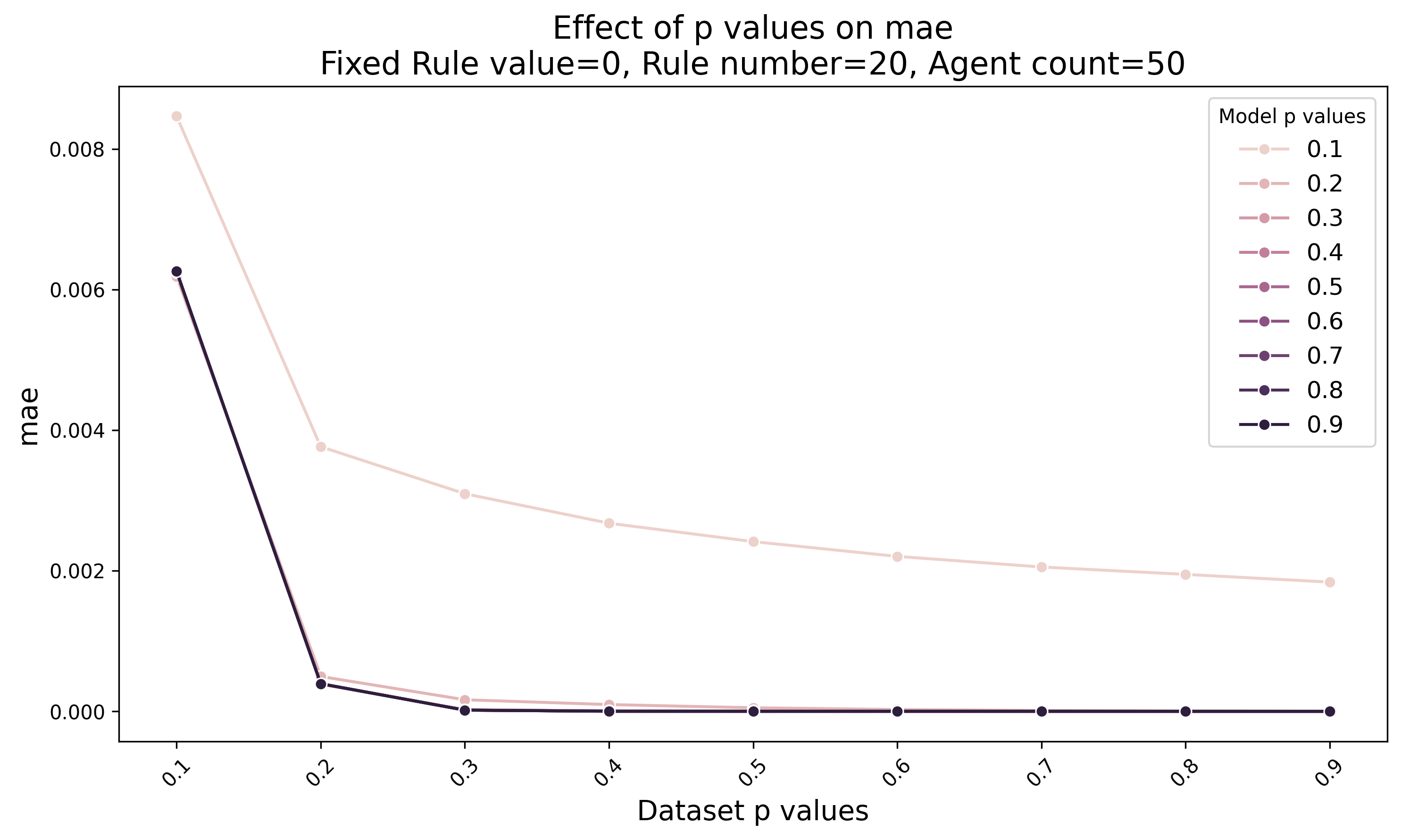}
        \caption{n=50 agents, uniform rules}
        \label{fig:coinflip_subfig3}
    \end{subfigure}
    \\[\baselineskip]
    \begin{subfigure}[b]{0.3\textwidth}
        \includegraphics[width=\textwidth]{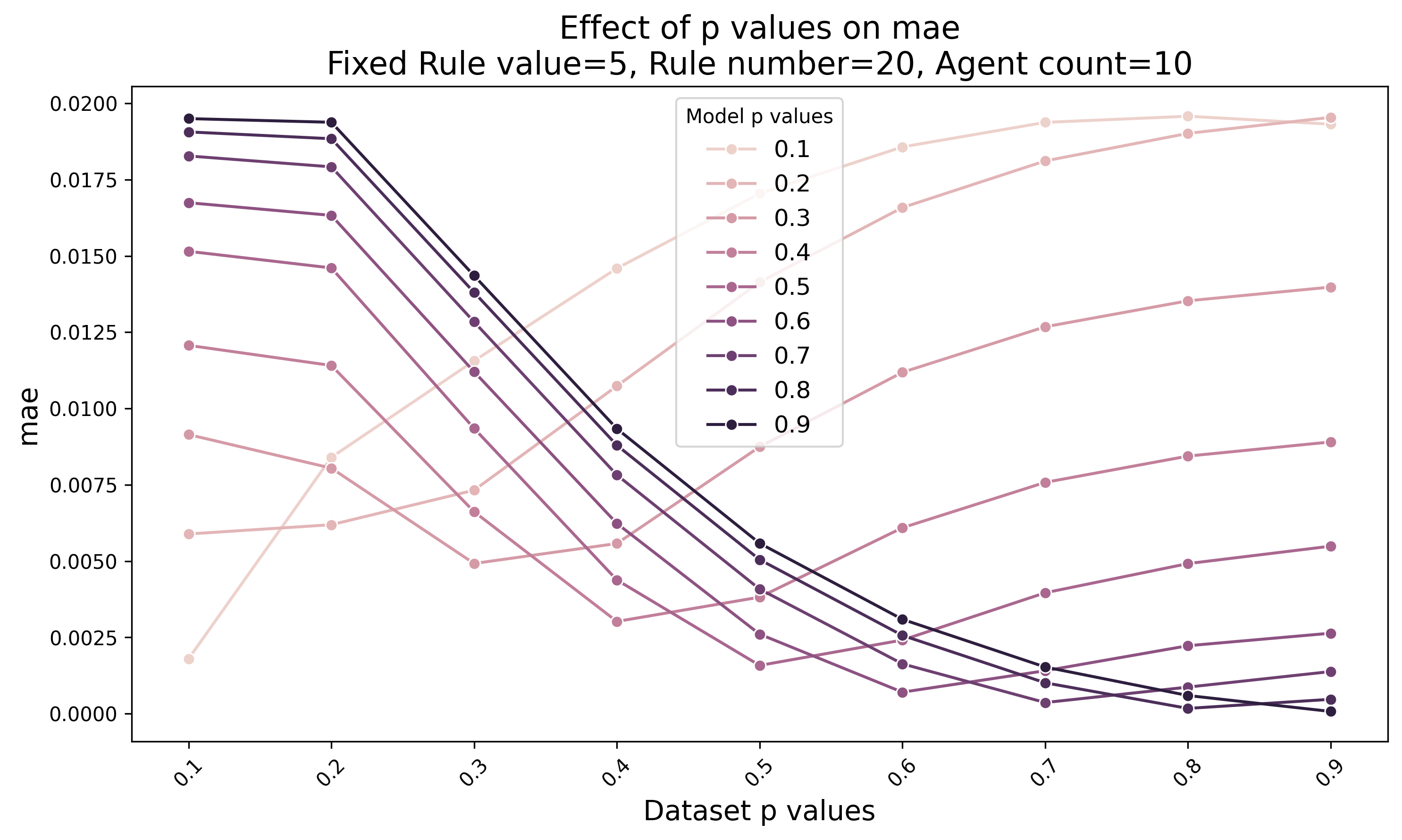}
        \caption{n=10 agents, Low-variance Gaussian rules ($\mu = 5$)}
        \label{fig:coinflip_subfig4}
    \end{subfigure}
     \hfill
    \begin{subfigure}[b]{0.3\textwidth}
        \includegraphics[width=\textwidth]{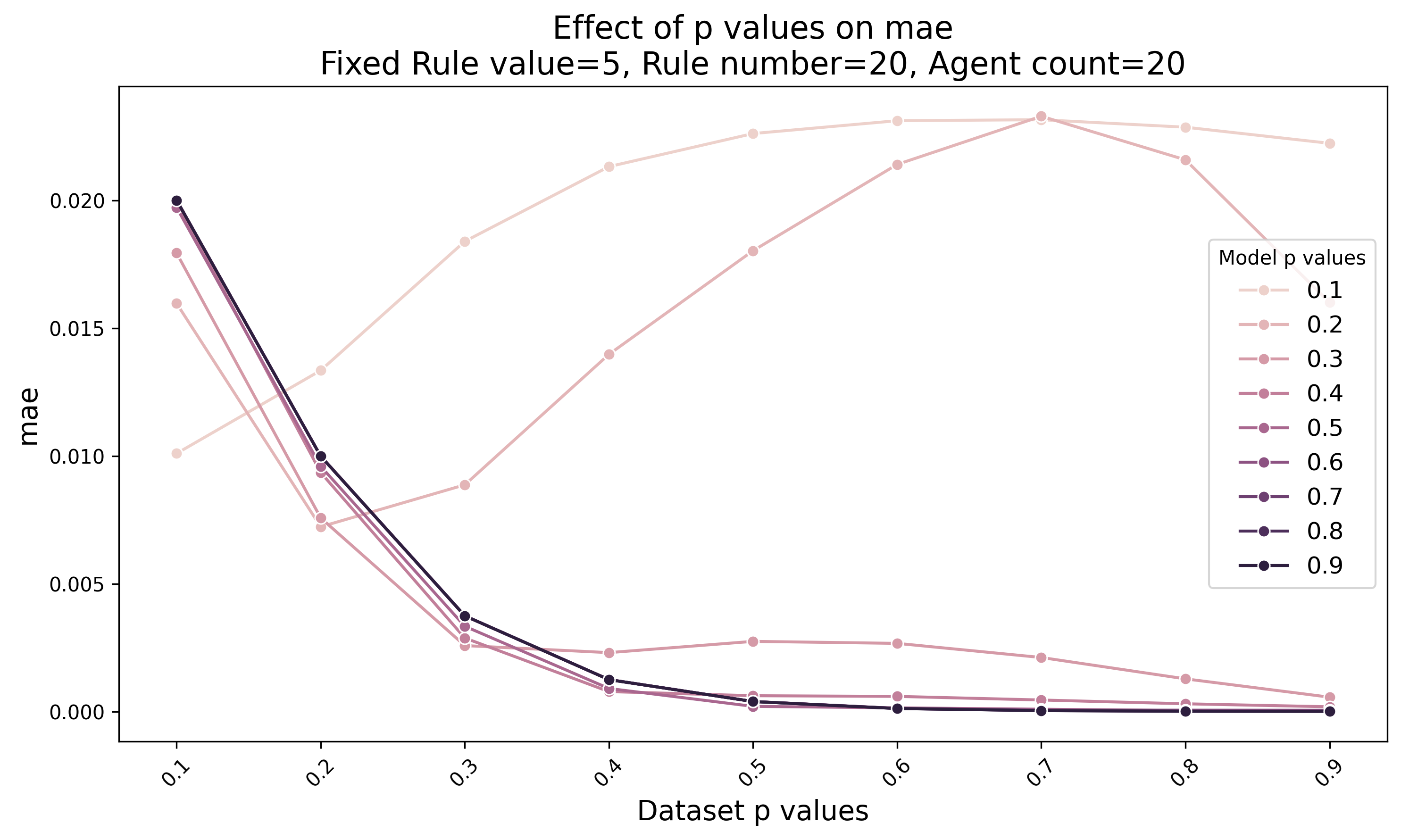}
        \caption{n=20 agents, Low-variance Gaussian rules ($\mu = 5$)}
        \label{fig:coinflip_subfig5}
    \end{subfigure}
     \hfill
    \begin{subfigure}[b]{0.3\textwidth}
        \includegraphics[width=\textwidth]{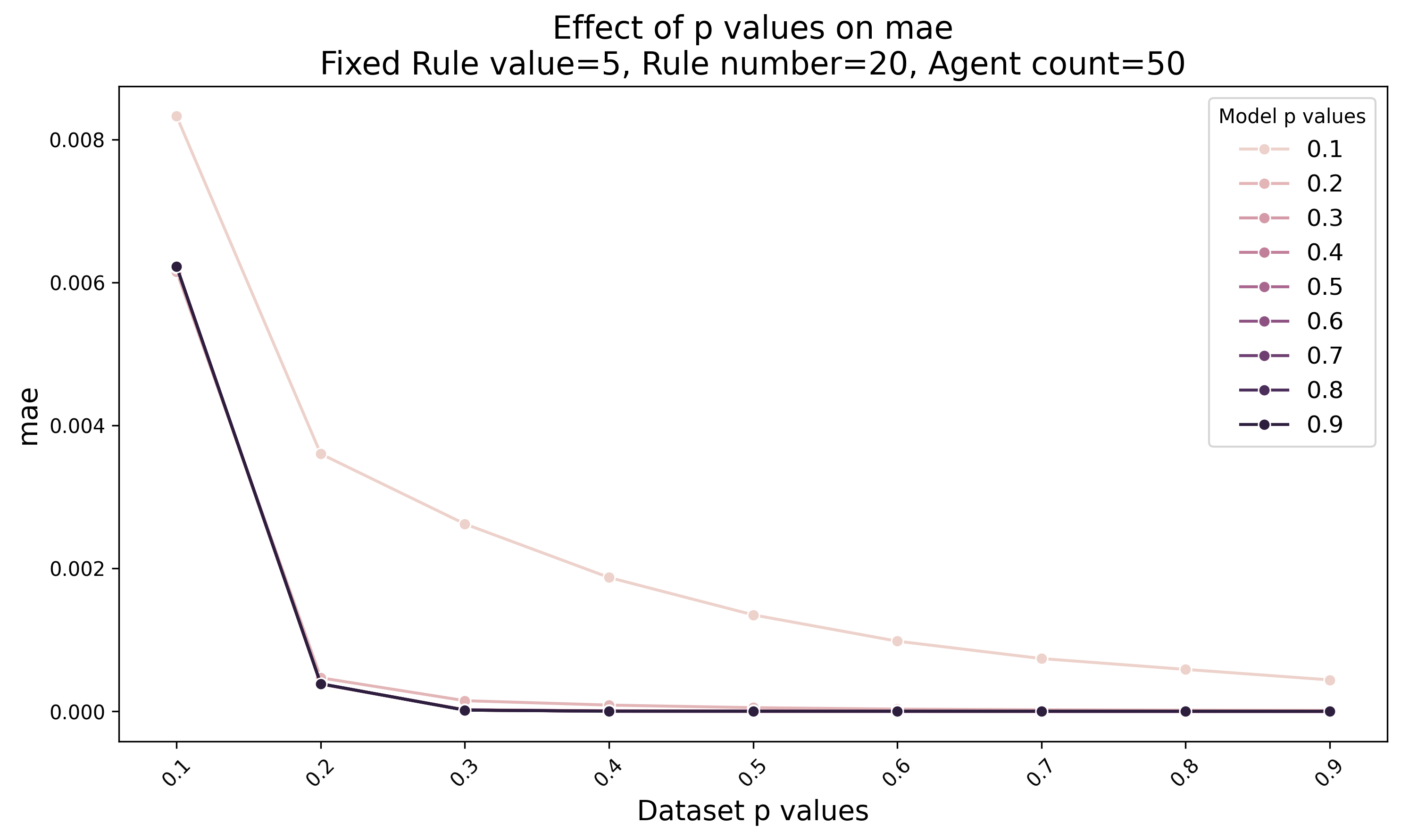}
        \caption{n=50 agents, Low-variance Gaussian rules ($\mu = 5$)}
        \label{fig:coinflip_subfig6}
    \end{subfigure}
    \\[\baselineskip]
    \begin{subfigure}[b]{0.3\textwidth}
        \includegraphics[width=\textwidth]{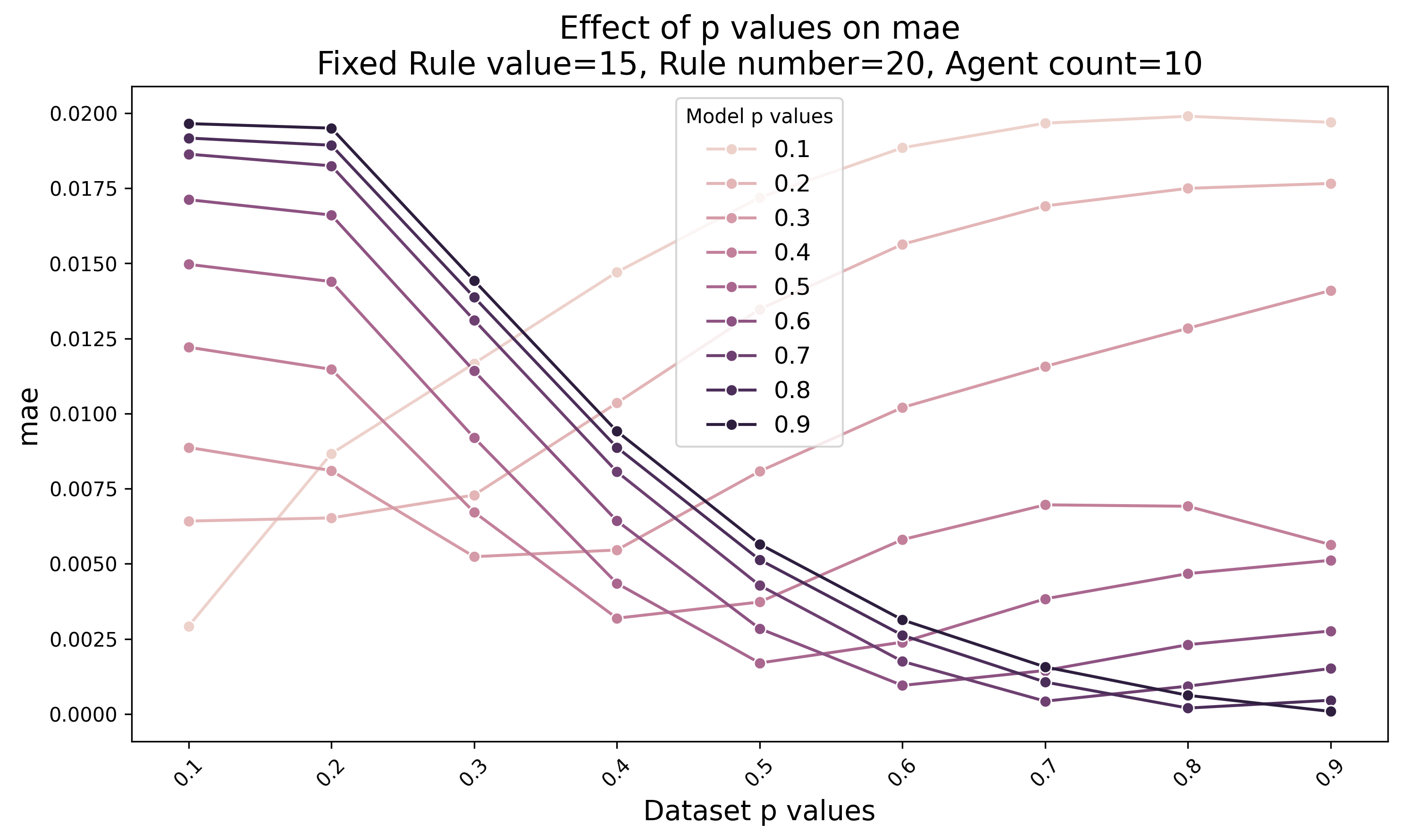}
        \caption{n=10 agents, High-variance Gaussian rules ($\mu = 15$)}
        \label{fig:coinflip_subfig7}
    \end{subfigure}
     \hfill
    \begin{subfigure}[b]{0.3\textwidth}
        \includegraphics[width=\textwidth]{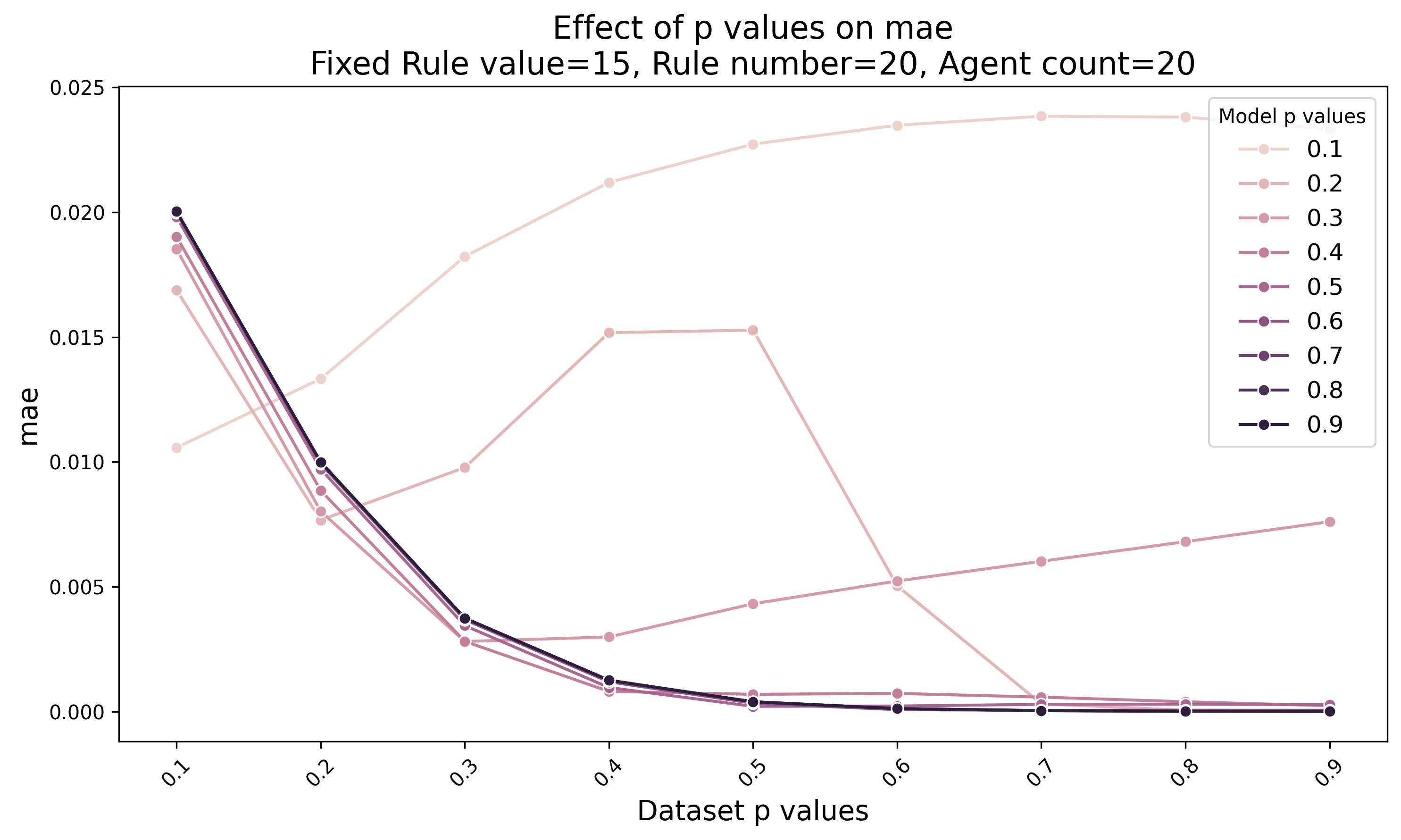}
        \caption{n=20 agents, High-variance Gaussian rules ($\mu = 15$)}
        \label{fig:coinflip_subfig8}
    \end{subfigure}
     \hfill
    \begin{subfigure}[b]{0.3\textwidth}
        \includegraphics[width=\textwidth]{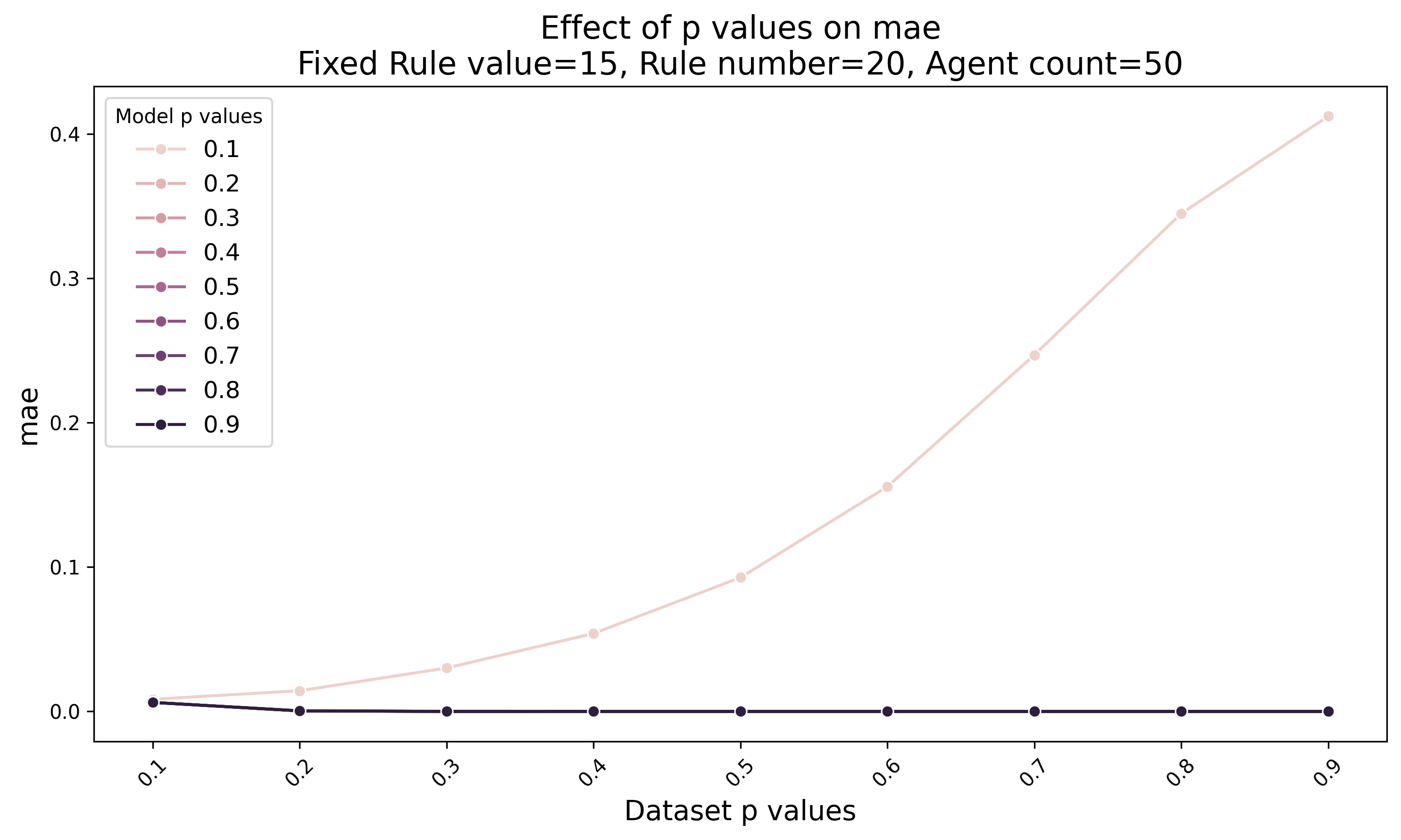}
        \caption{n=50 agents, High-variance Gaussian rules ($\mu = 15$)}
        \label{fig:coinflip_subfig9}
    \end{subfigure}
    \caption{Models performance on coin-flip random datasets Mean Absolute Error (MAE) across different dataset configurations. The x-axis represents the sparsity threshold (p), and the y-axis shows the MAE values. Subplots show variations across: (A-C) uniform rule values with 20 rules and 10, 20, and 50 agents respectively; (D-F) low-variance Gaussian distribution ($\mu=5$) with 20 rules across 10, 20, and 50 agents; (G-I) high-variance Gaussian distribution ($\mu=15$) with 20 rules for 10, 20 and 50 agents. Each line within subplots represents a model trained with a different p value while maintaining consistent coalition parameters.}
    \label{fig:coinflip_cross_p_data_model_same_config_MAE}
\end{figure}

Cross p-threshold comparisons across all coin-flip datasets are presented in Figure ~\ref{fig:coinflip_cross_p_data_model_same_config_MAE}. For n=50 agents(right column, subfigures C, F and I) the model comparison behaviour mirrors that observed in the models trained on uniformly distributed random datasets. However, smaller agent populations exhibit distinct patterns. 

With n=10 agents, (left column, subfigures A, D and G) we observe a V-shaped pattern:  models perform optimally with the p-threshold datasets they were trained on, but show diminishing performance at both higher and lower p-values, with particularly poor performance on lower p0value datasets. 

For n=20 (middle column, subfigures B, E and H), models trained on higher p-values demonstrate robust performance, while those trained on lower p-values ($p\leq0.3$) maintain the V-shaped performance pattern observed in the n=10 case.

\begin{figure}
    \centering    
    \begin{subfigure}[b]{0.32\textwidth}
        \includegraphics[width=\textwidth]{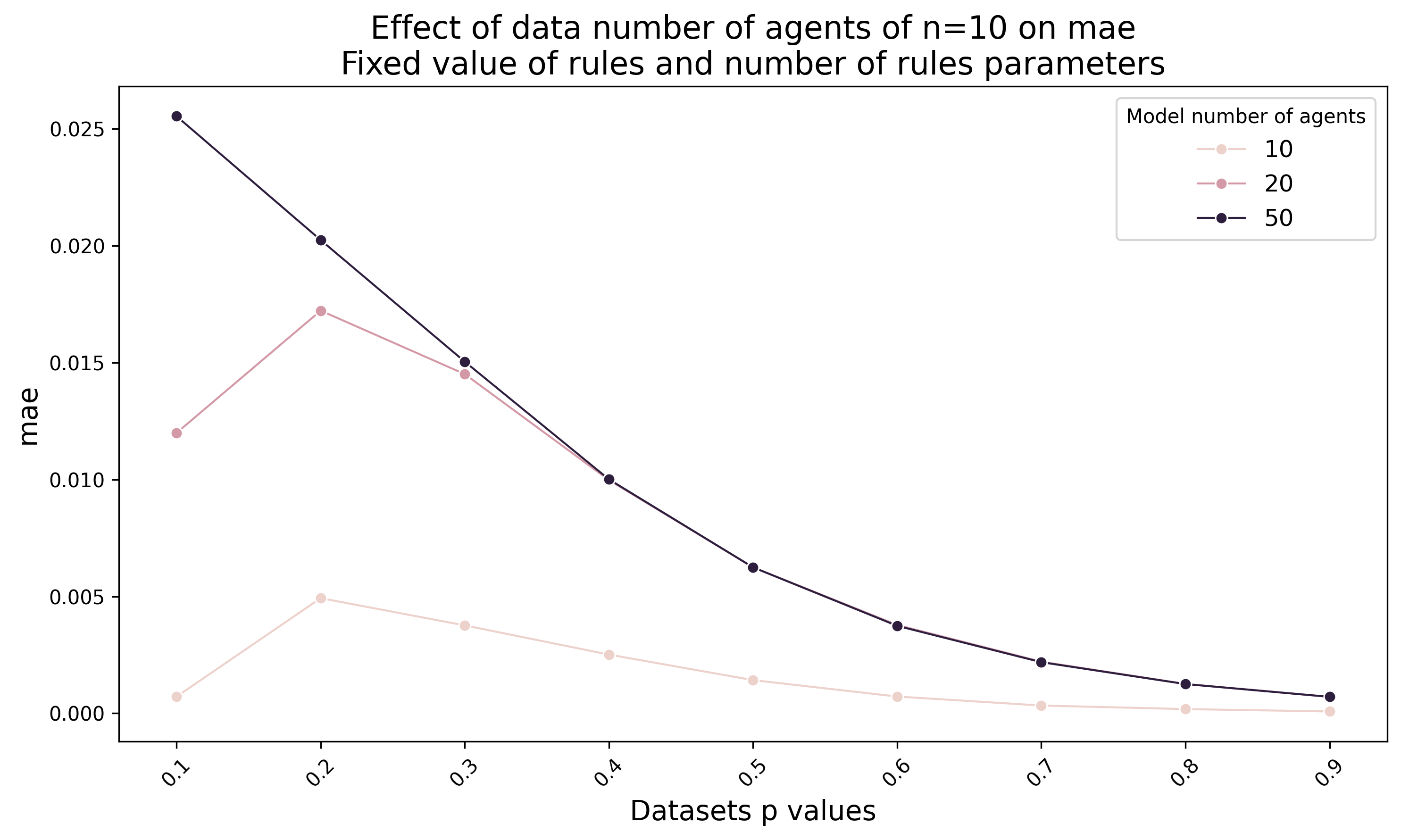}
        \caption{Performance on 10-agent datasets}
        \label{fig:delta_agent_subfig1}
    \end{subfigure}
         \hfill
    \begin{subfigure}[b]{0.32\textwidth}
        \includegraphics[width=\textwidth]{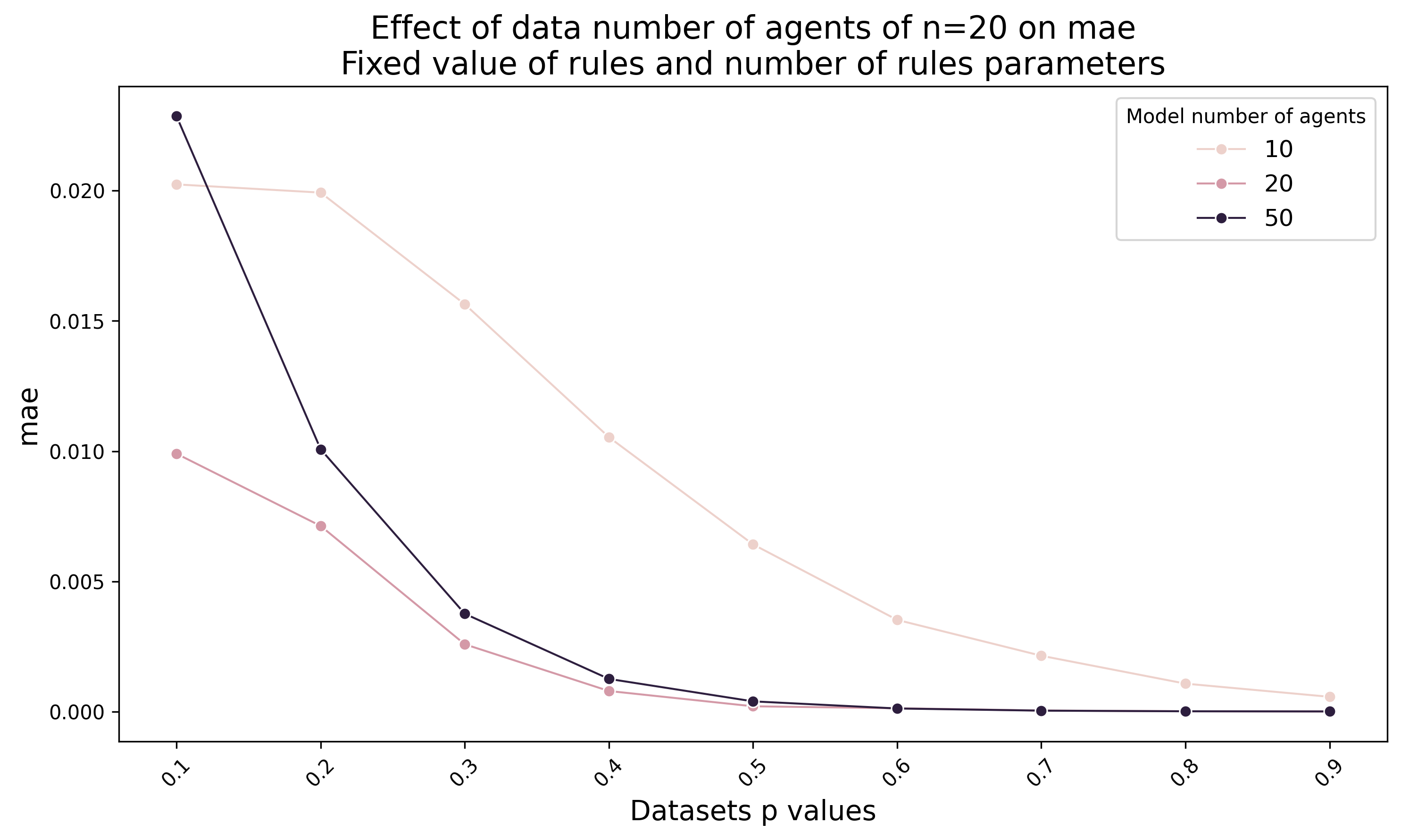}
        \caption{Performance on 20-agent datasets}
        \label{fig:delta_agent_subfig2}
    \end{subfigure}
     \hfill
    \begin{subfigure}[b]{0.32\textwidth}
        \includegraphics[width=\textwidth]{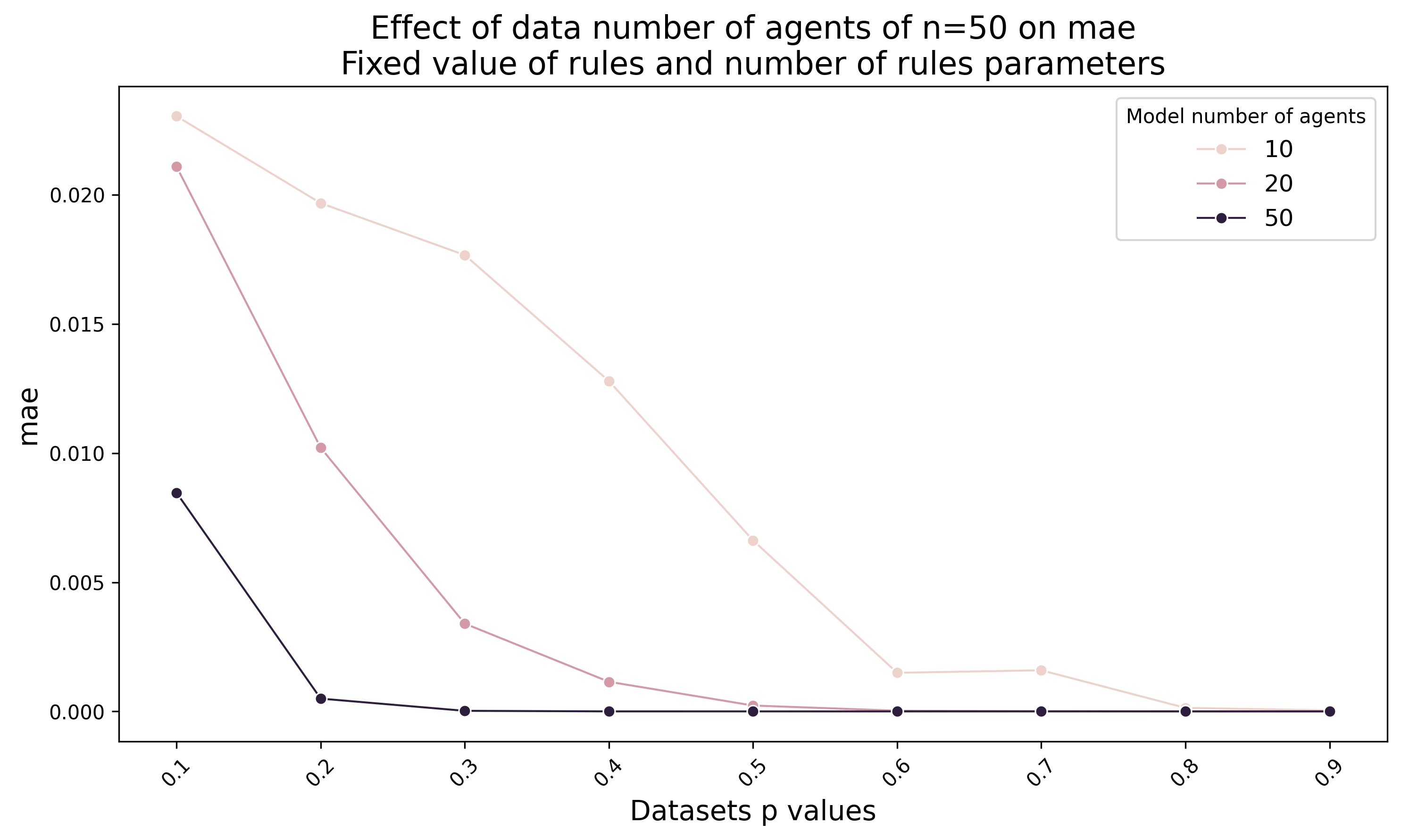}
        \caption{Performance on 50-agent datasets}
        \label{fig:delta_agent_subfig3}
    \end{subfigure}
    \caption{Cross-model comparison of Mean Absolute Error (MAE) for different numbers of agents. Results show model performance on datasets with (A) n=10 agents, (B) n=20 agents, and (C) n=50 agents, while maintaining consistent rule values and number of rules across all configurations. The x-axis represents the sparsity threshold (p), and the y-axis shows MAE values.}
    \label{fig:coinflip_data_model_differnet_number_of_agents_MAE}
\end{figure}

Further analysis of model performance across varying agent populations is illustrated in Figure ~\ref{fig:coinflip_data_model_differnet_number_of_agents_MAE}. Models achieve peak performance when tested on populations matching their training conditions, with performance degrading proportionately to the deviation from the training population size. For instance, models trained on n=50 agents show greater performance deterioration when tested on n=10 datasets compared to models trained on n=20 agents.

\begin{figure}
    \centering
    \begin{subfigure}[b]{0.32\textwidth}
        \includegraphics[width=\textwidth]{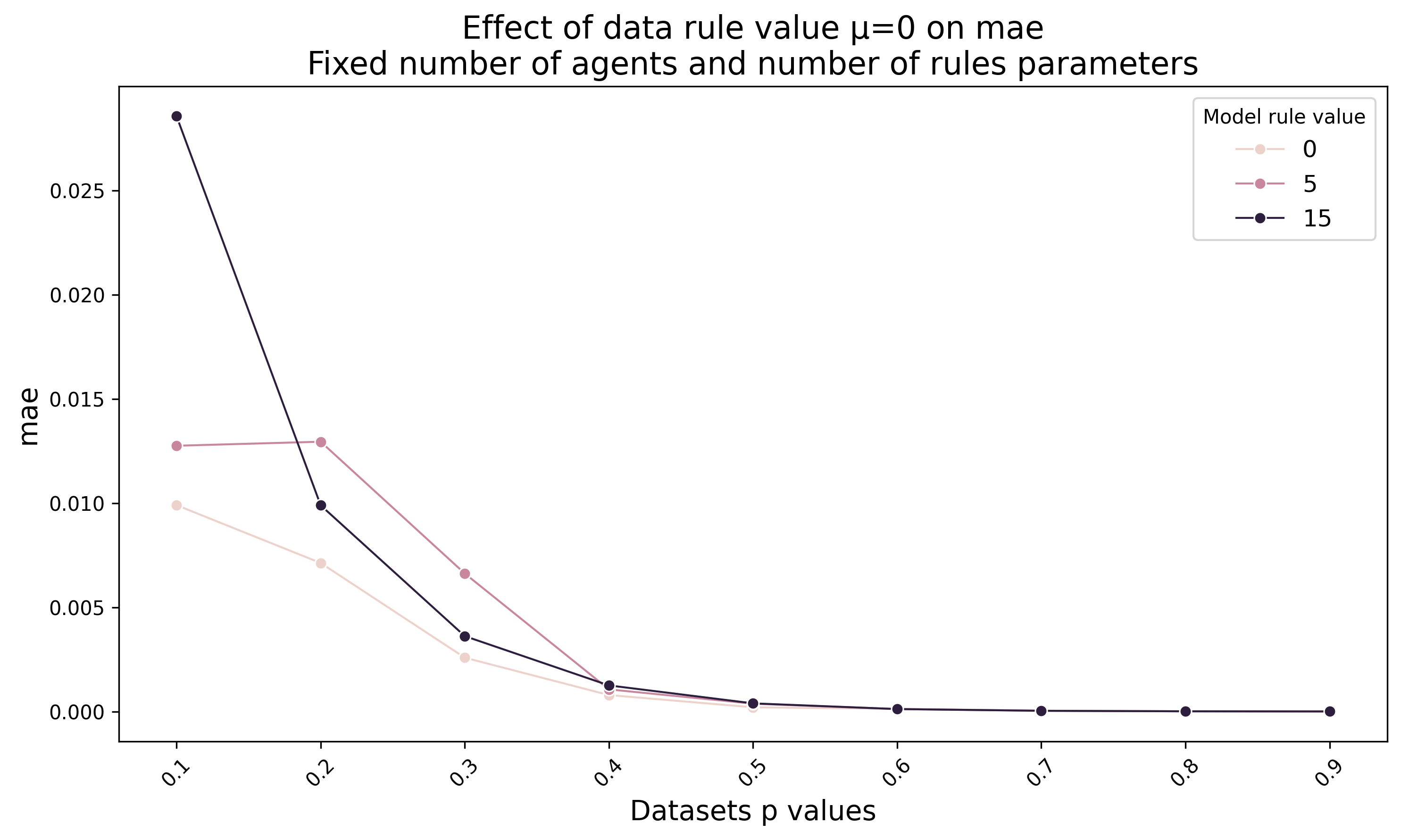}
        \caption{Uniform rule value distribution}
        \label{fig:delta_rule_value_subfig1}
    \end{subfigure}
         \hfill
    \begin{subfigure}[b]{0.32\textwidth}
        \includegraphics[width=\textwidth]{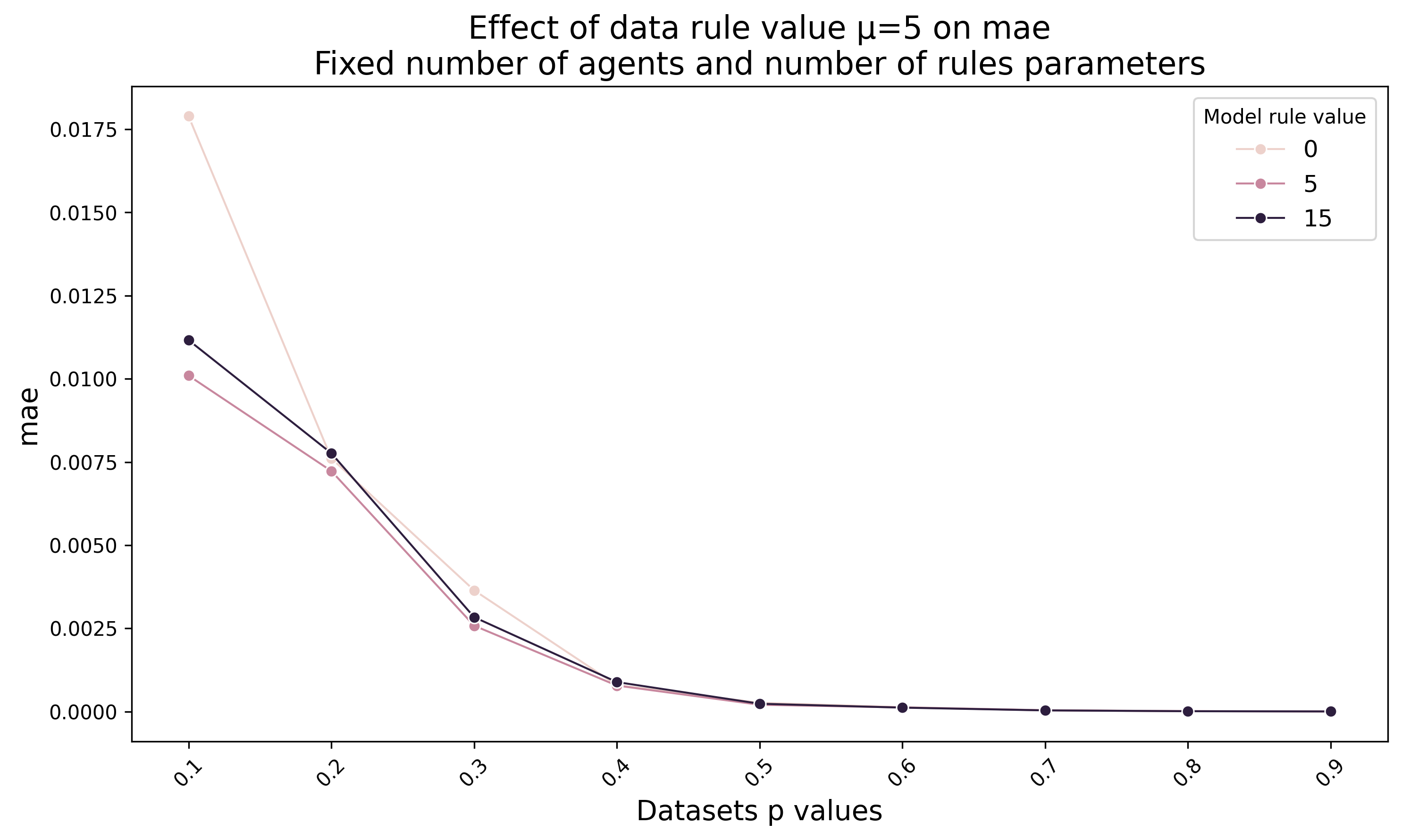}
        \caption{Low-variance $(\mu=5)$ rule value distribution}
        \label{fig:delta_rule_value_subfig2}
    \end{subfigure}
         \hfill
    \begin{subfigure}[b]{0.32\textwidth}
        \includegraphics[width=\textwidth]{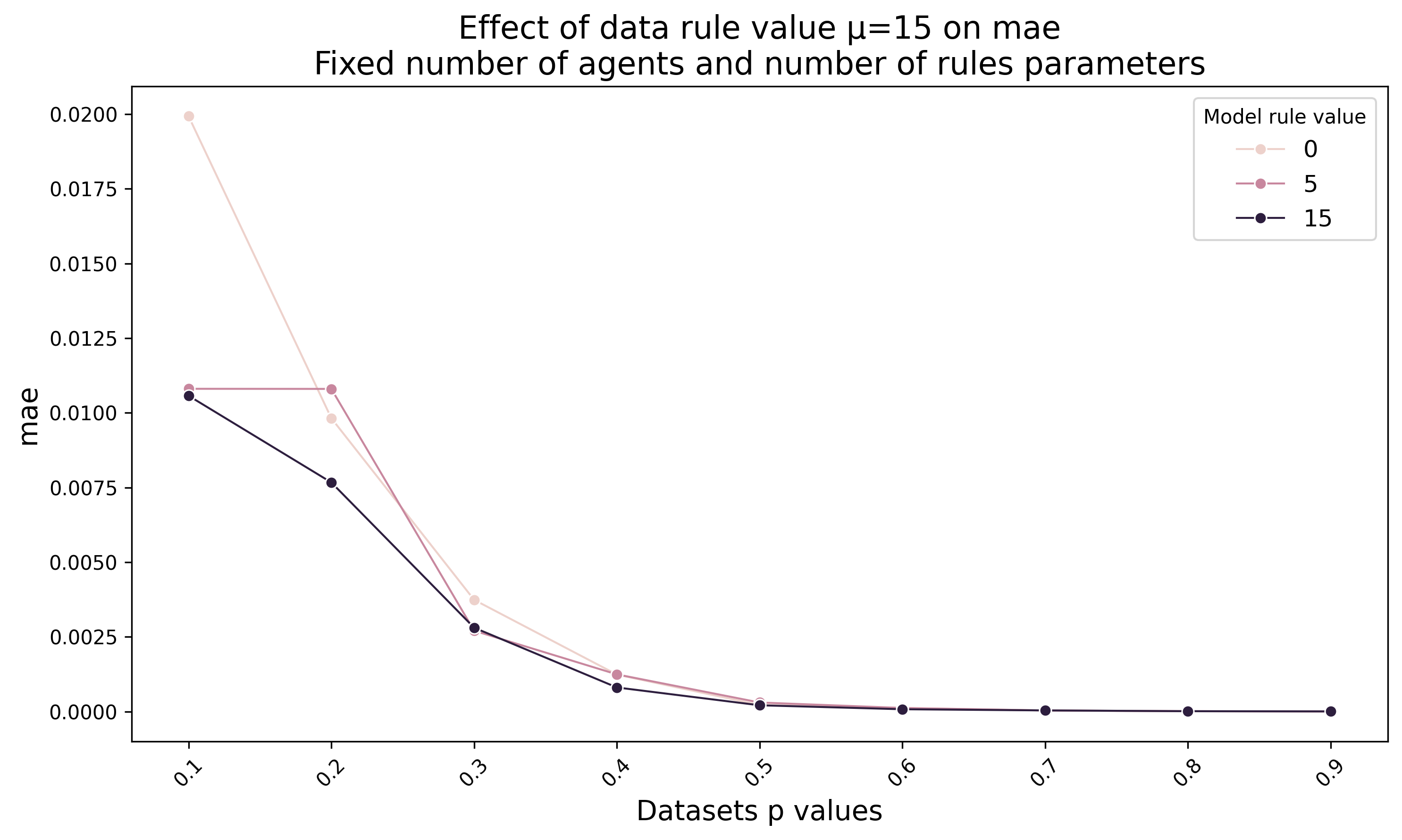}
        \caption{High-variance $(\mu=15)$ rule value distribution}
        \label{fig:delta_rule_value_subfig3}
    \end{subfigure}
    \caption{Impact of rule value distribution on model performance using Mean Absolute Error (MAE). Comparison across (A) uniform rule value, (B) low-variance Gaussian rule value, and (C) high-variance Gaussian rule value distributions. All configurations maintain consistent number of rules and agents. The x-axis represents the sparsity threshold (p), and the y-axis shows MAE values.}
    \label{fig:coinflip_data_model_different_rule_value_MAE}
\end{figure}

Analysis of rule value variations, presented in Figure ~\ref{fig:coinflip_data_model_different_rule_value_MAE}, reveals distinct patterns across dataset types. In uniform random dataset, models trained on both high and low variance rule values show reduced performance, with high variance conditions yielding the poorest results. However, in the remaining two datasets, models trained on high and low variance rule values perform comparatively, while models trained on uniform random conditions show consistently inferior performance. This suggest that while the transition from uniform to alternating rule values significantly impact model performance, the magnitude of rule-value variance has limited influence, with models demonstrating robust performance under both high and low variance conditions.

\FloatBarrier

\section{Mixture of Gaussian results}
\label{appendix:mog}
\begin{figure}[h!] 
    \centering
    \begin{subfigure}[b]{0.32\textwidth}
        \includegraphics[width=\textwidth]{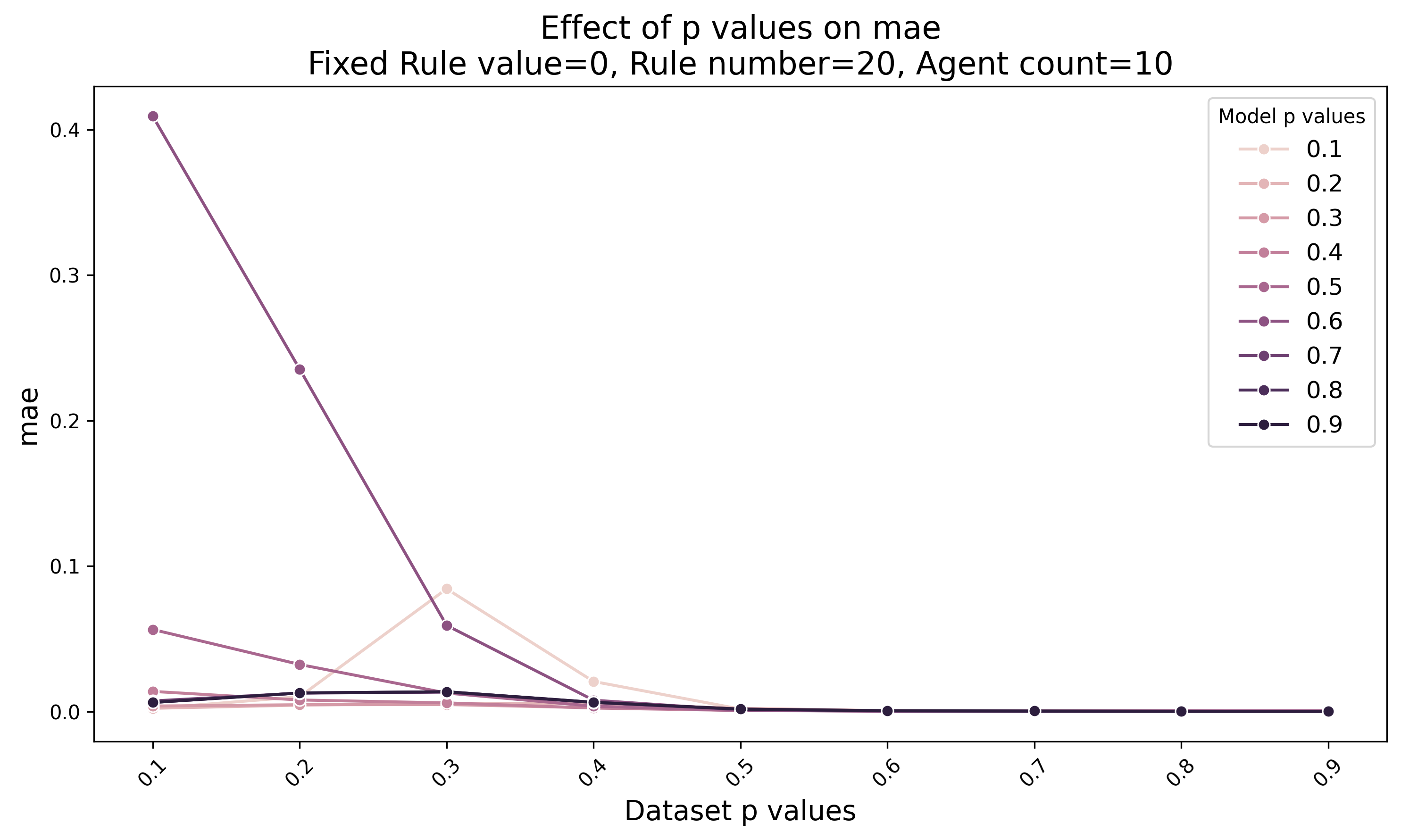}
        \caption{n=10 agents, uniform rules}
        \label{fig:coinflip_subfig1}
    \end{subfigure}
     \hfill
    \begin{subfigure}[b]{0.32\textwidth}
        \includegraphics[width=\textwidth]{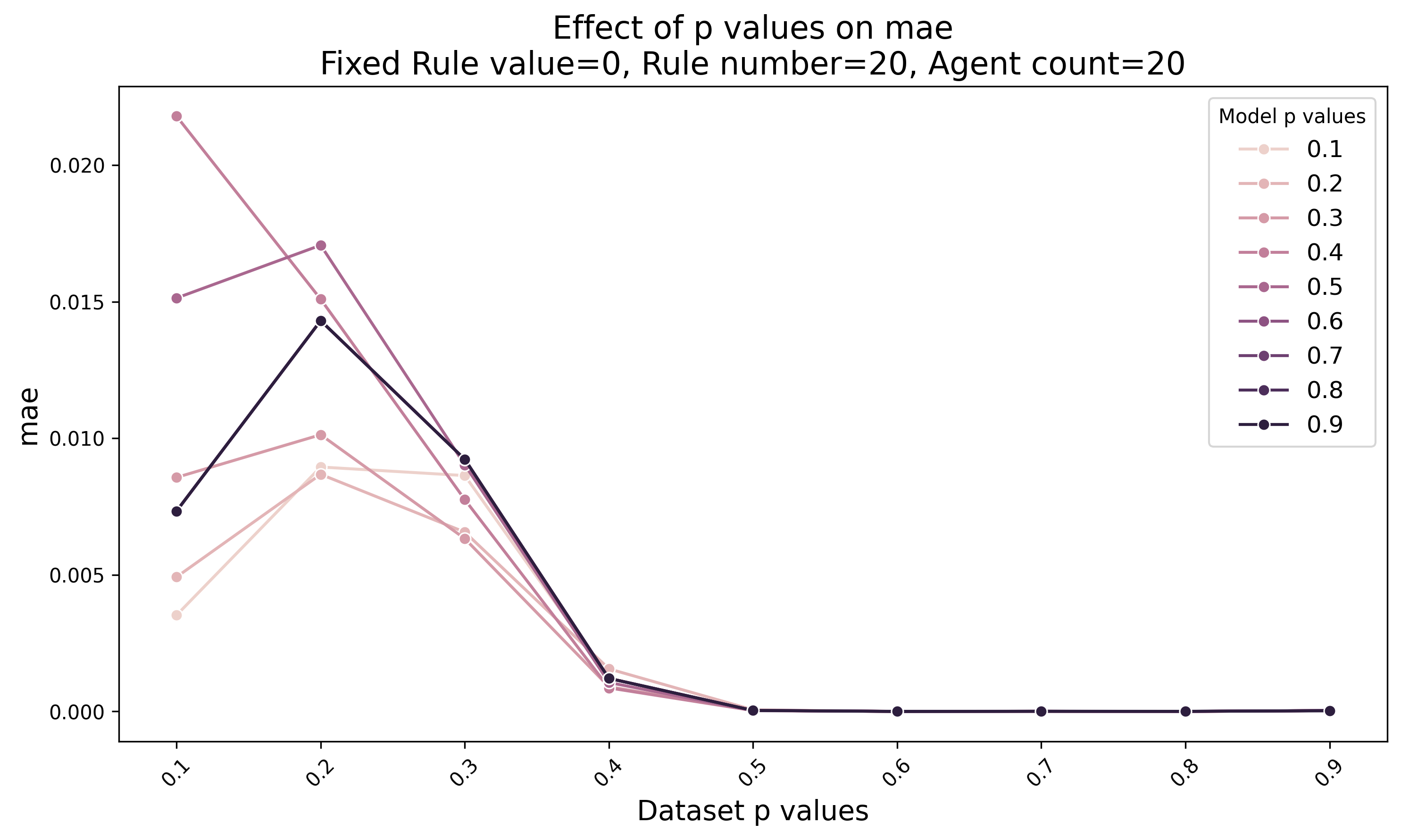}
        \caption{n=20 agents, uniform rules}
        \label{fig:coinflip_subfig2}
    \end{subfigure}
     \hfill
    \begin{subfigure}[b]{0.32\textwidth}
        \includegraphics[width=\textwidth]{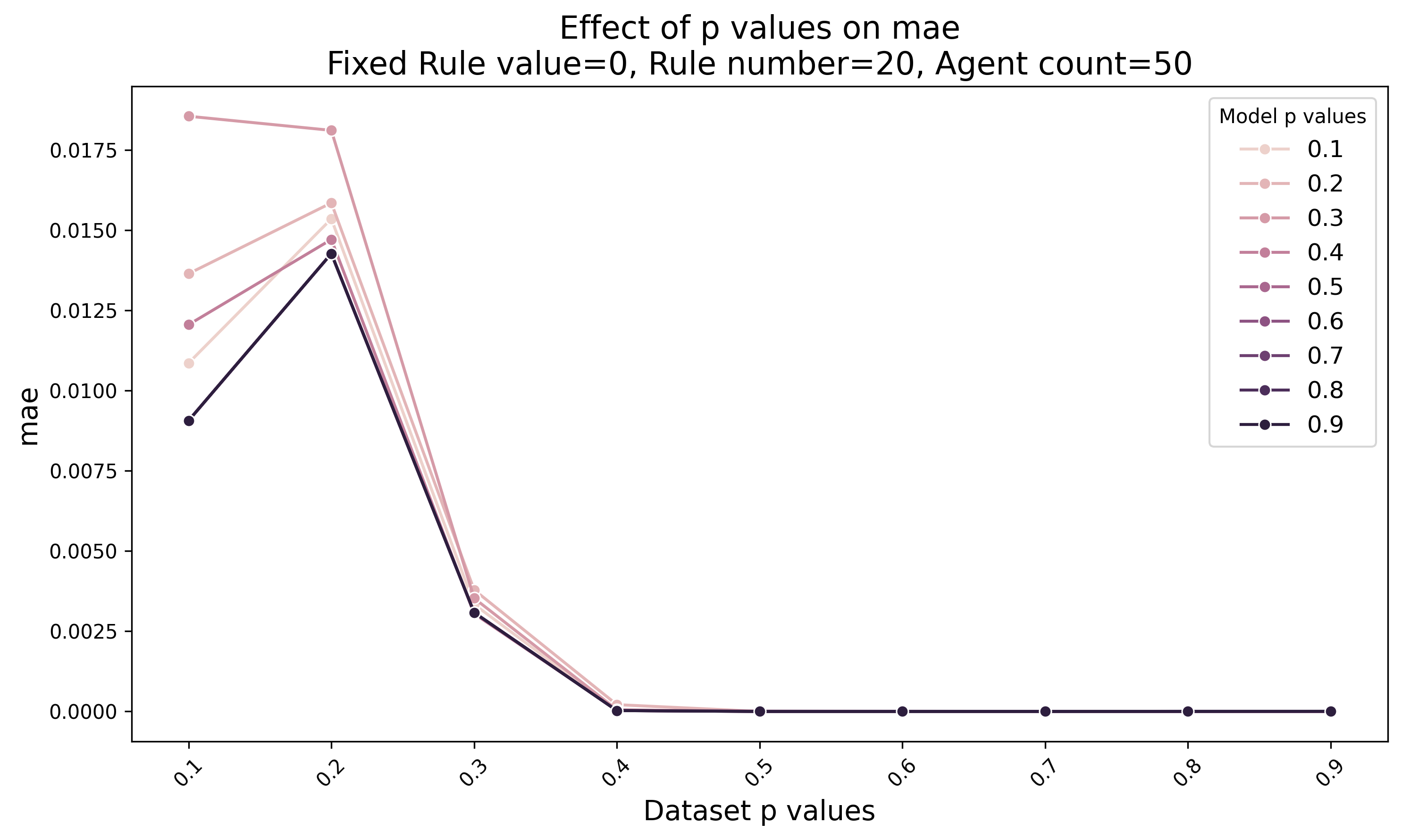}
        \caption{n=50 agents, uniform rules}
        \label{fig:coinflip_subfig3}
    \end{subfigure}
    \\[\baselineskip]
    \begin{subfigure}[b]{0.32\textwidth}
        \includegraphics[width=\textwidth]{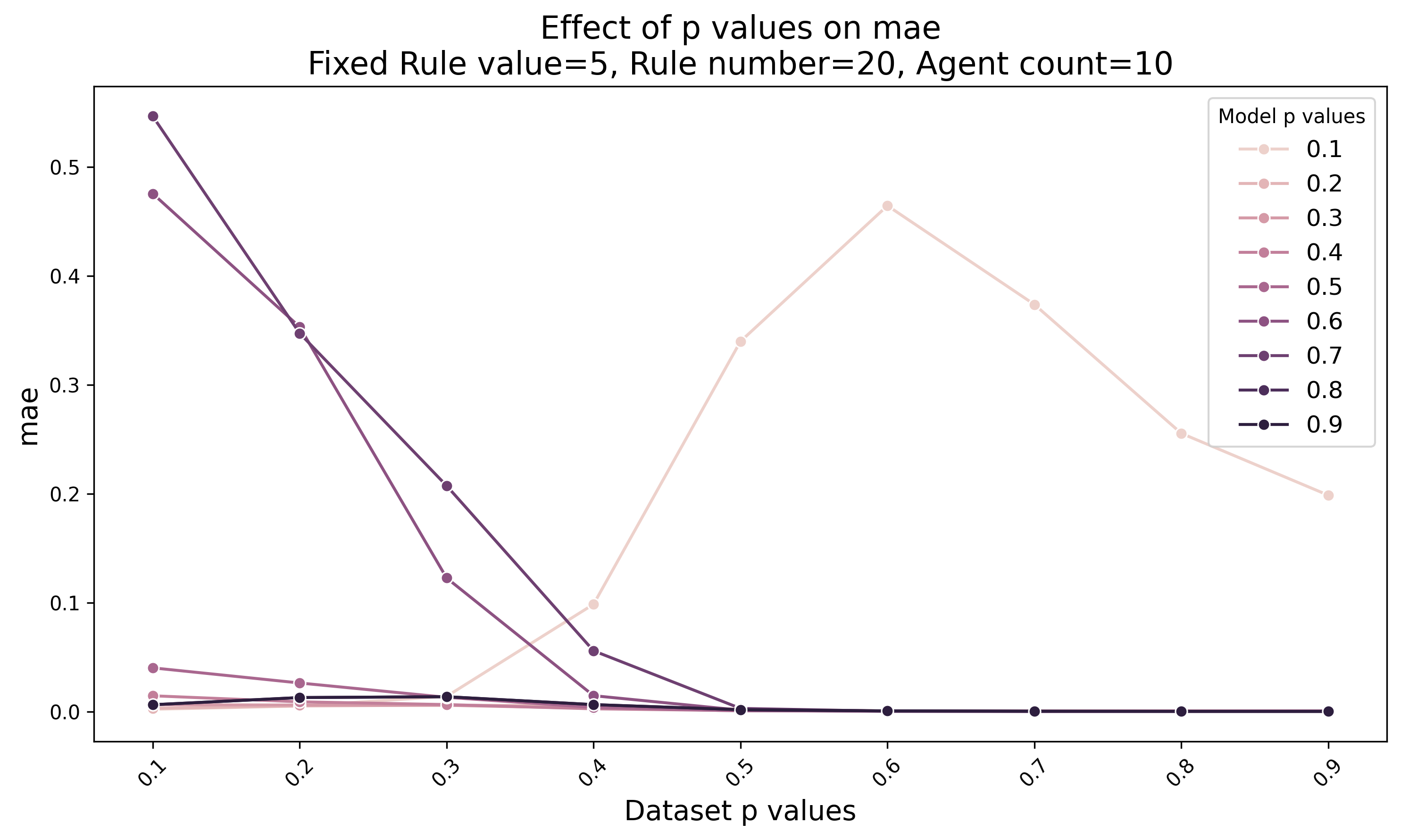}
        \caption{n=10 agents, Low-variance Gaussian rules ($\mu = 5$)}
        \label{fig:coinflip_subfig4}
    \end{subfigure}
     \hfill
    \begin{subfigure}[b]{0.32\textwidth}
        \includegraphics[width=\textwidth]{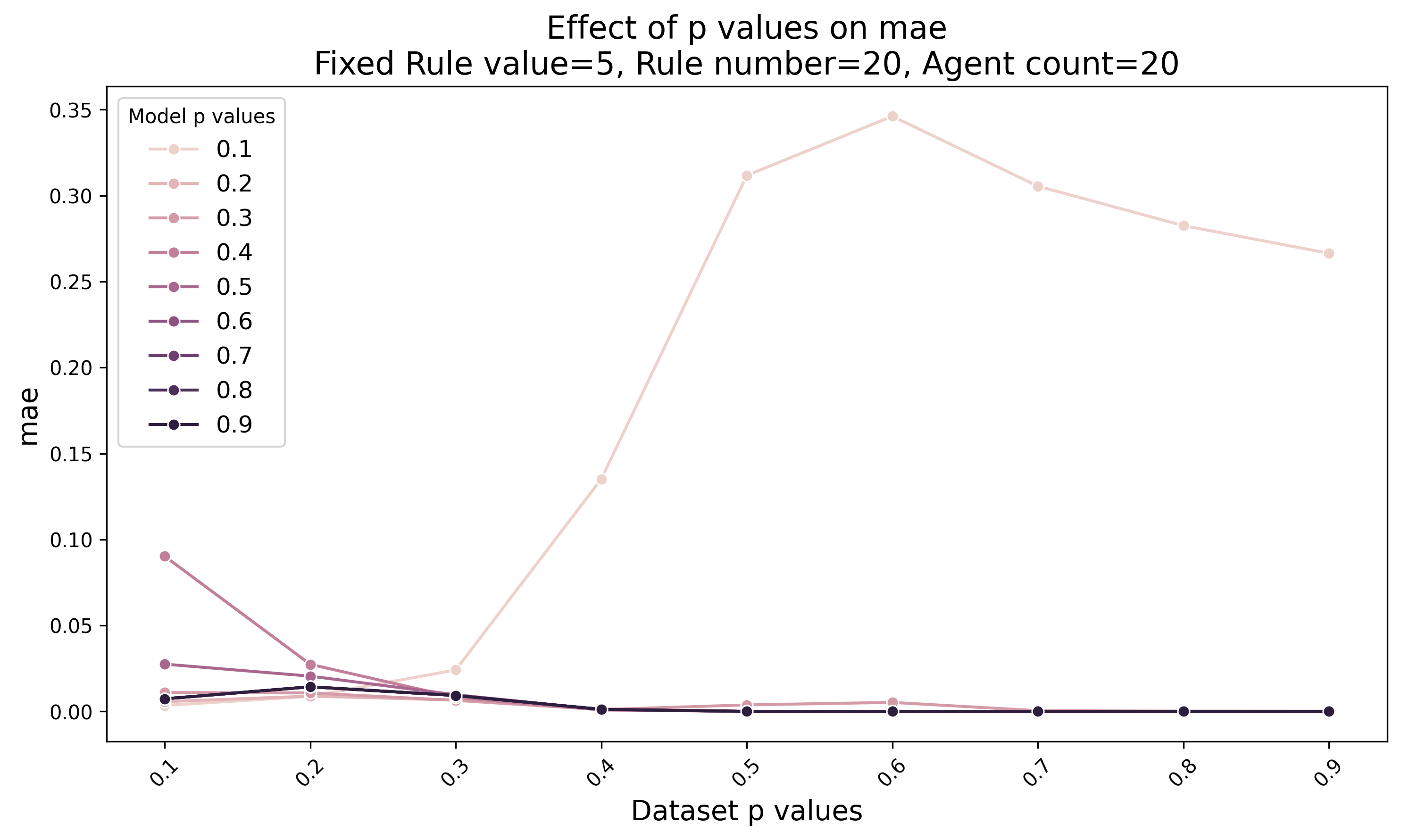}
        \caption{n=20 agents, Low-variance Gaussian rules ($\mu = 5$)}
        \label{fig:coinflip_subfig5}
    \end{subfigure}
     \hfill
    \begin{subfigure}[b]{0.32\textwidth}
        \includegraphics[width=\textwidth]{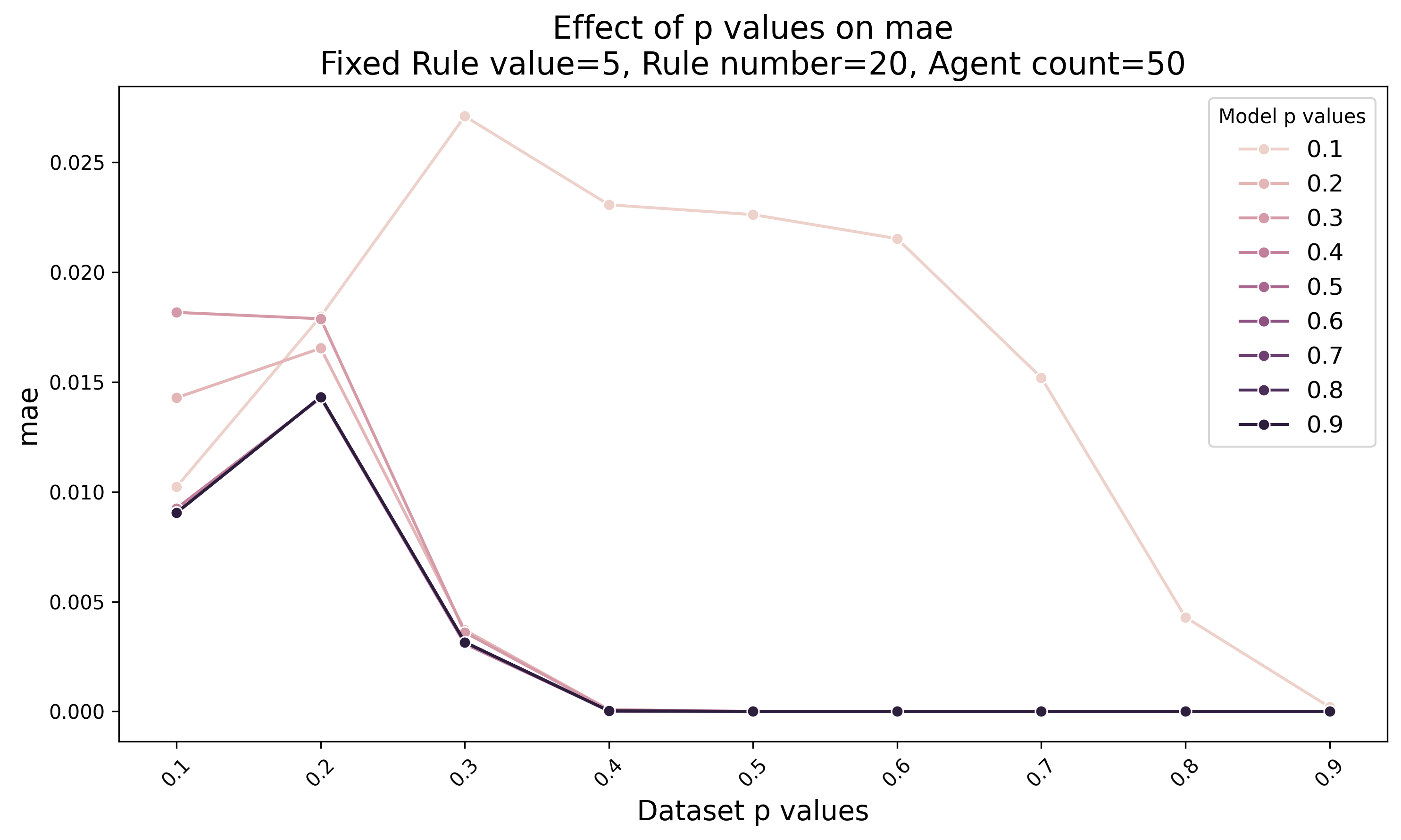}
        \caption{n=50 agents, Low-variance Gaussian rules ($\mu = 5$)}
        \label{fig:coinflip_subfig6}
    \end{subfigure}
    \\[\baselineskip]
    \begin{subfigure}[b]{0.32\textwidth}
        \includegraphics[width=\textwidth]{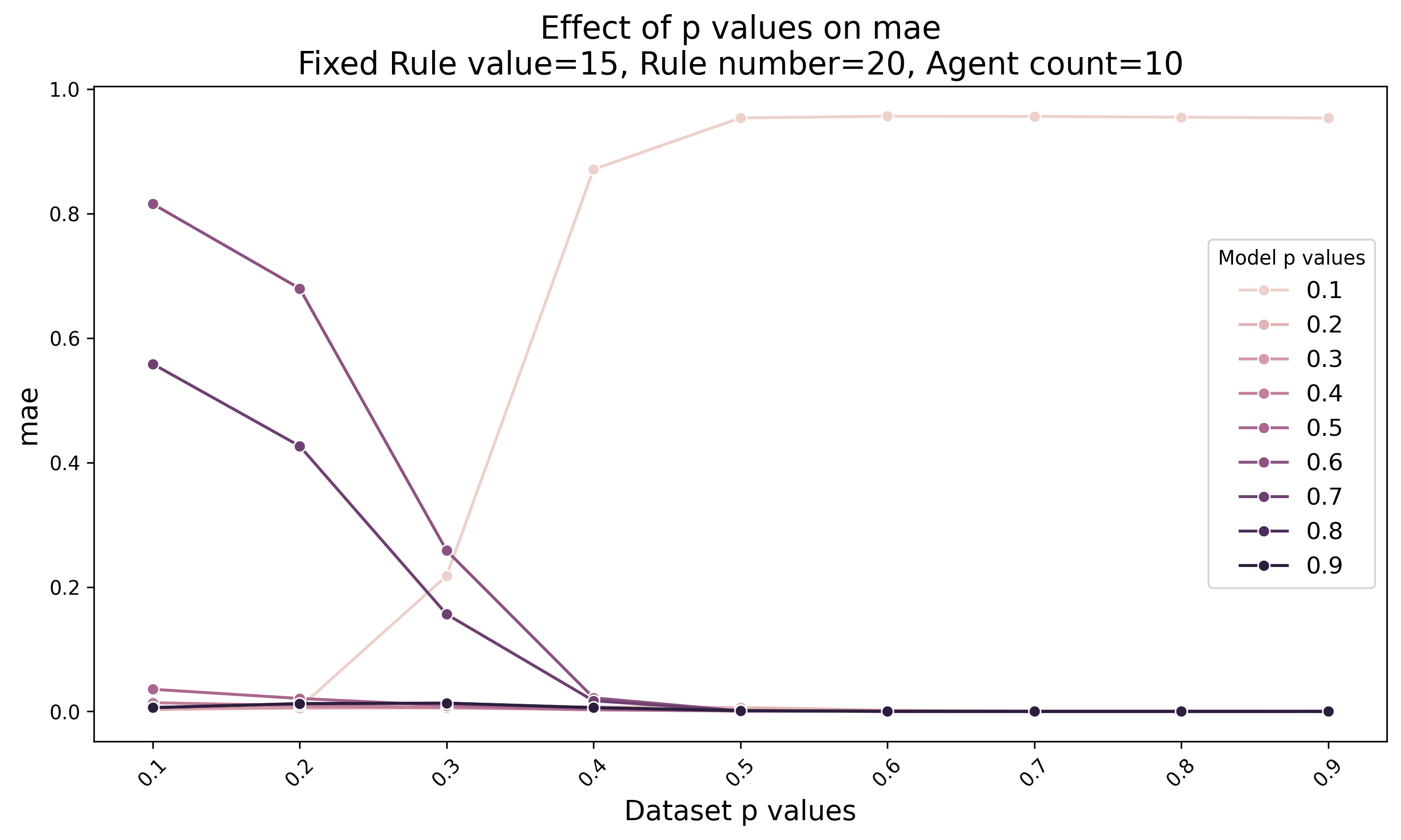}
        \caption{n=10 agents, High-variance Gaussian rules ($\mu = 15$)}
        \label{fig:coinflip_subfig7}
    \end{subfigure}
     \hfill
    \begin{subfigure}[b]{0.32\textwidth}
        \includegraphics[width=\textwidth]{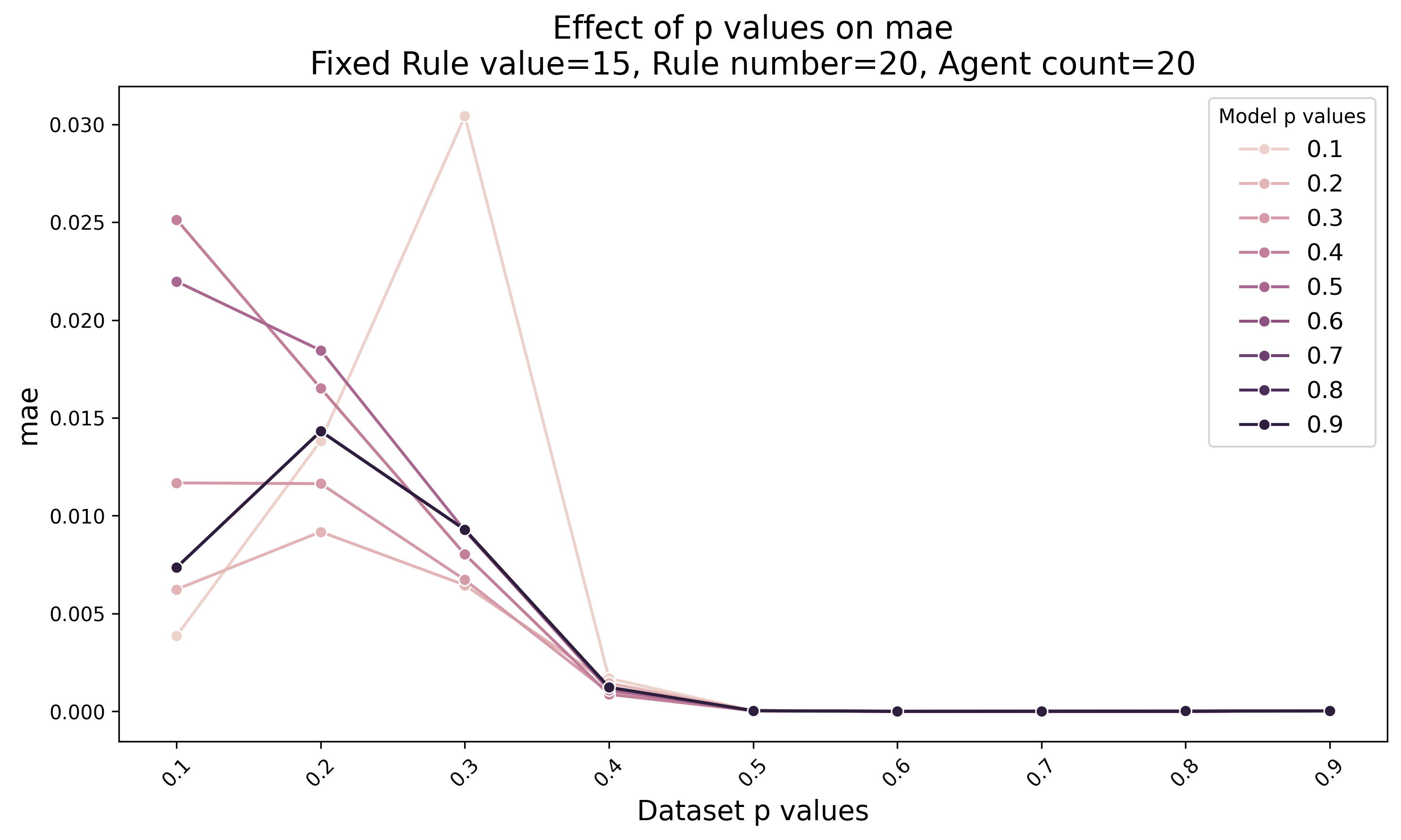}
        \caption{n=20 agents, High-variance Gaussian rules ($\mu = 15$)}
        \label{fig:coinflip_subfig8}
    \end{subfigure}
     \hfill
    \begin{subfigure}[b]{0.32\textwidth}
        \includegraphics[width=\textwidth]{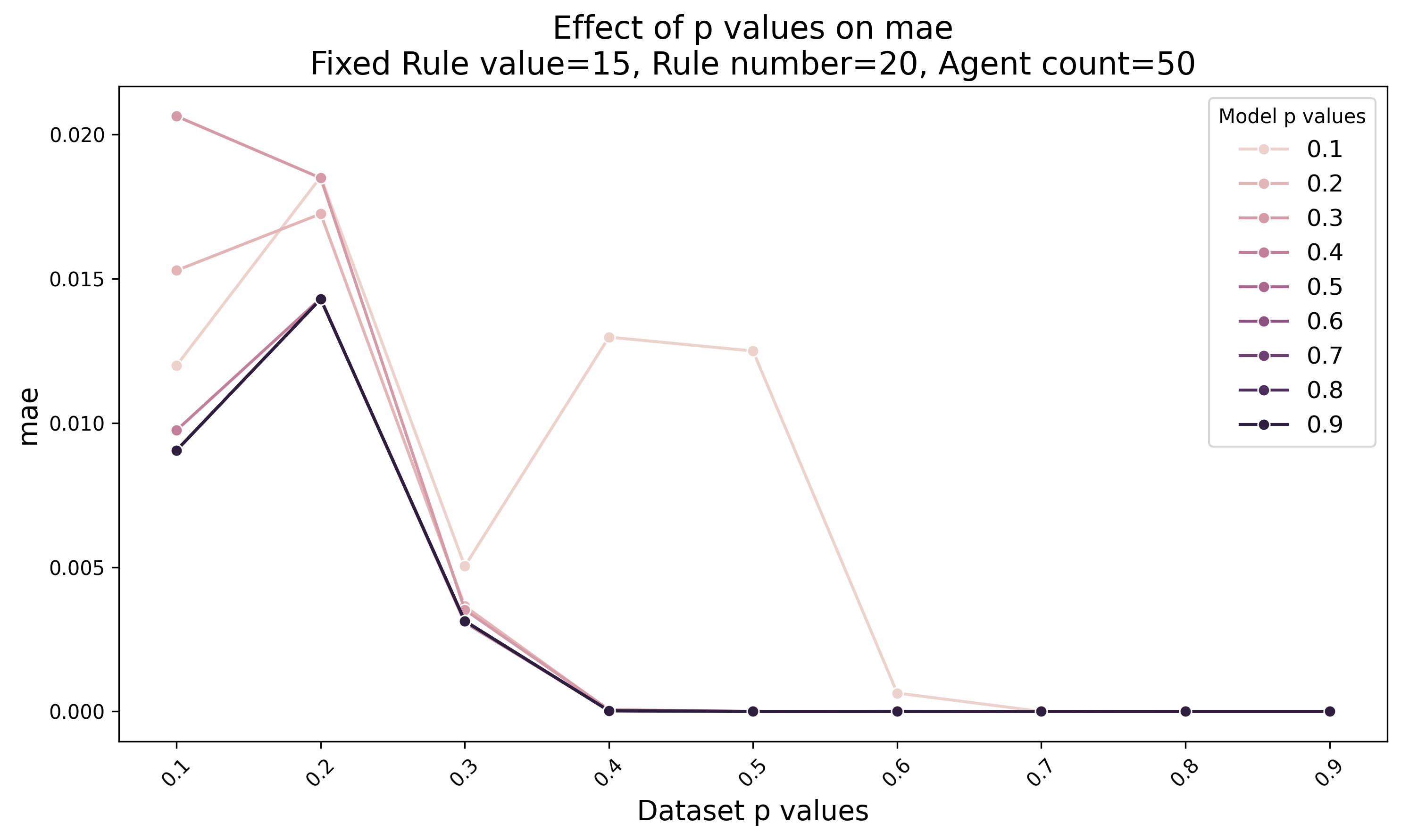}
        \caption{n=50 agents, High-variance Gaussian rules ($\mu = 15$)}
        \label{fig:coinflip_subfig9}
    \end{subfigure}
    \caption{Models performance on Mixture of Gaussian datasets using Mean Absolute Error (MAE) across different dataset configurations. The x-axis represents the sparsity threshold (p), and the y-axis shows the MAE values. Subplots show variations across: (A-C) uniform rule values with 20 rules and 10, 20, and 50 agents respectively; (D-F) low-variance Gaussian distribution ($\mu=5$) with 20 rules across 10, 20, and 50 agents; (G-I) high-variance Gaussian distribution ($\mu=15$) with 20 rules for 10, 20 and 50 agents. Each line within subplots represents a model trained with a different p value while maintaining consistent coalition parameters.}
    \label{fig:mog_cross_p_data_model_same_config_MAE}
\end{figure}

Cross p-threshold comparisons across all MoG datasets are presented in Figure ~\ref{fig:mog_cross_p_data_model_same_config_MAE}. Due to the shifting distributions, the models underperformed, at times significantly, when trying to predict on any conditions outside their trained upon datasets. The high error rate is noted across a wide range of threshold values, especially noticeable around $p= 0.2-0.3$.  
These results indicate how our NN cannot reliably learn multiple changing distributions, which is an integral part of the MoG method. Future research would try using more advanced NN architectures that can handle this kind of dynamic system better.

\begin{figure}
    \centering    
    \begin{subfigure}[b]{0.32\textwidth}
        \includegraphics[width=\textwidth]{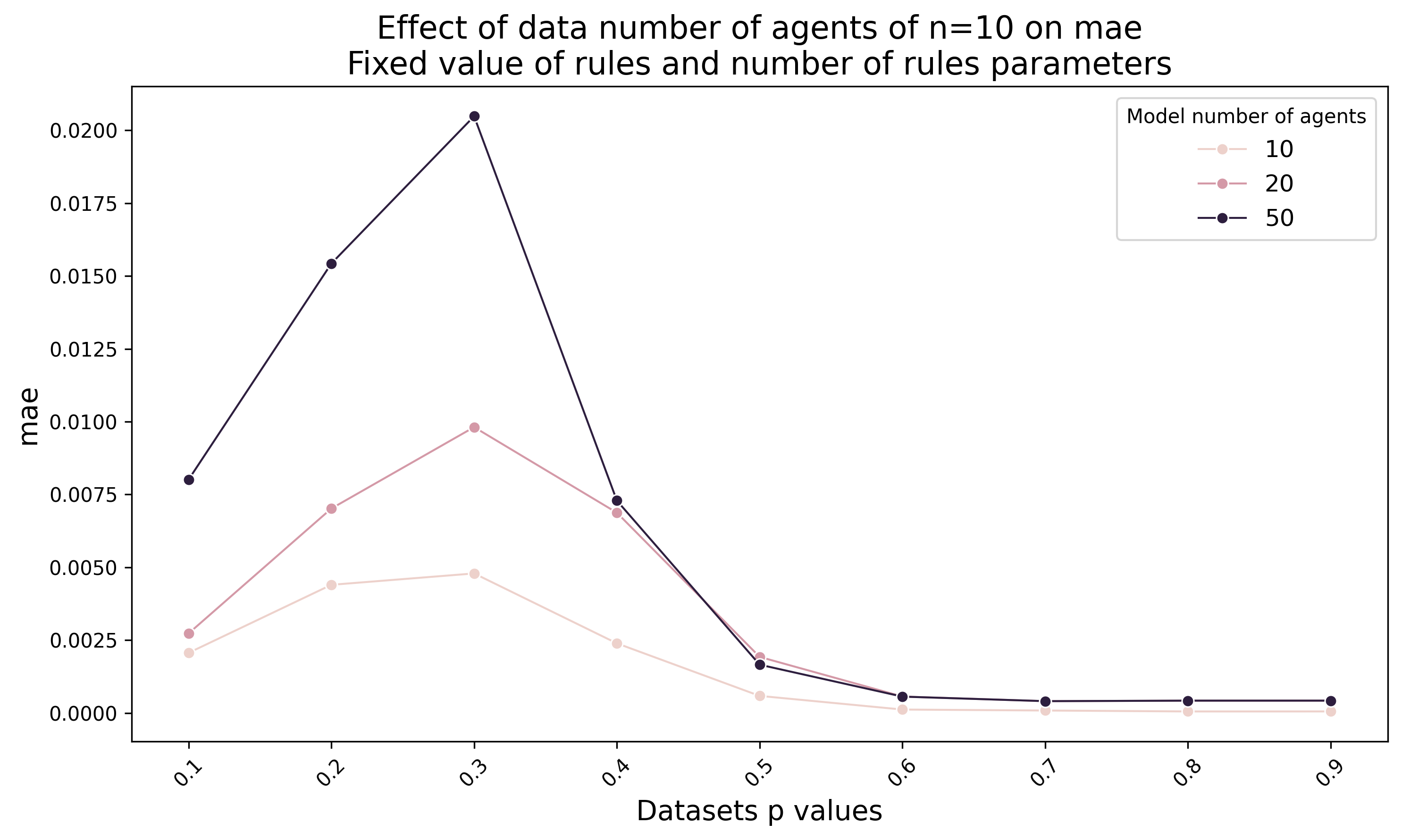}
        \caption{Performance on 10-agent datasets}
        \label{fig:delta_agent_subfig1}
    \end{subfigure}
         \hfill
    \begin{subfigure}[b]{0.32\textwidth}
        \includegraphics[width=\textwidth]{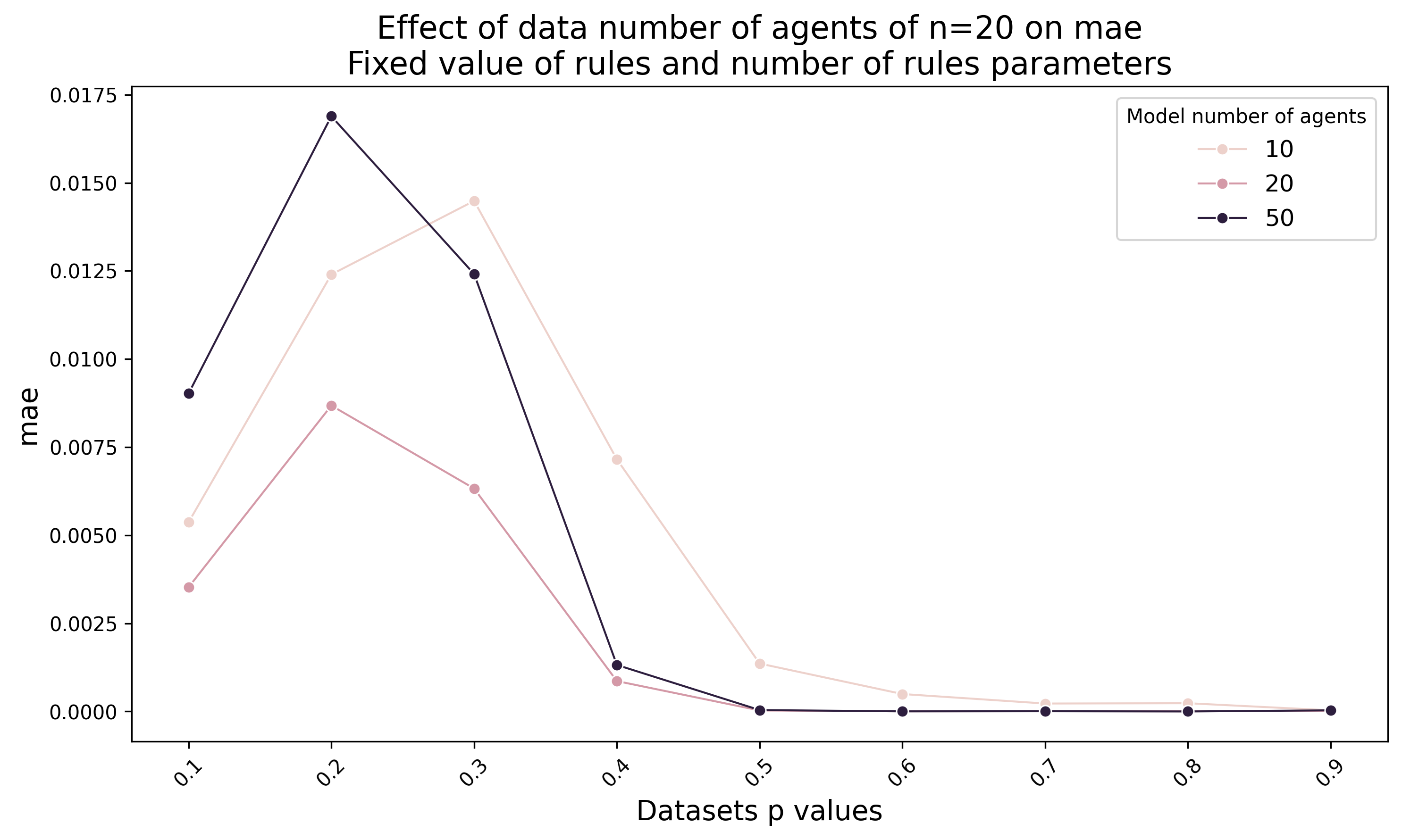}
        \caption{Performance on 20-agent datasets}
        \label{fig:delta_agent_subfig2}
    \end{subfigure}
     \hfill
    \begin{subfigure}[b]{0.32\textwidth}
        \includegraphics[width=\textwidth]{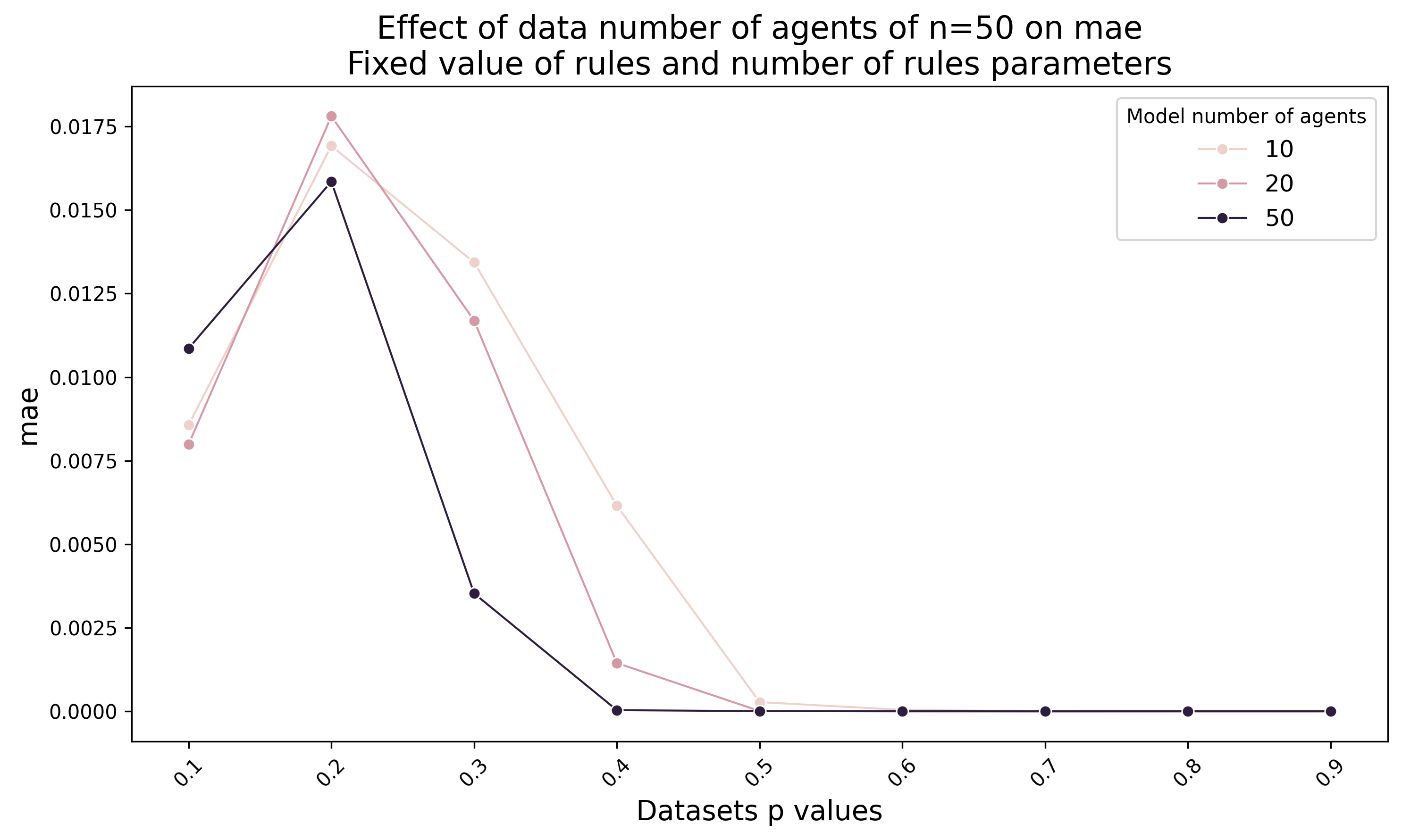}
        \caption{Performance on 50-agent datasets}
        \label{fig:delta_agent_subfig3}
    \end{subfigure}
    \caption{Cross-model comparison of Mean Absolute Error (MAE) for different numbers of agents. Results show model performance on datasets with (A) 10 agents, (B) 20 agents, and (C) 50 agents, while maintaining consistent rule values and number of rules across all configurations. The x-axis represents the sparsity threshold (p), and the y-axis shows MAE values.}
    \label{fig:mog_data_model_differnet_number_of_agnets_MAE}
\end{figure}

When looking at further analysis of the done to evaluate the number of agents on model performance, we see in Figure ~\ref{fig:mog_data_model_differnet_number_of_agnets_MAE} that the same pattern emerges as in the other randomisation method: one that states how the further away the dataset is in number of agents from the number of agents the model was trained on, the worse are the results of the model. This is most evident when comparing the n=10 agent dataset, where error delta is in a factor of 10 at its worse. 

\begin{figure}
    \centering
    \begin{subfigure}[b]{0.32\textwidth}
        \includegraphics[width=\textwidth]{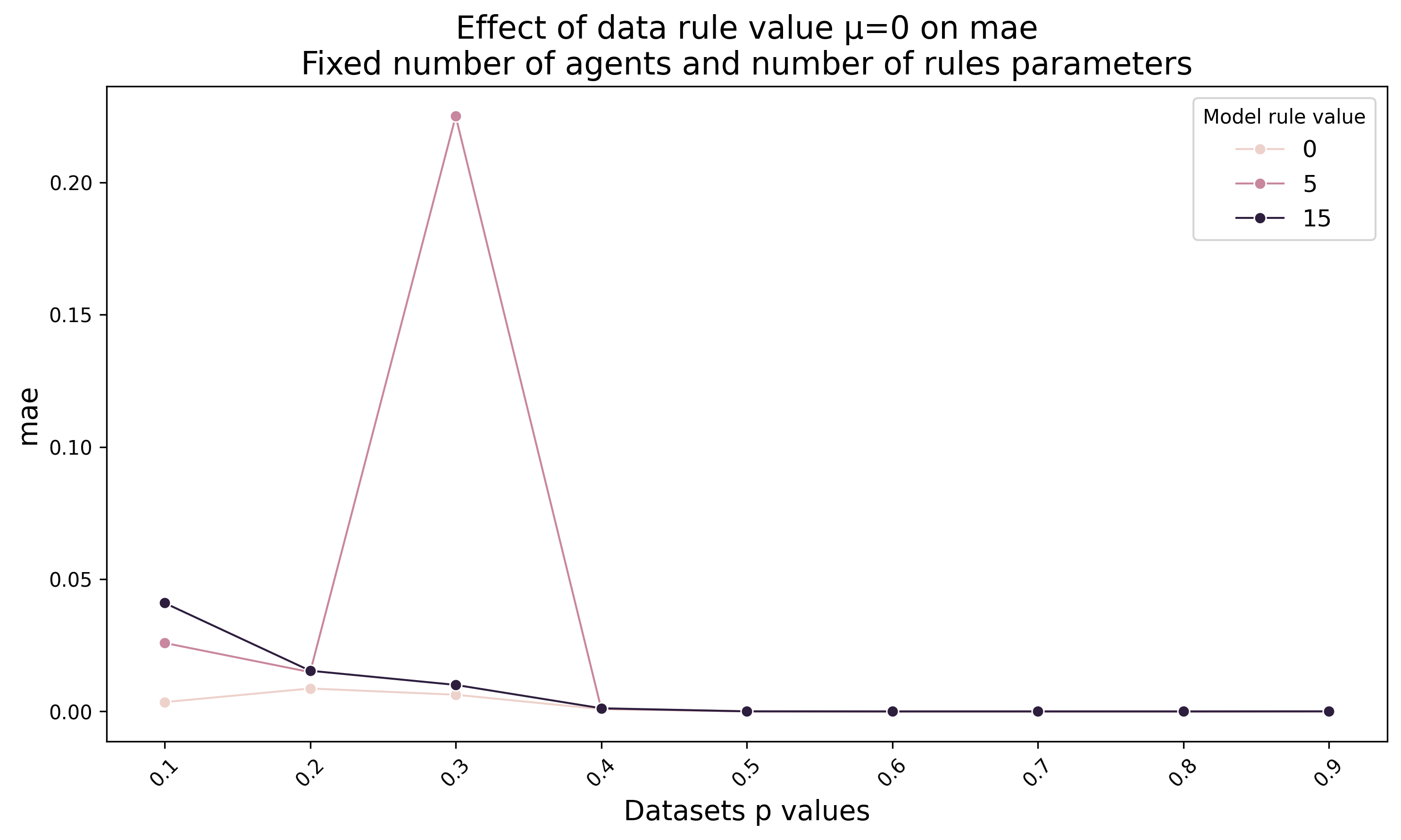}
        \caption{Uniform rule value distribution}
        \label{fig:delta_rule_value_subfig1}
    \end{subfigure}
         \hfill
    \begin{subfigure}[b]{0.32\textwidth}
        \includegraphics[width=\textwidth]{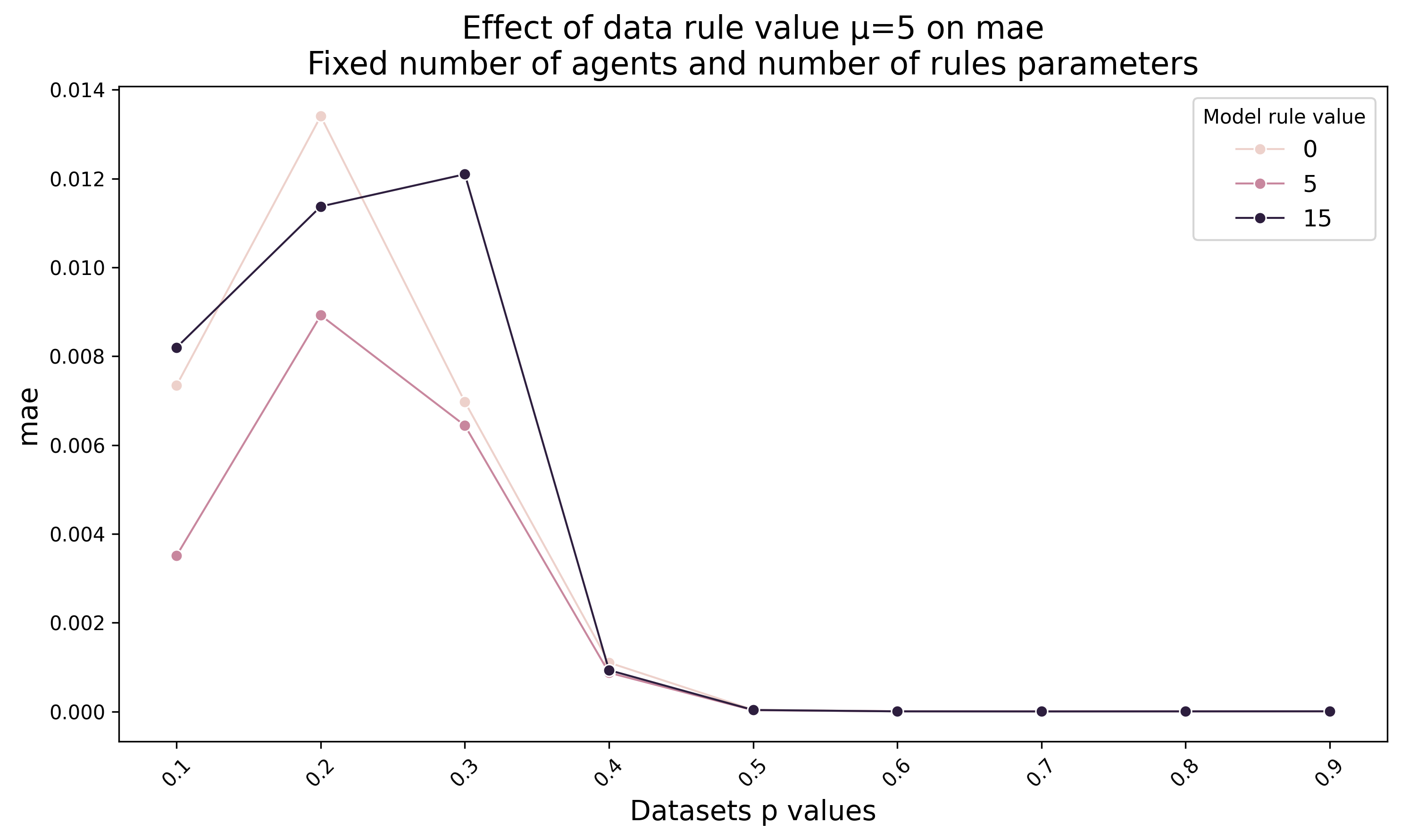}
        \caption{Low-variance $(\mu=5)$ rule value distribution}
        \label{fig:delta_rule_value_subfig2}
    \end{subfigure}
         \hfill
    \begin{subfigure}[b]{0.32\textwidth}
        \includegraphics[width=\textwidth]{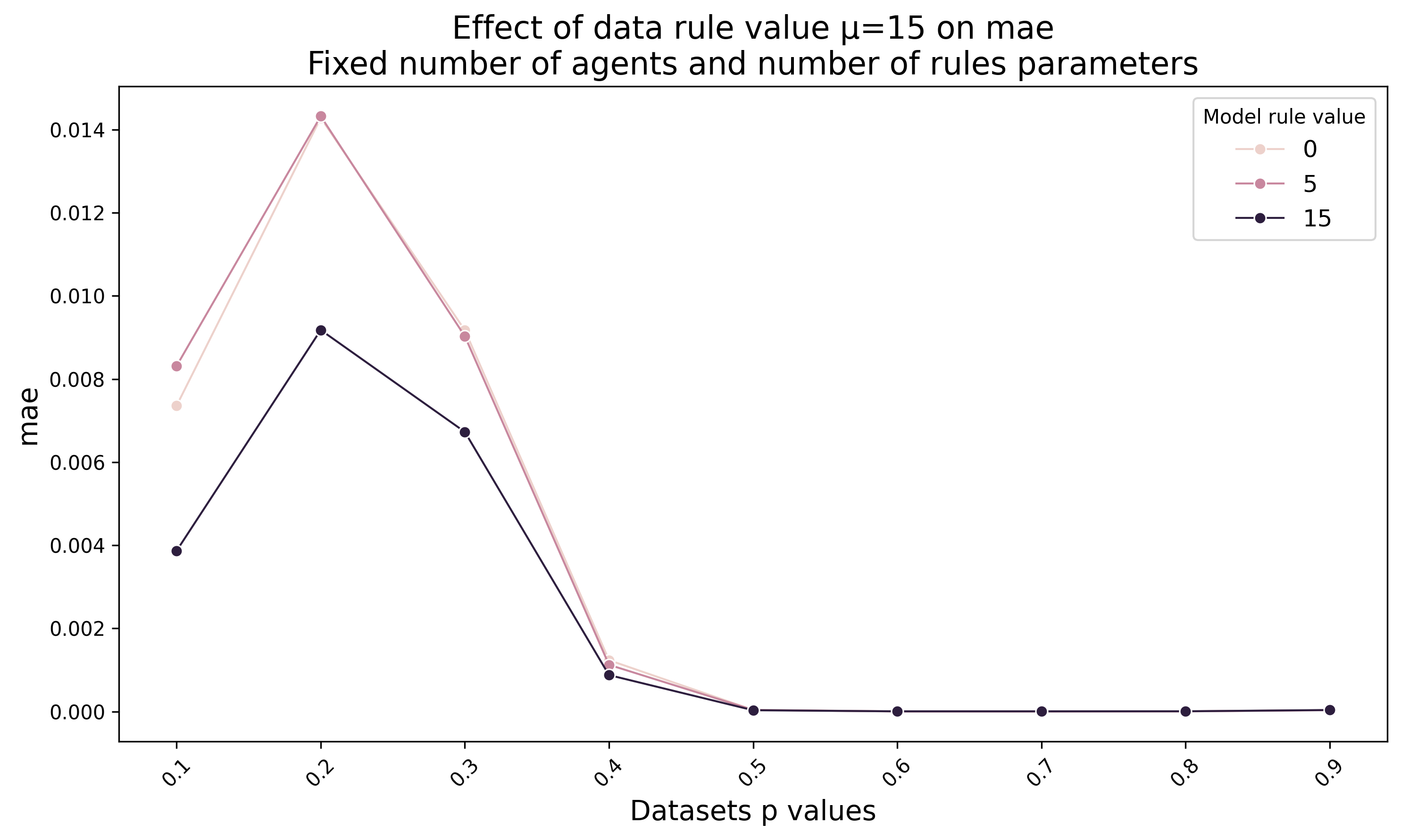}
        \caption{High-variance $(\mu=15)$ rule value distribution}
        \label{fig:delta_rule_value_subfig3}
    \end{subfigure}
    \caption{Impact of rule value distribution on model performance using Mean Absolute Error (MAE). Comparison across (A) uniform rule value, (B) low-variance Gaussian rule value, and (C) high-variance Gaussian rule value distributions. All configurations maintain consistent number of rules and agents. The x-axis represents the sparsity threshold (p), and the y-axis shows MAE values.}
    \label{fig:mog_data_model_different_rule_value_MAE}
\end{figure}

Analysis of rule value variations, presented in Figure ~\ref{fig:mog_data_model_different_rule_value_MAE}, the strong model limitations at the $p=0.2-0.3$ band. Models performed best when evaluated on conditions they were trained on. Error rate is more significant when compared to other randomisation methods, emphasizing how the models struggle with learning the different distributions of the datasets themselves, and more so in the variance in rule scores.

\end{document}